# Agricultural Policy in Ukraine

## English version of the Chapter 2 in the Textbook of Kvasha, S., A. Dibrova, O. Nivievskyi, and P. Martyshev (2022). Agricultural policy. NUBiP, Kyiv[1]

by Oleg Nivievskyi, Pavlo Martyshev, and Sergiy Kvasha

Building upon the theory and methodology of agricultural policy developed in the previous chapter, in Chapter 2 we analyse and assess agricultural policy making in Ukraine since the break-up of Soviet Union till today. Going from top down to the bottom, we begin by describing the evolution of state policy in the agri-food sector. In the beginning, we describe the major milestones of agricultural policy making since independence, paving the way to the political economy of the modern agricultural policy in Ukraine. Then we describe the role of agri-food sector in the national economy as well as globally in ensuring food security in the world. After, we dig deeper and focus on a detailed performance of agricultural sector by looking at farm structures, their land use, overall and sector-wise untapped productivity potential. Modern agricultiral policy and institutional set-up is contained and analysed in details in the section 2.4. A review of the agricultural up- and downstream sectors wraps up this chapter.

**2.1 Major milestones of agricultural policy since independence**

Building upon and continuing previously established milestones since its Independence in von Cramon-Taubadel and Nivievskyi (2009), we identify five main phases of agricultural policy making in Ukraine[2].

**Phase I: 1991-1994.** In this period few market reforms were undertaken. Most key elements of the Soviet system (state procurement of key agricultural products, state provision of inputs, administrative control of product flows, prices and margins) were maintained. In 1992, the collective and soviet farms (kolhozpes and radhozpes) were transformed into so-called collective agricultural enterprises (CAEs). This largely formal change led to little real restructuring in the farm sector. Input supply and food processing remained firmly in state hands. In 1991, a law made private farming possible. By 1994, 32,000 private farms had emerged. This number increased to roughly 43,000 by 2002. However, the private farms remained small (with an average size of under 30 ha in the 1990s, increasing to 66 ha in 2002), and have proven much less potent as a force shaping agricultural policy than the roughly 12,000 CAEs and their successor enterprises.

Altogether, policy followed a very conservative course in this first phase, largely maintaining Soviet-style ownership structures, budget transfers and state regulation of markets. Prices continued to be regulated and on average roughly 10% of corresponding world market prices and this made very large profits to a handful of individuals by purchasing agricultural products such as grain and livestock at very low prices, and selling them on world markets for considerably more. Rents of roughly 4.1 bUS$, or 20% of Ukrainian GDP in 1992 accrued to a handful of individuals who had access to goods and export opportunities. Policy makers responded with a flurry of administrative measures designed to stem such exports (or redirect the proceeds) including, in 1993, export quotas and licensing. Significant rents were also distributed in the form of budget subsidies, including those to agriculture, and subsidised credits to enterprises. In 1993, against a background of significant macroeconomic destabilization, when inflation exceeded 4,700%, state credits were granted at 20% rates of interest and, thus, essentially represented gifts to those who could quickly convert them in to currency or tradable commodities.

---

[1] The texbook production was supported by the **German-Ukrainian Agricultural Policy Dialogue**.
[2] Four phases of agricultural policy development were taken from von Cramon-Taubadel and Nivievskyi (2009) with some minor edits. We extended the Phase IV till 2013 with amendments that took place in agricultural policy making after 2006. Phase V is essentially a modern phase of agricultural policy making in Ukraine.

Agricultural production and especially livestock production also declined dramatically in this first phase, albeit at a slower rate than production in the rest of the economy. The reduction of subsidies led to a rapid increase in input prices and a corresponding deterioration in agriculture's terms of trade. As a result, input use and yields fell dramatically; between 1990 and 1996, mineral fertiliser applications fell from an average of 102.5 to 12.5 kg nitrogen equivalent/hectare, while average grain yields fell from 3.2 t/ha in 1988/90 to 2.3 t/ha in 1994/96. As the economy imploded, agriculture absorbed labour shed by contracting industrial production, and subsistence production of food on household plots became the only feasible survival strategy for many Ukrainians. Household production therefore remained more or less constant through 1994.

**Phase II: 1995-1998.** Several promising reforms were implemented in this period. These were mainly directed at achieving macroeconomic stabilisation by reducing fiscal deficits and their financing via monetary expansion. As a result of these efforts, budgetary transfers to agriculture in Ukraine contracted sharply after 1994, from as much as 11% to roughly 2% of GDP. A number of policy reforms specific to agriculture were also undertaken early in this phase; in late 1994, a legal basis for the distribution of land shares to CAE members was created, and by 1996 most quotas and licensing restrictions on agricultural exports had been eliminated.

Following this promising start, however, agricultural reforms lost momentum, and the years from 1996 to 1998 can accurately be described as wasted. The CAEs proved to be little more than the old kolhozpes and radhospes under new names. While members theoretically had rights to their individual land shares, they had few practical means of exercising these rights, as land sale and rental were forbidden and individual land parcels were not demarcated.

In the food processing industry, a privatisation mechanism that gave supplying farms and the state 51% and 25% shares, respectively, with the rest going to employees and open sales, was introduced in 1996. In so-called 'strategic' areas (for example grain marketing), however, the state's share was often larger, and key enterprises were often exempted from privatisation. As a result, much of the food processing and marketing sector remained monopolistic and inefficient. For key agricultural export products (e.g. grain and oilseeds), inefficient processing and marketing (i.e. transportation and storage) translated directly into depressed farm-gate prices. In 1999, it was estimated that inefficient grain marketing structures were leaving Ukrainian farmers with only roughly 40% of the f.o.b. export price.

In the area of trade policy, the elimination of quotas and licensing restrictions led to little effective liberalisation. Trade controls are valves that make it possible to channel trade flows and any associated rents. While export quotas and licences were eliminated to comply with IMF and World Bank conditionality in 1996, those who had benefited from these restrictions quickly developed alternatives. For example, so-called 'indicative' and 'recommended' prices (minimum export prices) were implemented for many products. Even if these were not officially binding, local customs officials could, depending on who was asking, insist on their application. To avoid costly delays, traders either had to 'resolve' disputes locally with the customs officials in question, or they had to cultivate high-ranking contacts in Kyiv who could 'facilitate' transactions. Beginning with the 1996 harvest, some regional (oblast) authorities declared bans on grain exports, ostensibly to secure payment for inputs that had been delivered in the spring and for tax debts. While the regional authorities had no right to impose such bans, the response of the central government in Kyiv was ambiguous; repeated statements that such bans were illegal were coupled with references to the need to keep the state reserves supplied and to collect taxes and debts. In each of the following three years (1997-99), regional export bans and confiscation of grain and oilseeds were employed in a similar manner.

Under these conditions, private input suppliers were unable to secure payment for their deliveries (foreign agricultural chemical firms had accumulated receivables of roughly 200 mUS$ by late 1999), and private input supply stagnated at very low levels. Together with the government's inability to supply the right inputs at the right time to the right farms, and the low farm-gate prices mentioned above, this caused a rapid decline in crop production in Ukraine in the second half of the 1990s. Livestock production also continued to contract, and by 1999 agricultural output had fallen to 50% of its pre-Independence level. Household production remained more or less constant, but production on the CAEs fell by more than 70% in the 1990s.

Altogether, this second phase of agricultural policy developments was characterised by an imbalance between macroeconomic and sectoral reforms. While a resemblance of macroeconomic stability was regained in the mid-1990s as inflation rates dropped and economic contraction decelerated, macroeconomic reforms were not supported by structural reforms in agriculture and other sectors. Hence, macroeconomic stability formed a thin crust over a rotten core. These imbalances culminated in a financial crisis in September 1998. This crisis was triggered by international developments (Southeast Asia, Russia, Latin America), but the extreme vulnerability of the Ukrainian economy was home-made and some correction was inevitable. The Hryvnia devalued by roughly 45% vis-à-vis the US dollar between the third and fourth quarters of 1998, and by roughly 100% by the fourth quarter of 1999. This provided agriculture with an important impetus, setting the stage for the next phase in the evolution of agricultural policy in Ukraine.

**Phase III: 1999-2000.** The third phase in independent Ukrainian agricultural policy was brief but crucial. In the aftermath of the 1998 financial crisis and following his re-election in late 1999, President Kuchma recognised the need to speed up the reform process, including in agriculture. On December 3, 1999 he signed a Presidential Decree (No. 1529/99 "On Urgent Measures for Accelerating Reformation of the Agrarian Sector of the Economy") that stipulated that all CAEs distribute land shares and restructure to form new entities by no later than April 30, 2000. He entrusted Victor Yushchenko, a reform-oriented former Chairman of the National Bank of Ukraine, with the formation of a new government. One of Prime Minister Yushchenko's first measures was the January 17, 2000 Cabinet of Ministers Resolution "On New Approaches to Supply Inputs to Farms" which stipulated that the government would henceforth supply inputs to farms only on a cash payment basis and which essentially put an end to the state order for grain and other agricultural products.

In 1999 Government of Ukraine introduced substantial tax benefits for agricultural sector that have been the dominant element of the overall fiscal support to agriculture since then (see section 2.4.4 for more details). Tax benefits accrued from a so-called single tax (or Fixed Agricultural Tax before 2015 - FAT) and a special value-added tax regime in agriculture – AgVAT. The FAT is a flat rate tax that now replaces profit and land taxes, but it replaced about 12 other taxes and fees before 2012 (World Bank, 2013) and it left agricultural profits essentially untaxed. According to the AgVAT regime, farmers were entitled to retain the VAT received from their sales to recover VAT on inputs and for other production purposes. Both types of tax benefits are progressive by nature, since they favor or provide disproportionally more support to more productive larger farms thus implicitly favouring large-scale agriculture in Ukraine.

In March 2000, a further law wrote off the debts of farm enterprises that had fulfilled the terms of Decree No. 1529/99. Most former CAEs had done so, and in the process the number of collective farms fell as they adopted new legal forms, primarily partnerships and cooperatives. The distribution of land shares stipulated in Decree No. 1529/99 shifted the ownership structure of agricultural land in Ukraine in favour of private owners. By January 2002, only 4% of the arable land in the country remained in state hands; roughly 30% was privately owned and used by rural residents (private farms and household plots), and over 65% was owned by the members of the former CAEs. Altogether, almost 7 million Ukrainians became owners of land, with average land shares of 4.2 hectares. Accompanying measures to promote the development of a rental market for agricultural land (land rent had been formally legalised by a law passed in October 1998) led to the emergence of a rental market, providing land owners with a new source of income.

Together, these decisions generated considerable optimism in Ukrainian agriculture, and in 2000 much more capital flowed into farming than in earlier years. In 2000 and 2001, for the first time since 1995, Ukraine's agricultural enterprises generated an aggregate profit and agricultural output increased in these years, for the first time since Independence food processing industry also began to grow at this time. In both agriculture and food processing, employment began to fall and wages began to increase. The development of food processing – supported by significant inflows of foreign direct investment and with exports doubling in 5 years – is especially impressive.

**Phase IV: 2001-2013.** The third phase of key reforms was short-lived and gave way to fourth phase of stop-and-go reforms and dirigistic measures. These measures mainly represented attempts to regulate individual products markets such as those for grains, sugar and oilseeds. Decree No. 832 (June 2000) and Law No. 2238-14 (January 2001), for example, required the certification of grain exports, provided for mandatory crop insurance for grain producers, and enhanced the role of the state holding Khlib Ukrainy (Bread of Ukraine), which had been founded in 1996 and continued to control a strategic chunk of Ukraine's grain marketing infrastructure (e.g. elevators at key locations, harbour facilities). These measures were taken against the background of a poor wheat harvest in 2000, which led to a rapid jump in wheat prices from export parity to import parity levels. Due to the political sensitivity of wheat and bread prices, policy makers reverted to their planning ways and attempted to regulate prices and product flows. This pattern of market instability, dirigistic over-reaction and amplified instability was repeated following the very poor grain harvest in 2003, in response to increasing meat and sugar prices in 2005, and again multiple export restrictions on grain markets as world market prices started its high cycle in late 2006. Export restrictions in 2006/07, 2007/08, 2010/11 and in 2011/12 marketing years took the form of either quotas or export taxes. Since 2012/13 export VAT refund was cancelled essentially taxing grain exports (World Bank, 2013). On top of it, voluntary export quotas were introduced. Learning from past experience, grain traders took action to reduce the uncertainty of such restrictions and signed a Memorandum of Understanding with the Ministry of Agricultural Policy and Food where they voluntarily agreed to cap their grain exports at 80 percent of the grain exportable volumes.

Other measures taken in or after 2001 included minimum prices for sugar, and a pledge price system for grains modelled along the lines of the US loan rate system (that has been underfunded and therefore largely ineffective so far). In September 1999, the decision had been taken to introduce a 23% tax on sunflower seed exports, and neither the reform government under Yushchenko nor later governments showed any intention of eliminating this tax. A July 2001 amendment did reduce this export tax from 23% to 17%, but it also closed loopholes that had provided exemptions, thus increasing the effective export tax burden.

Finally, new Land Code that abolished collective land ownership was adopted by the Verkhovna Rada in 2001. At the same time it introduced a moratorium or ban on the purchase and sale of about 38.5 mln ha of agricultural land – a decision that put agriculture and entire economy at a lower development path. It was initially adopted for five years, and subsequently renewed multiple times and is still in place now but is expected to be lifted in July 2021 (KSE, 2021).

The Orange Revolution, which followed controversial presidential elections in late 2004, did not result in major changes in the stop-and-go, generally non-market orientation of agricultural policy since 2000. On the positive side, a number of important steps were taken towards Ukraine's WTO accession that took place in 2008. In particular, important changes in tariff schedules were introduced in mid-2005, reducing tariffs for non-sensitive food and agricultural products, unifying MFN and full tariff rates, increasing the uniformity of tariffs and dropping a number of mixed and specific tariffs. Besides, Ukraine signed a free trade agreement with EFTA States (Iceland, Liechtenstein, Norway and Switzerland) that came into force in 2012.

**Phase V: 2014 – today.** The fifth phase of agricultural reforms was entirely driven by the Assossiation Agreement (AA) agenda with the EU that involved a substantial market and institiution reform agenda. The trade-related content is defined in a Deep and Comprehensive Free Trade Area (DCFTA), which is an important part of the overall Agreement. The Agreement was negotiated in the Phase IV presidency of Viktor Yushchenko and Viktor Yanukovych. The AA was initiated in March 2012 and it was due to be signed at the EU's Vilnius summit in November 2013. Contrary to Ukraine's people expectations, it was not signed at the last minute, thereby triggering the Maidan uprising and resulting in Russia's annexation of Crimea and its hybrid war in the eastern Donbas region. Muddling through various hurdles, the AA was eventually signed

after the Revolution of Dignity and came into force on Septmeber 1st, 2017 (Emerson and Movchan, 2018)

DCFTA is designed to deepen Ukraine's access to the European market and encourage further European investment in Ukraine, facilitating closer economic integration in the overall context of a political association. DCFTA foresee some trade liberalization, but the major component relevant for agri-food sector include institutional approximation (World Bank, 2013; a detailed guide and description of the EU-Ukraine relations framework is available in Emerson and Movchan, 2018), such as:

- technical regulations on industrial products, standards and conformity assessment procedures (standardization, conformity assessment, market surveillance, metrology and accreditation concerning the provisions, regulating circulation of industrial products in line with the EU acquis);

- sanitary and phytosanitary (SPS) measures (ensuring a gradual approximation of the Ukrainian SPS food and feed, animal health and welfare legislation and practice to that of the EU, including legislative approximation, capacity building and implementation, among others in the area of food and feed safety, animal health and welfare, traceability, and audits exercised by the controlling bodies); and

- agriculture and rural development (through enhanced agricultural policy dialogue, improved competitiveness and quality improvement schemes).

In addition, Ukraine signed a free trade agreement with Canada and Israel (Movchan, 2020). The main Ukrainian food products benefiting from the free access to Canadian market are sunflower oil, sugar and confectionery, baked goods and alcohol. On the other side, Canada can increase the export of canola oil, beef, frozen fish and various types of processed foods.[3]

Revolution of Dignity in 2014, resulting Crimea annexation by Russia and a war with Russia in the East of Ukraine put a country under a substantial macroeconomic and fiscal pressure in 2014 that required a support from international partners and donors to ensure macroeconomic stability of the country. On the other hand it opened a window of reforms' opportunities that the support of donors (primarily from IMF) was conditioned upon. As far as agri-food sector is concerned, the major steps were adoption of the flexible exchange rate policy, inflation targeting policy, drastic cleaning up of the domestic banking sector and abolishing AgVAT tax regime in 2016.

These developments also paved the way to a substantial land reform agenda, culminating in lifting of the land sales moratorium (the land turnover law # 552-IX as of March 31, 2020), land governance decentralization and deregulation in 2021 (KSE, 2021)

---

[3] https://www.international.gc.ca/trade-commerce/trade-agreements-accords-commerciaux/agr-acc/ukraine/canada-ukraine.aspx?lang=eng

## 2.2 Ukraine's growing role in contributing to global food security

**Ukraine has been increasingly supplying the world with its agri-food products.** Over the last two decades, Ukraine's agri-food exports increased by more than 10 times in nominal terms (though with some short run fluctuations) and by 5.5 times in real terms (see Figures 2.2.1 and 2.2.2). At the same time, Ukraine's agri-food imports increased only by 4.5 times in nominal and by 2.4 times in real terms, leaving Ukraine as a net exporter of agri-food products. Since 2001, the average annual growth for agri-food exports was 14.6% in nominal and 10.5% in real terms.

**Figure 2.2.1. Nominal agri-food trade in Ukraine 2001-2018**

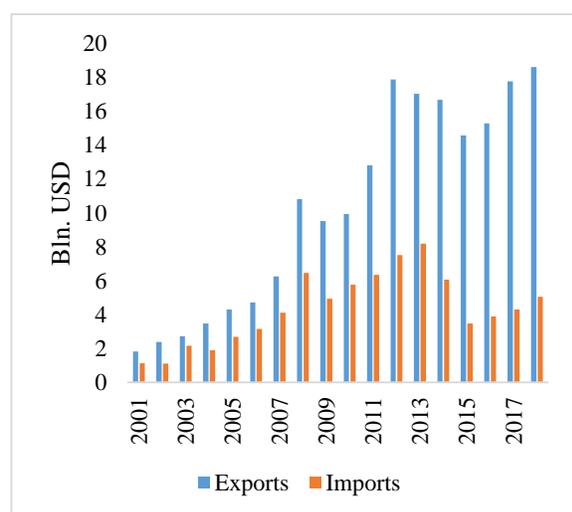

Source: own presentation using Ukrstat trade data

**Figure 2.2.2. Real agri-food trade in Ukraine 2001-2018 (in 2010 prices)**

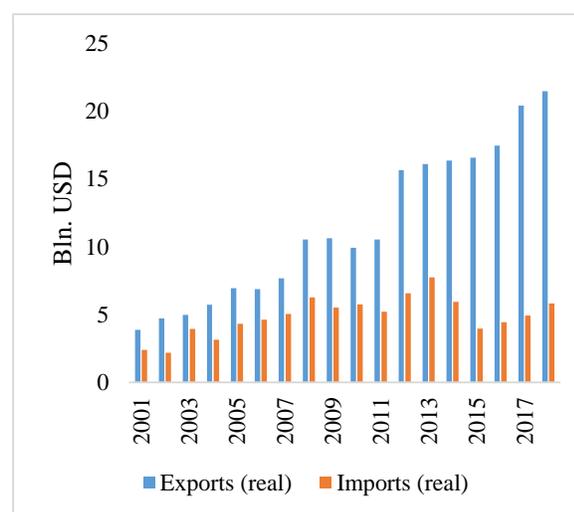

Source: own presentation using Ukrstat and WDI commodity agricultural price indexes data

**The structure of Ukraine's agri-food exports has been increasingly dominated by agricultural raw products.** For instance, the share of food items (commodity group codes 16-24 in the Harmonized Commodity Description and Coding System - HS ) in total agri-food exports decreased from the average 27% over 2001-2003 to 16% over 2016-2018. Over the same time period, the shares of the main commodities increased from 6.6% to 11% for oilseeds, from 27.7% to 38.4% for grains (with increased domination of wheat and maize), and from 15.7% to 25.3% for sunflower oil (see Figure 2.2.3).

**Ukraine's exports of its top agricultural products (oilseeds, vegetable oil, and grains) capture increasing shares of world trade in declining markets**. Ukraine's oilseeds and grains in particular, and vegetable oils on a margin, are all product groups classified as "winners in declining sectors" (see Figure 2.2.4), that is, products for which Ukraine has growing shares in world exports, which themselves are declining. This might be a bad signal for further growth perspectives for these agricultural commodities.

**Figure 2.2.3. Top agri-food export products in Ukraine 2001-2018, nominal**

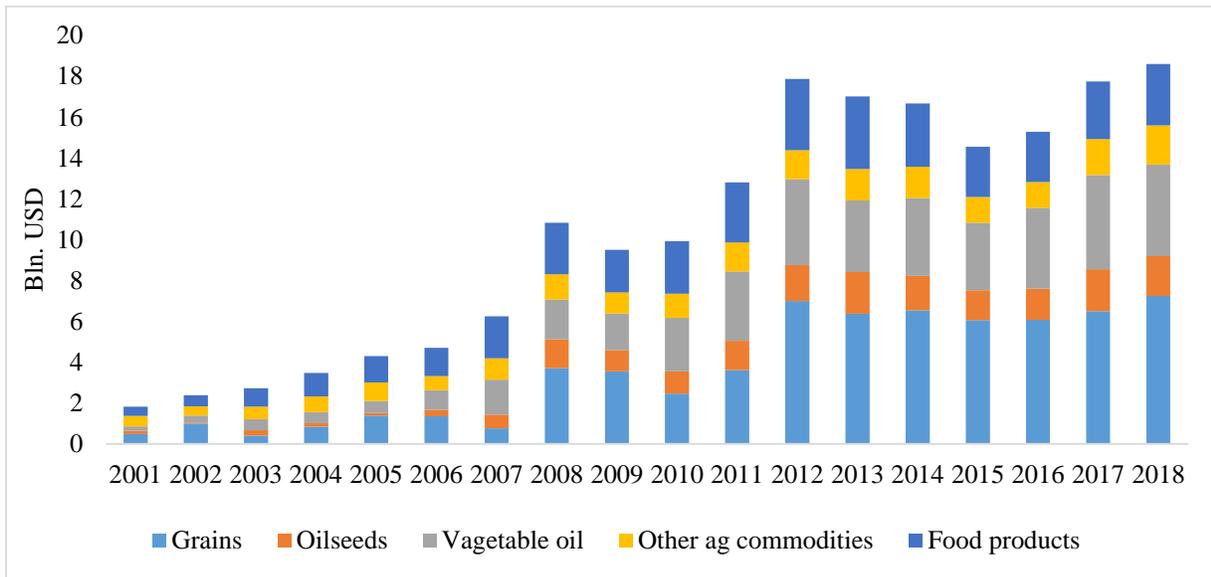

Source: own presentation using Ukrstat data

**Figure 2.2.4. Ukraine's export versus world import growth on world markets**

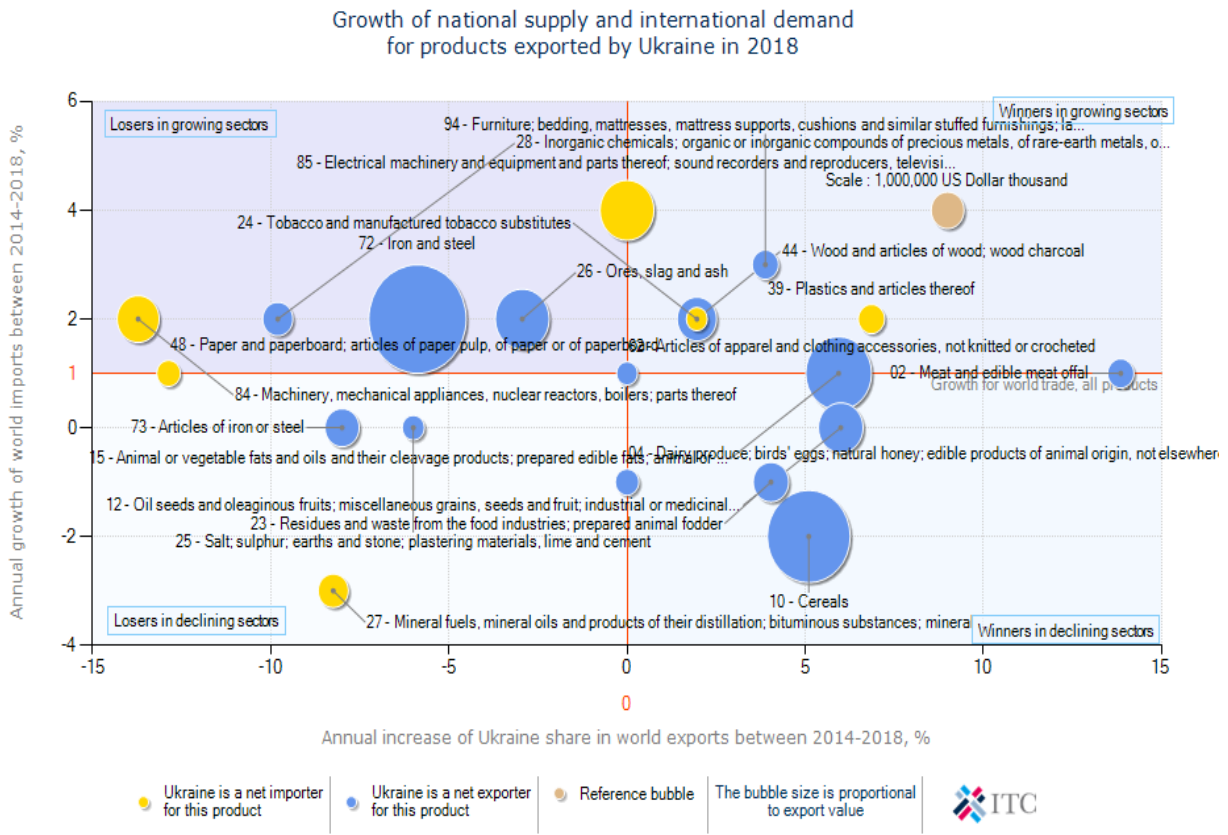

Source: ITC calculations based on UN COMTRADE and ITC statistics. Note: the graph shows the annual increase of Ukraine's share in world exports against the annual growth of world imports, 2014-2018, product groups in 2-digit Harmonized System

**For its leading export commodities (grains and sunflower oil), Ukraine has emerged as an important supplier of global markets**. Although Ukraine's grain exports to world markets showed some notable fluctuations, they were typically found among the top global suppliers. For oilseeds, Ukraine captures even larger shares of world exports, and for sunflower oil, Ukraine has captured the position of the world's leading exporter (see Figure 2.2.5).

**Figure 2.2.5. Ukraine's leading export commodities: export volumes and global export shares**

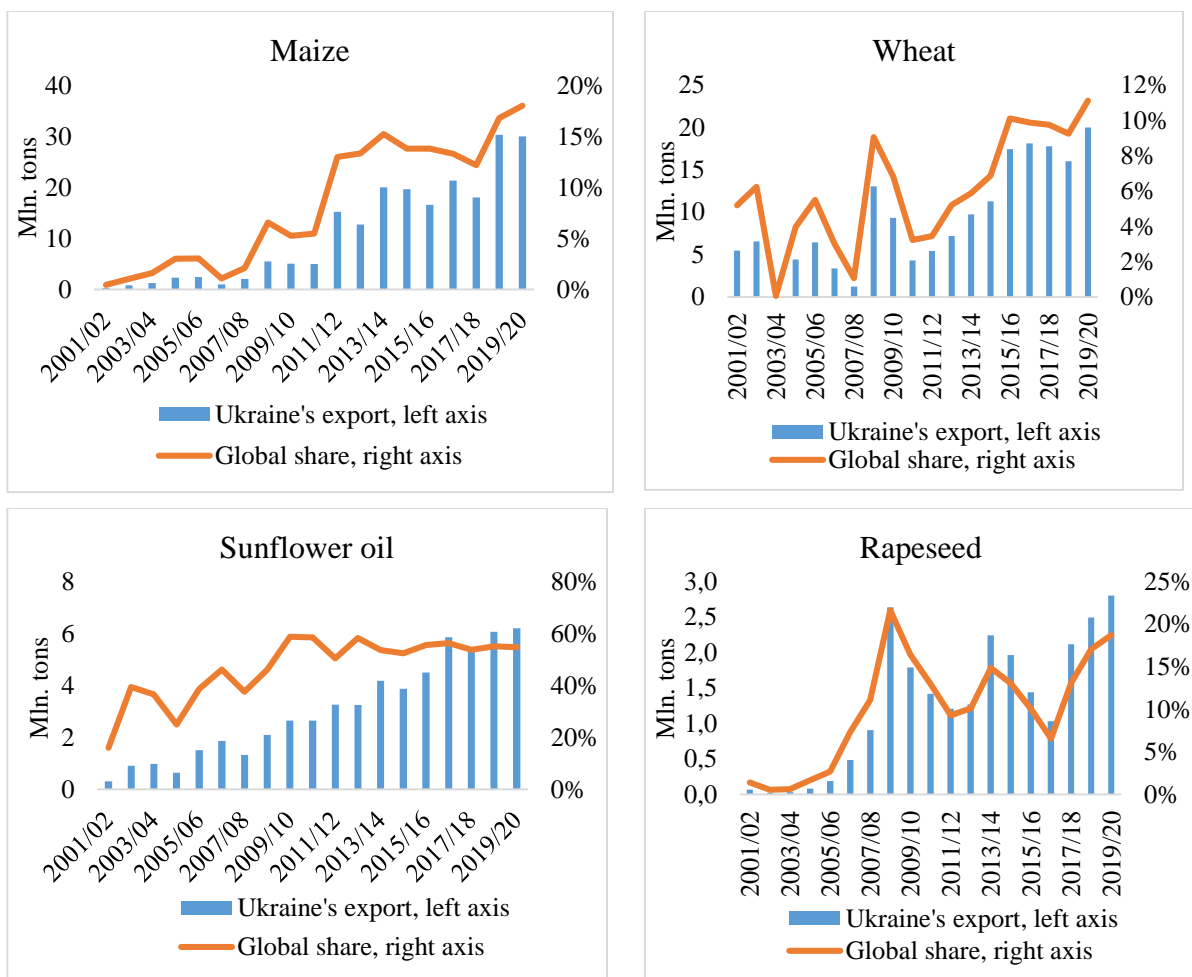

Source: own presentation using USDA PSD data

**Ukraine supplies some of the largest and most dynamic markets with its key agricultural commodities.** Asia has been the main exporting market for Ukrainian agri-food products, providing more than 45% of agri-food export revenues for Ukraine. The main partners in Asia are India, Turkey and China, almost a half of sunflower oil is supplied to India (see Figure 2.2.6). Russia has substantially lost its weight as a trading partner due to the conflict at the east of Ukraine.

EU is the second top export market for Ukraine, earning around 32% of agri-food export revenues. The main partners are the Netherlands, Spain and Italy. The third biggest export market is Middle East and Northern Africa, making up 14% of Ukraine's agri-food export revenues. The main partners are Egypt, Tunisia and Morocco.

**Globally, Ukraine's agri-food sector plays significant and growing role in ensuring global food security.** In terms of grains exports, its contribution to global food security is equivalent to feeding about **332 million people** (in addition to its own population), which is comparable to current US population. Only about two decades ago Ukraine could feed only about **40 million** people[4]. The figures are certainly simplifications and do not account for other important Ukraine's agri-food commodities[5] (vegetable oils, dairy products, oilseeds). Still taking into account that cereals are at the core of human nutrition and that additional 776 million people will be living on the planet in 2023 (OECD/FAO, 2014), the role of Ukraine for global food security is difficult to underestimate.

**Figure 2.2.6. Major export markets supplied by top Ukraine's export commodities in 2018, in % of total export value**

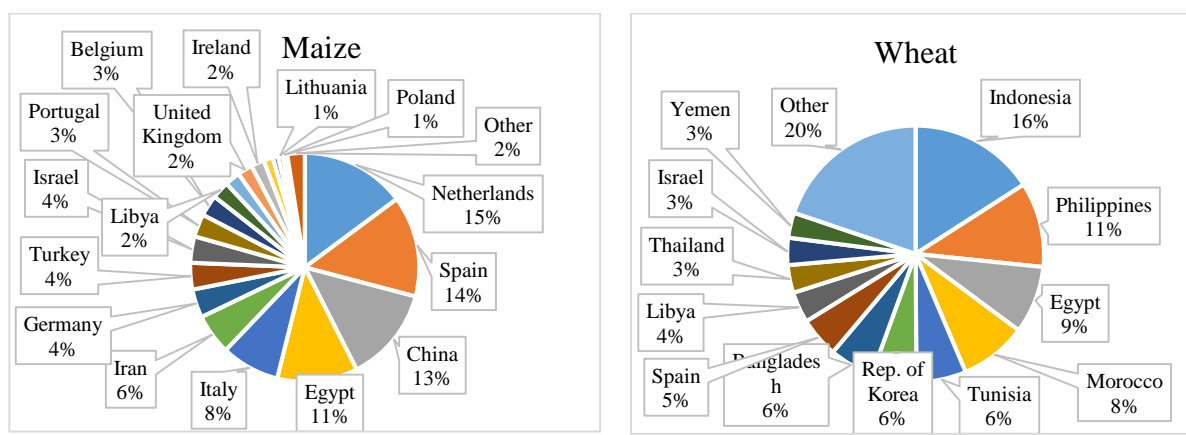

---

[4] Exports of the main Ukraine's cereals (wheat, barley, and maize) were on average 56.5 m tons in 2019/20 MY, and 7 m tons in 2001/02 MY. Based on the destinations of Ukraine's cereal exports and cereals consumption per capita (FAOSTAT), we assumed that average consumption of Ukraine's cereals was 170 kg/person/year. This converts 56.5 m tons of cereals' exports (7 m tons on average in 2001/02 MY) into 332 m people in 2019/20 (40 m people in 2001/02 MY).

[5] Agri-food commodities are the commodities contained in the groups 1-24 in the Harmonized System (HS) of tariff nomenclature (see http://comtrade.un.org/db/mr/rfCommoditiesList.aspx).

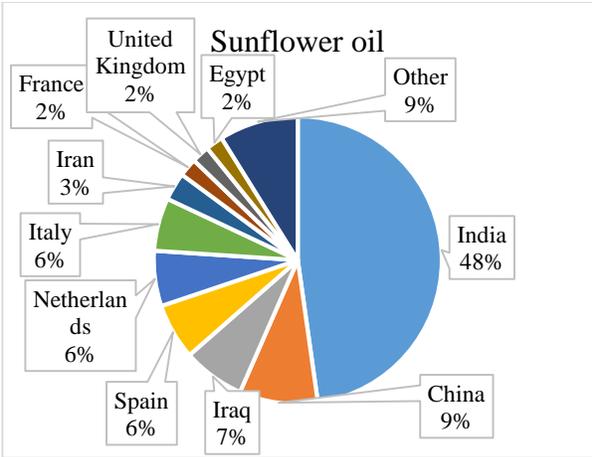

Source: own presentation using UN COMTRADE trade statistics

## 2.3 Role of agriculture in the economy of Ukraine

### 2.3.1 Macroeconomic developments and context

**Ukraine's economic growth has been uneven since the country's independence in 1991 and was influenced by its strong dependence on external factors, the global financial crisis of 2008, and the recent internal conflict that began in 2014 (World Bank, 2020).** Over the last two decades, Ukrainian economy experienced the periods of boom (2004/08), bust (2008/09), tepid recovery (2010/11), stagnation (2012-2013), recession (2013/2015), deep drop (2014/2015) and tepid recovery (2015/today). During the 2000-2007 period, Ukraine sustained high annual growth rates of seven percent on average. Capital inflows surged and credit growth boomed, fueled by the external borrowing of the commercial banks. This enabled an expansionary fiscal policy that resulted in the accumulation of a structural deficit. Growth was almost entirely driven by the favorable external conditions but was not supported by the structural reforms needed to sustain long-term growth. With the onset of the global financial crisis in 2007, the Ukrainian economy contracted rapidly. During 2008-2013, the dynamic of economic growth was negative, averaging -0.7% annually. The Maidan revolution of 2013/2014, the events in Crimea in March 2014, and the armed conflict in the east of the country since 2014 resulted in a continued severe recession, and the economy contracted by 16% during 2014-2015. The Government of Ukraine subsequently undertook a series of fiscal and business environment-related reforms that helped stabilize the economy. Economic growth resumed albeit at a modest rate. Real gross domestic product (GDP) grew by 2.5 percent in 2017 and by 3.4 percent in 2018 (Table 2.3.1). The reasons of this upturn are the increase of remittance inflows from the labor migrants in EU countries, higher public sector wages and pensions, revamp of consumer lending.

**Table 1.3.1. Main macroeconomic indicators in Ukraine**

|  | 2009 | 2010 | 2011 | 2012 | 2013 | 2014 | 2015 | 2016 | 2017 | 2018 |
|---|---|---|---|---|---|---|---|---|---|---|
| Average population (mln. persons) | 46.1 | 46 | 45.8 | 45.6 | 45.6 | 45.4 | 42.9 | 42.8 | 42.6 | 42.4 |
| Real GDP growth rate (% change on previous year)* | - | - | 5.5 | 0.2 | 0.0 | -6.6 | -9.8 | 2.4 | 2.5 | 3.4 |
| GDP at current prices (bln. UAH) | 913 | 1,083 | 1,302 | 1,411 | 1,455 | 1,567 | 1,979 | 2,383 | 2,983 | 3,559 |
| GDP per capita at current prices (thds. UAH) | 19.8 | 23.6 | 28.5 | 31.0 | 32.0 | 35.8 | 46.2 | 55.8 | 70.2 | 84.2 |
| GDP per capita at purchasing power (current prices; thds. USD) | 7.3 | 7.7 | 8.3 | 8.5 | 8.7 | 8.8 | 8 | 8.3 | 8.7 | 9.3 |
| GVA at current prices (bln. UAH) | 831 | 950 | 1,127 | 1,214 | 1,285 | 1,384 | 1,685 | 2,019 | 2,519 | 3,016 |
| GDP deflator (%) | 13.0 | 13.8 | 14.3 | 8.1 | 4.3 | 15.9 | 38.9 | 17.3 | 22.1 | 15.4 |
| Inflation (annual average, % change on previous year) | 15.9 | 9.4 | 8.0 | 0.6 | -0.3 | 24.9 | 43.3 | 12.4 | 13.7 | 9.8 |
| Total employment (mln. persons) | 20.2 | 20.3 | 20.3 | 20.4 | 20.4 | 19 | 17.4 | 17.3 | 17.2 | 17.3 |
| Unemployment rate (%) | 9.6 | 8.8 | 8.6 | 8.1 | 7.7 | 9.7 | 9.5 | 9.7 | 9.9 | 9.1 |
| Current account balance (% of GDP) | -11.7 | 3.7 | -1.5 | -2.4 | 1.1 | -1 | 0.9 | 1.4 | 2.3 | 2.2 |
| General government balance (% of GDP) | -3.9 | -5.9 | -1.8 | -3.8 | -4.4 | -5 | -2.3 | -2.9 | -1.6 | -1.7 |
| External debt (% of GDP) | 88.2 | 86 | 77.4 | 76.8 | 77.5 | 95.8 | 131 | 121.7 | 103.9 | 87.7 |
| Exchange rate, annual average (UAH/EUR) | 10.9 | 10.5 | 11.1 | 10.3 | 10.6 | 15.7 | 24.2 | 28.3 | 30 | 32.1 |
| Exchange rate, annual average (UAH/USD) | 7.8 | 7.9 | 8.0 | 8.0 | 8.0 | 11.9 | 21.8 | 25.5 | 26.6 | 27.2 |
| Consolidated government spending (bln. UAH) | 242 | 304 | 333 | 396 | 403 | 430 | 577 | 685 | 839 | 986 |

Sources: own presentation using data from Ukrstat, National Bank of Ukraine, World Bank WDI.
Note: * - at constant prices of 2010.

**The most prominent of the recent economic reforms was the monetary policy and banking sector transformation conducted by the National Bank of Ukraine (NBU).** The NBU has been a leader of reforms in Ukraine since 2014 by adopting a number of policies that lay foundation for economic prosperity of Ukraine (Gorodnichenko and Bilan, 2020). First, it was adopting inflation targeting with a flexible exchange rate. Tough and transparent monetary policy helped to reduce inflation to the target level of 5% in 2019 providing the background for the following decrease of interest rate. Flexible exchange rate regime made Ukrainian economy more resilient to domestic and external shocks. Second, NBU cleaned up the banking sector and made it more transparent and resilient. Despite the massive bankrupt of commercial banks and panic of their clients, this led to the consolidation of the sector and improved the allocation of financial resources. Finally, NBU changed the approach of policymaking by ensuring higher transparency and independence from the political pressure. In 2019, the institution was awarded as one of the most transparent central banks (Gorodnichenko and Bilan, 2020).

**The reform of fiscal policy and public finance contributed to the essential contraction of budget deficit.** By 2013 fiscal and public finance policies put the economy of Ukraine on the brink of the crisis (Marchak, 2020). To avoid a default, in 2014 the government negotiated a stand-by program with the IMF (replaced by a four-year Extended Financing Program in 2015) forced Ukraine to adhere the serious fiscal discipline. During the following years, the country made a number of steps towards the balancing of the state budget. These steps included: the restructuring of government debt, the reduction of Naftogaz deficit and a transparent scheme of utility subsidies, automatic VAT refund, increased ability to attract foreign investments via the government bonds, the reduction of tax exemptions. In addition, the amendments of Budget Code in 2015 implied much shorter list of cases when changes in the state budget can be made (Marchak, 2020).

**Trade policy reform stimulated the opening of new export markets and growth of trade volumes.** The trigger for this reform was the launch of DCFTA with the EU that has been the most ambitious trade liberalization project of Ukraine (Movchan, 2020). Besides duty-free access to european market, DCFTA contributed to the reforms of regulations related to sanitary and phytosanitary measures (SPS) and technical barriers to trade (TBTs). This allows to export products not only to EU market, but also to markets of other developed countries. The other achievements related to trade policy include the foreign currency liberalization, easing of administrative barriers for trade in services, internationalization of public procurement, changes in trade policy framework (Movchan, 2020).

**The progress in the above mentioned directions is closely related to the governance reforms which ensure the improvements of institutional capacity.** Particularly, the public administration reform attracted talent persons to well-paid government positions, partially sponsored by EU. The other improvements concerned public accounting and tax administration. The judicial reform was aimed to increase the level of protection of basic rights that are crucial for economic growth. The reform includes enhancing the role of the Supreme Court, introduction of e-declarations for judges, re-attestation of judges, the establishment of Higher Court on Intellectual Property and the Higher

Anti-Corruption Court, the strengthening of institution of attorneys and other changes. Finally, the deregulation reform created more favorable economic environment for business through the simplified registering property rights for firms and real estate, easing the procedure of starting business, protecting the rights of investors, abolition or simplification for a number of sector-specific regulation. These measures resulted in improving of Ukraine's ranking in the Doing Business from 112$^{th}$ in 2014 to 71$^{th}$ in 2019.

**Despite the progress in key reforms after 2014, a few essential barriers for economic recovery are still existing.** First, the high level of corruption and dominance of oligarchy groups in politics contribute to unequal rules of doing business, which are reflected in the negative sentiments of investors. Coupled with the regulatory drawbacks and a large number of state-owned enterprises, this can create the serious market distortions and monopolization in key sectors such as energy and transport. Second, the banking sector suffers from the high share of state-owned banks; their dominance lead to incompetent credit allocations and a large number of non-performing loans. Third, the absence of farmland market decreases the credits and investments in agricultural sector.

### 2.3.2 Major performance indicators of agriculture

**Overall Ukrainian agriculture shows a remarkable and resilient growth.** Since 2000, the sector experience a recovery after almost a decade of a deep transition recession. In 2018, it generated value added and output above the 1990 level (**Error! Reference source not found.**.3.1 and 2.3.2). This shows that agriculture increasingly contributes a value to Ukrainian economy by constantly increasing the value added of its produce. Overall Ukraine's agriculture grew by 71% since 2001, demonstrating a remarkable resilience even in times of lower global commodity prices and deep crisis. The rest of the economy sectors, however, grew at more modest pace or even contracted: services grew by 45% and manufacturing contracted by 8% since 2001. Nevertheless, Ukraine's agriculture is still performing well below its potential. Given its fertile black soils and favorable climate, Ukraine is capable to reach the same average crop yields as in the EU, i.e. to increase them by about two times. By closing this productivity gap Ukraine's agriculture could make a much larger contribution to economy and country's welfare. This will require more capital-intensive agriculture, financed by potential domestic and foreign investments into the sector.

**The role of agri-food sector to Ukraine's economy is critical.** Its share in the GDP (including forestry and fishing) has been floating around 10% since 2001: being at 14% in 2001, then dropping to its minimum 6.5% in 2007, bouncing back to 12% in 2015 and stabilizing at 10% since then (Figure 2.3.2). The share of agriculture in total employment in Ukraine has decreased to some extent from 18% in 2000 to 15% in 2018. Rural population (14 million people) constitutes 31% of the total population. Food industry accumulates another 4% for Ukraine's GDP and further 4% of all employed. If upstream and downstream industries of agriculture (input supply, food processing, trade) are also considered, the contribution of the sector to the Ukrainian economy increases roughly to 20% of GDP. The importance of agri-food sector is also confirmed by the high share of food expenditures in households' budgets. Due to the permanent economic growth over the last two decades, this proportion declined slowly from 64 to 47,7% (Figure 2.3.3).

**Figure 2.3.1. Contribution of agriculture, industry and services in GDP (in 2010 USD prices)**

**Figure 2.3.2. Agricultural value added (in 2010 USD prices) and in % of GDP**

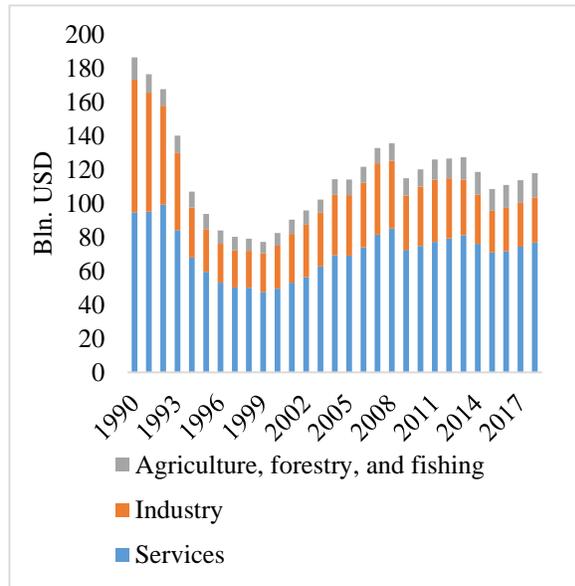
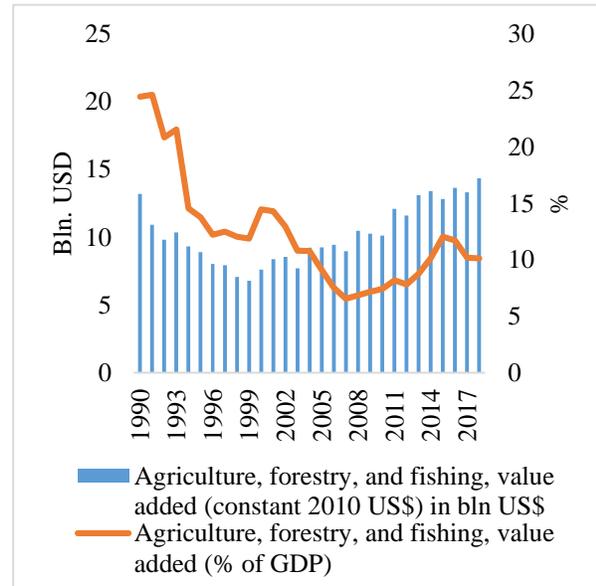

Source: own presentation using the World Bank WDI statistics on Ukraine

Source: own presentation using the World Bank WDI statistics on Ukraine

**Figure 2.3.3. The proportion of food products in expenditures of Ukrainian households**

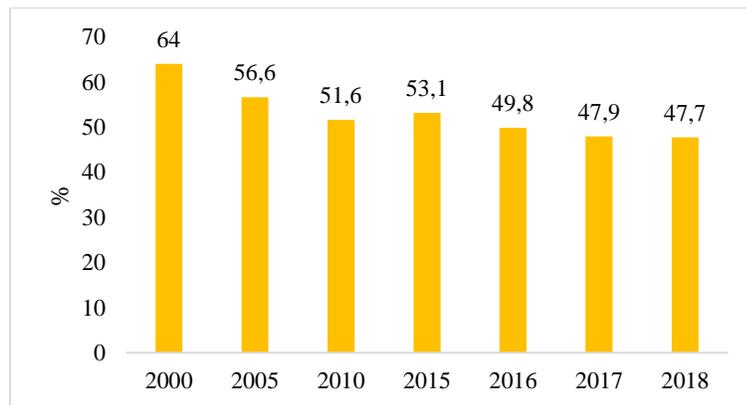

Source: own presentation using Ukrstat data

**Agri-food sector is important for country's trade balance and earning foreign exchange.** The share of agri-food exports in total exports increased from 11% in 2001 to almost 50% in 2020 (Figure 2.3.4). Meanwhile, the proportion of agri-food products in total imports remains low, not exceeding 10%. In the future, the share of agriculture may increase further as services usually grow slowly and the agricultural productivity in Ukraine is far from potential. The increasing role of agri-food sector seems especially likely against the current recession in the industry and ever-growing global demand for food.

**Figure 2.3.4. Share of agriculture in export and import of Ukraine**

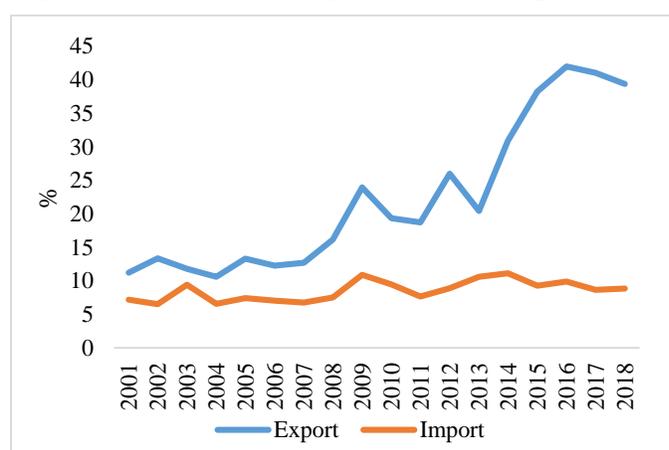

Source: own presentation using Ukrstat data

## 2.3.3 Farm structures

**The development of agricultural sector is characterized by a strong consolidation process.** Despite the essential growth of production volumes, the total number of agricultural holdings dropped from 50,648 in 2008 to 40,333 in 2018 (Table 2.3.2). The most shrinking category is the small farmers below 20 ha. The area controlled by medium-size producers (up to 5000 ha) showed a moderate decline. At the same time, the number and total area of large enterprises with more than 10,000 ha increased twofold over the last decade. The concentration of farmland was driven by different factors, namely depopulation in rural territories, the lack of proper governmental support for small farmers and the specialization on crop production that increases the importance of economies of scale.

**Table 2.3.2. Agricultural holdings by size classes**

|  | 2008 | | 2012 | | 2018 | |
|---|---|---|---|---|---|---|
|  | Area (1000 ha) | Number of agricultural holdings | Area (1000 ha) | Number of agricultural holdings | Area (1000 ha) | Number of agricultural holdings |
| Total | 21,229 | 50,648 | 21,914 | 55,866 | 20,005 | 40,333 |
| 0 |  | 8,411 | - | 8,214 | - | 8,875 |
| < 5.0 | 19 | 5,965 | 17 | 5,332 | 9 | 2,972 |
| 5.1–10.0 | 33 | 4,213 | 30 | 3,809 | 19 | 2,496 |
| 10.1–20.0 | 80 | 5,170 | 74 | 4,795 | 59 | 3,811 |
| 20.1–50.0 | 536 | 14,118 | 504 | 13,334 | 417 | 11,076 |
| 50.1–100.0 | 349 | 4,892 | 361 | 5,016 | 354 | 4,909 |
| 100.1–500.0 | 1,832 | 7,572 | 1,777 | 7,261 | 1,851 | 7,573 |
| 500.1–1000.0 | 2,050 | 2,846 | 1,884 | 2,624 | 1,933 | 2,704 |
| 1000.1–2000.0 | 4,090 | 2,863 | 3,684 | 2,565 | 3,513 | 2,447 |
| 2000.1–3000.0 | 3,338 | 1,362 | 3,102 | 1,270 | 2,594 | 1,063 |
| 3000.1–4000.0 | 2,481 | 721 | 2,179 | 632 | 1,612 | 467 |
| 4000.1–5000.0 | 1,659 | 372 | 1,482 | 334 | 1,110 | 250 |

| 5000.1–7000.0 | 1,823 | 313 | 1,959 | 337 | 1,497 | 258 |
| 7000.1–10000.0 | 1,216 | 148 | 1,504 | 179 | 1,057 | 127 |
| > 10000.0 | 1,721 | 93 | 3,356 | 164 | 3,978 | 180 |

Sources: own presentation using Ukrstat data

**The gross agricultural output (GAO) in Ukraine is generated by two groups of producers**, i.e. legally registered commercial enterprises and not legally registered individual family farms – households. There are more than four million small households (having 2.8 ha of land each on average) producing food both for subsistence needs and for the markets, but managing 37% of the Ukraine's total agricultural land and accounting for nearly 41% of the country's GAO in 2018 and their share in GAO is decreasing (see Figure 2.3.5). The rest of agricultural output was generated mainly by private agricultural enterprises, since the state-owned agricultural enterprises generated only less than 1% of the GAO in 2018.

**Figure 2.3.5. Gross agricultural output (GAO) in Ukraine**

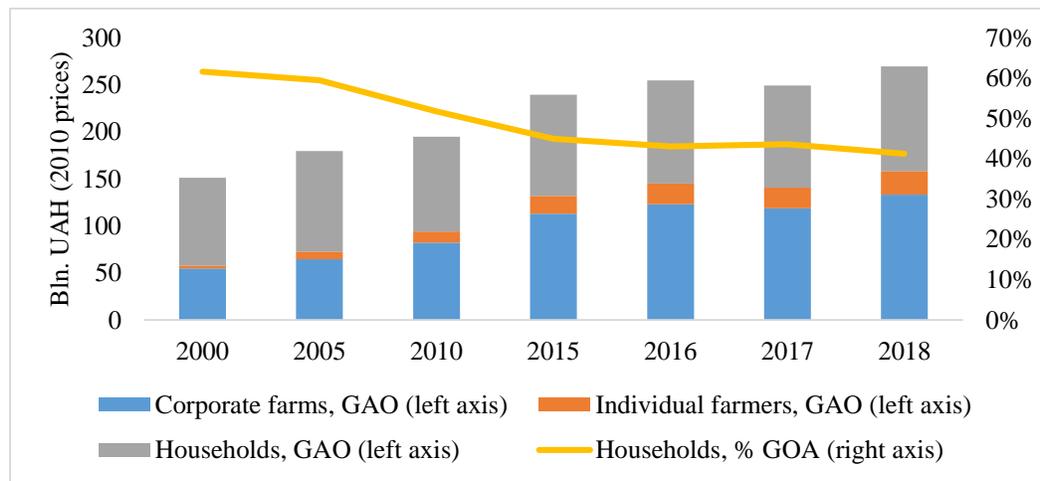

Source: own presentation using Ukrstat data

**Agricultural enterprises in Ukraine are of two types: corporate farms and individual commercial farms.** These farms, unlike households, are registered legal entities. There are about 9,892 corporate farms (mainly the successors of the former collective and state farms) each cultivating about 1,650 ha of arable land on average (Table 2.3.3) and generating almost 60% of the GAO in 2018. There are about 30,441 much smaller individual farmers with an average of 105 ha of arable land each, altogether cultivating only about 13% of the Ukraine's arable land and generating 9% of the total GAO in 2018.

**The land consolidation process in agriculture has led to the emergence of large, vertically-oriented agri-holdings.** The number of corporate farms has shrunk from roughly 17,700 in 2004 to 9,892 in 2018. An increasing number of corporate farms is coming under the control of agri-holdings, which were created with different purposes, in different sizes, shapes, and organizational forms but share some common characteristics. Agri-holdings usually consist of a mother company that, in most cases, is not involved in primary agricultural production but adopts overall strategy, production orientation and investments, and manages access to production factors, including input and output markets, land and finance. This mother company is typically "holding" 5-50 individual

corporate farms of about 2,000-15,000 ha each, with the size of the agri-holdings varying from 30,000 to 700,000 ha. In 2018, agricultural holdings farmed five to six million ha of agricultural land in Ukraine. The largest agri-holding is the Kernel with the land bank of about 700,000 ha. These super large agri-holdings produced about 23% of the GAO.

**Table 2.3.3. Land use by farm type in 2018**

|  | Number of units | Land area, total (1000 ha) | Average area of agricultural land (ha) |
|---|---|---|---|
| Agricultural holdings, private | 40,333 | 20,746 | - |
| Corporate farms | 9,892 | 16,294 | 1,650 |
| *Incl. Agriholdings* | - | 5,000 – 6,000 | 30,000-700,000 |
| Individual farms | 30,441 | 4,452 | 105 |
| Agricultural holdings, state-owned | 278 | 937 | 2,963 |
| Households/Individual farms | 4.6 million | 15,706 | 3 |

Source: own presentation using Ukrstat data

## 2.3.4 Land use

**Ukraine's territory consists primarily from agricultural land.** The territory of Ukraine is 60.4 million hectares, 71% (42.7 million hectares) of which is agricultural land (see Table 2.3.4). About 5% (2.6 million hectares) of the agricultural area has been developed for irrigation. Most of the irrigated area is concentrated in the southern part of the country. The area actually irrigated has declined from 2.2 million ha in 2003 to about 0.5 million ha in 2018. About 31 million ha of agricultural land is arable. More than half of the Ukraine's arable land is the world's most productive black soil, providing an excellent and high quality basis for the production of crops, livestock and energy crops (Figure 2.3.6).

**Table 2.3.4. Agricultural land use in Ukraine 2000–2018, thds. ha**

|  | 2000 | 2008 | 2010 | 2012 | 2015 | 2018 |
|---|---|---|---|---|---|---|
| Land area, total | 60,355 | 60,355 | 60,355 | 60,355 | 60,355 | 60,355 |
| Agricultural land, total | 48,817 | 41,626 | 41,576 | 41,536 | 41,511 | 41,489 |
| Arable land | 32,537 | 32,473 | 32,477 | 32,518 | 32,531 | 32,544 |
| Permanent crops in bearing age | 924 | 304 | 291 | 291 | 893 | 894 |
| Permanent grassland | 7,924 | 7,918 | 7,893 | 7,870 | 7,848 | 7,820 |
| *Hayfields* | 2,407 | 2,416 | 2,411 | 2,411 | 2,407 | 2,399 |
| *Pastures* | 5,517 | 5,502 | 5,482 | 5,460 | 5,441 | 5,421 |
| Fallow land owned by enterprises and individuals | 431 | 192 | 180 | 147 | 239 | 229 |

Source: own presentation using Ukrstat data

**Figure 2.3.6. Inherent land quality map**

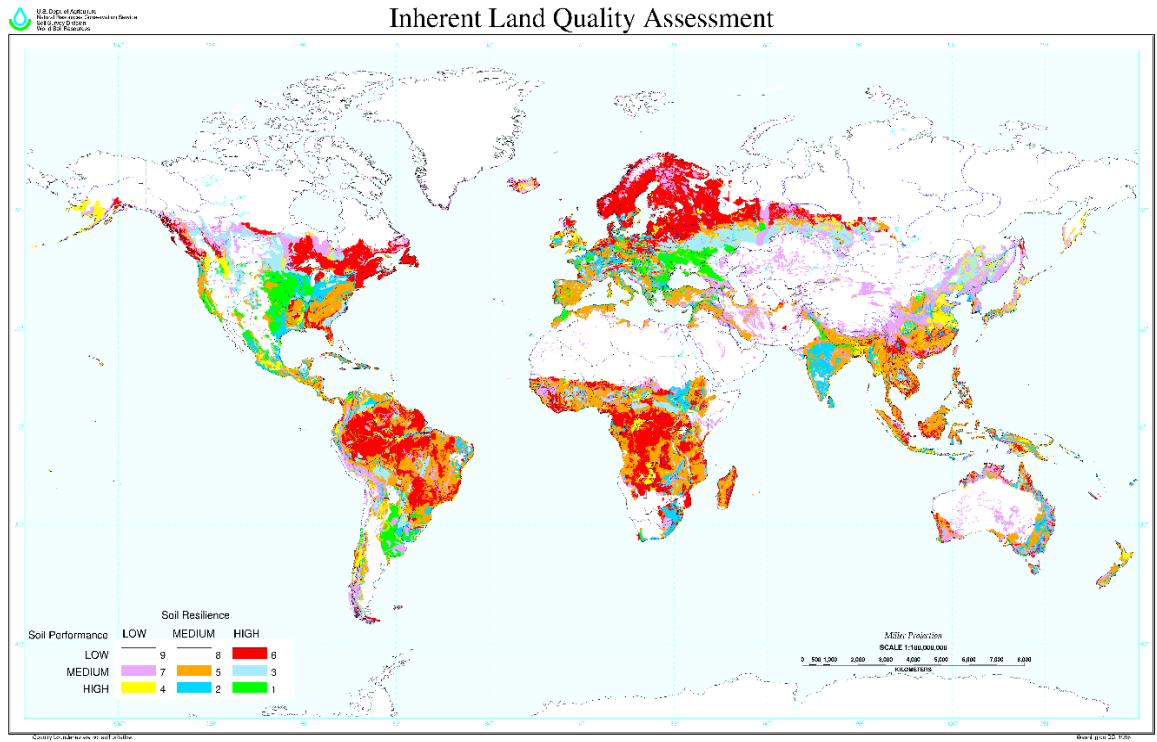

Source: Natural Resource Conservation Service, USDA

**Overall, the area of agricultural land slightly decreased in Ukraine under the pressure of urbanization.** About 137 thousand hectares ceased to be agricultural land during 2008-2018 due to urbanization. Even more important factor is soil erosion, since more than 500 million tons of soil is annually eroded in the country. Every year, erosion is causing a loss of soil fertility over the entire arable land that can be valued currently at around five billion USD, in nutrient equivalent (World Bank/FAO, 2014). So the potential for the land expansion and resulting extensive growth in crop production is not significant in Ukraine.

## 2.3.5 Untapped potential

**Ukrainian agriculture is experiencing a recovery after almost a decade of the transitional recession after the break-up of the USSR in 1991**. Only in 2018 its output reached the pre-independence levels (see Figure 2.3.7). Agricultural enterprises' output drastically dropped during the transition period. Households, on the other hand, used agriculture for the subsistence during the period of transformation hardships and increased their level of output. Over the last decade the share of agricultural enterprises in the total output has been persistently increasing.

**Agricultural productivity in Ukraine, however, is still far from its potential.** Agricultural value added per hectare is just a fraction of that in other European countries and its competitors on world agricultural markets. In 2018 it was 440 USD in Ukraine, compared to 1,100 USD in

Poland, 1,400 USD in Brazil, 1,700 USD in Germany, and 2,450 USD in France[6]. The primary reason for this is that agricultural production in Ukraine increasingly focuses on lower value-added products (such as grains or oilseeds).

**Figure 2.3.7. Index of the Gross Agricultural Output (GAO) in Ukraine**

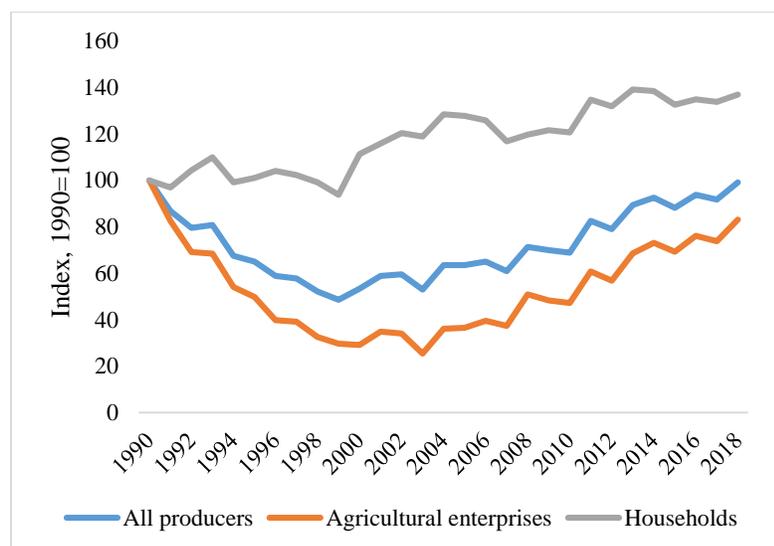

Source: own presentation using Ukrstat data

**Agricultural growth is driven primarily by crop production sector.** Since 2000, crop production output increased by more than two times, while animal production output increased only by 20% and its share in GAO contracted from about 40% to 26%. Moreover, animal output stagnated since 2010, while crop production output increased by 60% (Figure 2.3.8).

**Households yet dominate in the production of the animal products**, although their share substantially declined from nearly 80% in 2000 to 53% in 2018. Generally speaking, agricultural enterprises are taking over households shares in total animal production and households' animal output is shrinking (Figure 2.3.10). As of 2018, households' share was 73% in raw milk, 75% in beef and veal, 49% in pork and 14% in poultry output. Households also prevail in the production of potatoes, vegetables and fruits, taking 98%, 86%, and 78% respectively in 2018.

**Agricultural enterprises (including individual farmers) play a leading role mainly in the cultivation of export-oriented crops**, producing more than 62% of the crop production output in 2018 (Figure 2.3.9). They produced 80% of grains, 86% of sunflower seeds, 99% of rapeseeds, and 95% of sugar beets in 2018. Individual small farmers specialize mainly in crops rather than in livestock (Figure 2.3.11), employing the same cropping patterns as corporate agricultural enterprises yet produce at similar or lower rates of intensity. Individual farmers accounted for about 14% of the total grains, 7% of sugar beets, 20% of sunflower seeds, 17% of soybeans and 20% of rapeseeds, but only 3.8% of the total meat and 7.3% of raw milk produced in 2018.

---

[6] http://pubdocs.worldbank.org/en/204821574084103184/Ukraine-special-focus-note-Fall-2019-en.pdf

**Figure 2.3.8. Gross agricultural output (all producers)**

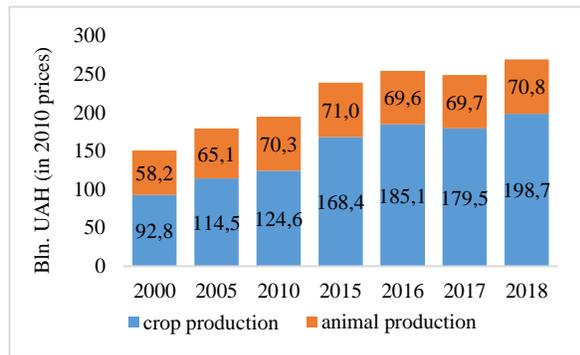

Source: own presentation using Ukrstat dataUkrstat

**Figure 2.3.9. Crop production output**

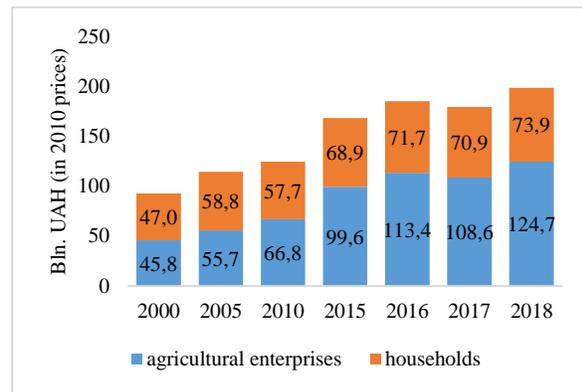

Source: own presentation using Ukrstat dataUkrstat

**Figure 2.3.10. Animal production output**

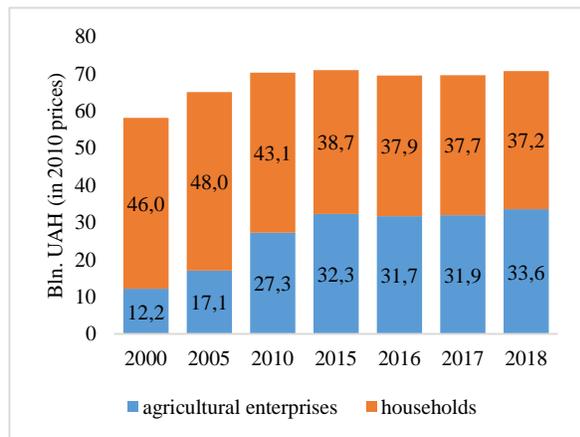

Source: own presentation using Ukrstat dataUkrstat

**Figure 2.3.11. Gross agricultural output (individual farmers)**

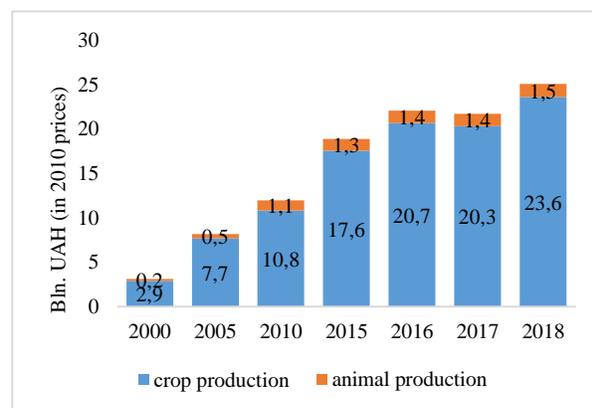

Source: own presentation using Ukrstat dataUkrstat

**The untapped potential of agricultural production can be measured as the gap between yields in Ukraine and the counterparts in the developed countries.** In particular, grain yields have increased by almost 2.5-fold since 2000 (with some short-term fluctuations). Better technologies, farm practices, management, production and post-harvest logistics investments have been the main reasons for improved yields. Still, the current grain yield levels in Ukraine (about 4.7 tons per ha) are far from the potential and from the countries with higher specific intensities, i.e. in Western Europe in particular (Figure 2.3.12).

**Figure 2.3.12. Grains yields in Ukraine and EU**

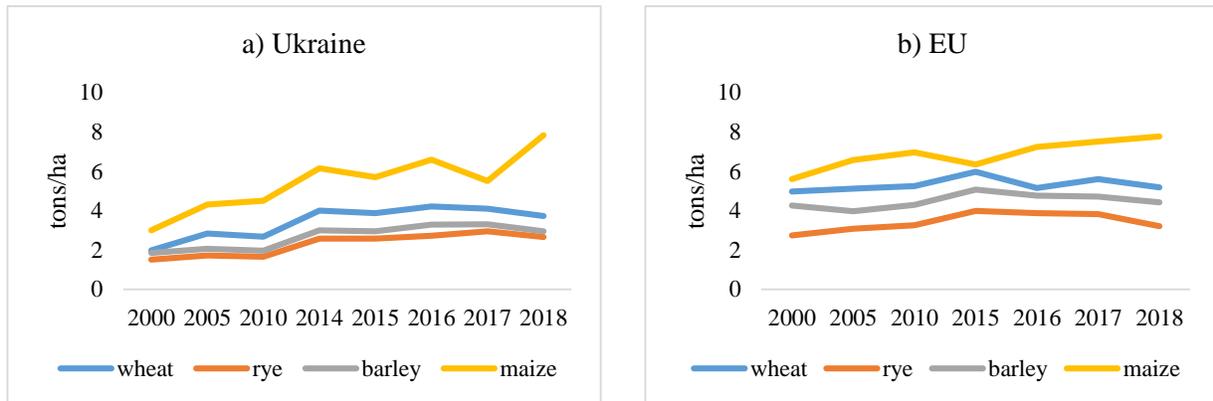

Source: own presentation using data from Ukrstat, USDA

**Oilseed yields experienced impressive growth since 2000**. They increased from average 1 t/ha in 2004 to 2.4 t/ha in 2018 (Figure 2.3.13), approaching to EU benchmarks. Nevertheless, this numbers can be increased even more. In particular, modern hybrids of sunflower allow to produce about 4-5 tons/ha (Shubravska et. al 2017). The growth of sunflower productivity will help to fill all the existing crushing capacities and expand the production of sunflower oil.

**Figure 2.3.13. Oilseeds yields in Ukraine and EU**

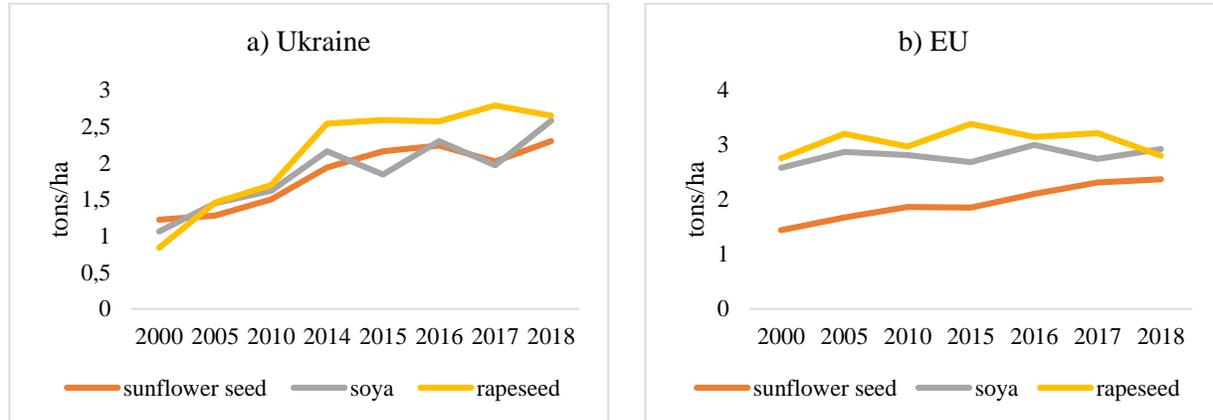

Source: own presentation using Ukrstat data

**The productivity gaps are also pronounced in the livestock sector.** Against the background of falling dairy herd numbers (from more than 8 million cows in 1990 to 2 million cows in 2018), the annual milk yield per cow, however, increased from 2.86 tons to almost 5 tons. This is still relatively low milk yield compared to western standards. For example, the average productivity of a cow is 6-7 tons in Germany and 11-12 in Israel. But milk yields at commercial dairy farms are already reaching the western milk yields (Figure 2.3.14).

**Figure 2.3.14. Milk yields in Ukraine**

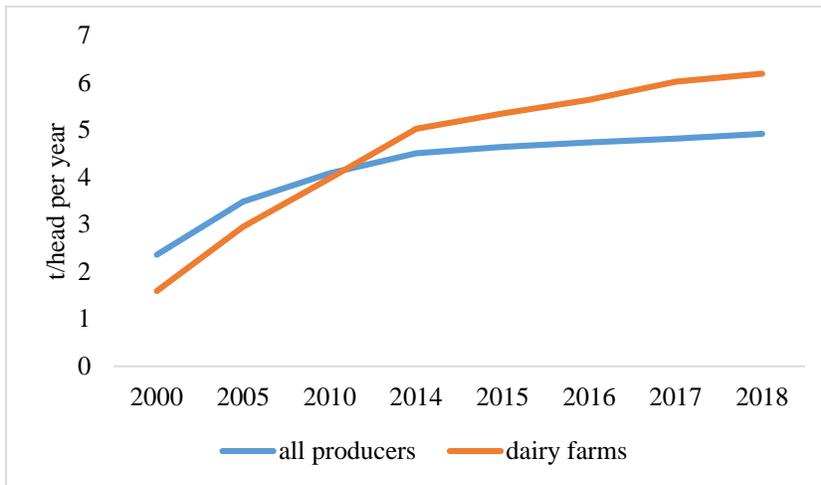

Source: own presentation using Ukrstat data

## 2.3.6 Development and structure of crop production

**Wheat, barley, maize and sunflower are the dominant crops**, covering about 62% of the Ukraine's total arable land (Figure 2.3.15). Over the last decade the structure of the harvested area has changed, mainly as Ukraine's response to the global market developments. In absolute and relative terms the harvested area of the main crops increased significantly, except barley. The most impressive expansion was recorded for rapeseed and soybean, followed by sunflower and maize. This expansion occurred at the cost of barley, rye, oats, millet, buckwheat and sugar beet. The fruit area was somewhat reduced, while vegetable area gained additional 30 000 ha.

**Figure 2.3.15. Structure of the harvested area of crops in 2005 and 2018**

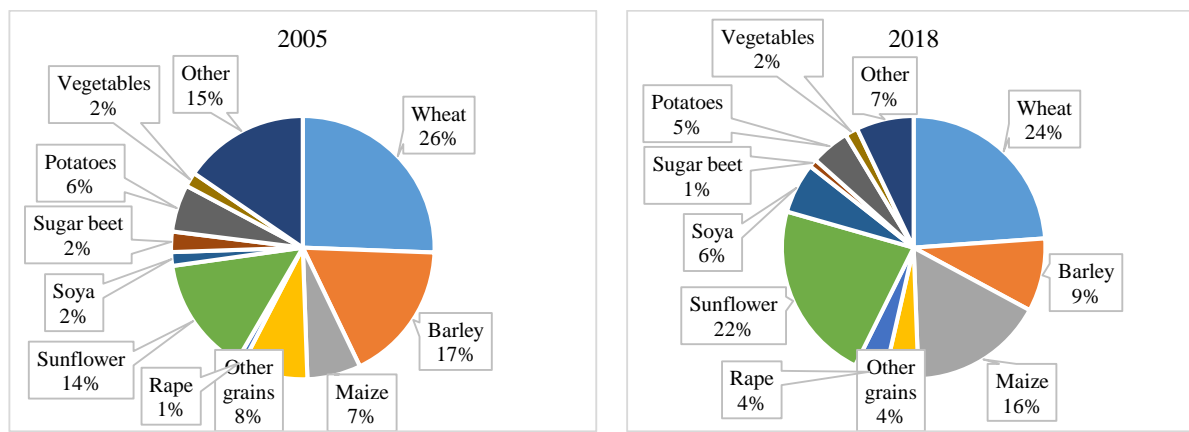

Source: own presentation using Ukrstat data

**Grains have traditionally been the leading crop in Ukraine**. Grain production already surpasses the pre-transition levels and reached a historical maximum of 72 million tons in 2019. Geographically, central Ukraine is the main region of grain production in Ukraine, gradually expanding to the northern and to the western regions (Table 2.3.5). Ukraine has emerged as one of the world's top grain exporters and continues to increase its production of exportable grains. Wheat and maize have been dominating the grain production, mainly driven by growing global demand for maize and by moderately growing commercial livestock sector (in poultry and pig production). The rest of grains (e.g. rye, oats and other grains) have been losing their share over the last decade.

**Table 2.3.5. Production of grains in Ukraine 2009–2018, 1000 tons**

|  | 2009 | 2010 | 2011 | 2012 | 2013 | 2014 | 2015 | 2016 | 2017 | 2018 |
|---|---|---|---|---|---|---|---|---|---|---|
| Ukraine | 46,028 | 39,271 | 56,747 | 46,216 | 62,997 | 63,859 | 60,125 | 66,088 | 61,916 | 70,056 |
| Center | 15,469 | 13,579 | 19,982 | 14,474 | 22,043 | 20,369 | 20,055 | 22,642 | 18,492 | 24,146 |
| East | 5,307 | 3,874 | 7,027 | 5,653 | 7,704 | 8,054 | 6,737 | 7,383 | 7,043 | 6,332 |
| North | 7,774 | 5,932 | 9,296 | 10,492 | 12,140 | 12,900 | 11,527 | 12,977 | 12,404 | 15,884 |
| South | 10,866 | 9,953 | 12,428 | 6,319 | 11,034 | 11,114 | 11,735 | 12,014 | 12,390 | 11,492 |
| West | 6,613 | 5,933 | 8,014 | 9,279 | 10,076 | 11,411 | 10,064 | 11,061 | 11,580 | 12,194 |

Sources: own presentation using Ukrstat data

**Sunflower seed dominates in the production structure of oilseeds in Ukraine.** The growth is especially pronounced in the last decade by responding to the demand from growing crushing industry. Ukraine emerged as a top sunflower oil exporter in the world. Rapeseed output expanded mainly as a response to the high demand from the EU (mainly for biodiesel production). Soybeans production was stimulated mainly by the growing global demand and to a less extent by the recovery of the domestic commercial livestock sector that increasingly consumes rich protein fodder. Sown area of soybeans increased almost 10-fold since 2000 (Figure 2.3.16). In regional perspective, Central and Southern regions of Ukraine dominate in the oilseed production (see Table 2.3.6).

**Figure 2.3.16. Structure of oilseed production in Ukraine in 2005 and 2018, 1000 tons**

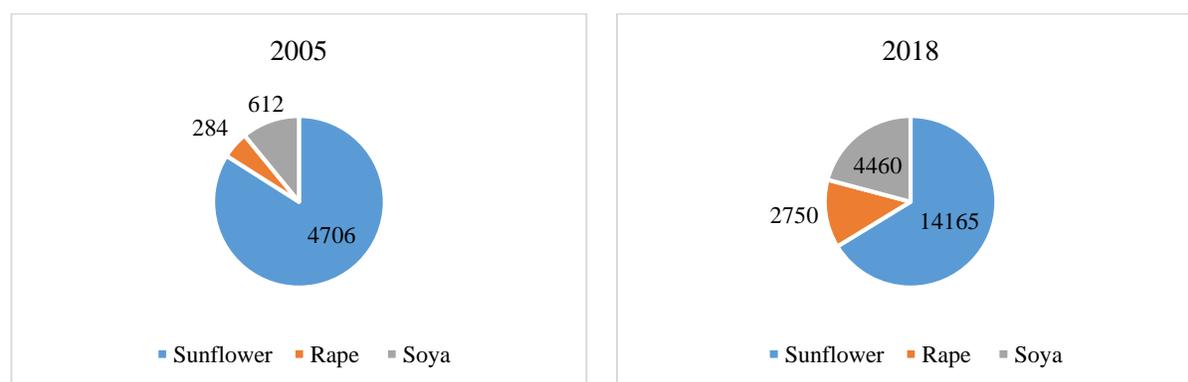

Source: own presentation using Ukrstat data

**Table 2.3.6. Production of oilseeds in Ukraine 2004–2018, 1000 tons**

|         | 2009  | 2010  | 2011   | 2012   | 2013   | 2014   | 2015   | 2016   | 2017   | 2018   |
|---------|-------|-------|--------|--------|--------|--------|--------|--------|--------|--------|
| Ukraine | 9,281 | 9,921 | 12,454 | 12,074 | 16,232 | 16,214 | 16,849 | 19,058 | 18,330 | 21,377 |
| Center  | 3,576 | 3,625 | 4,498  | 4,176  | 5,810  | 5,597  | 6,052  | 6,483  | 5,770  | 7,032  |
| East    | 1,777 | 1,748 | 2,374  | 2,273  | 2,642  | 2,533  | 2,284  | 2,760  | 2,376  | 2,844  |
| North   | 719   | 774   | 1,183  | 1,619  | 2,116  | 2,506  | 2,527  | 3,054  | 3,064  | 3,736  |
| South   | 2,447 | 3,001 | 3,382  | 2,769  | 4,003  | 3,300  | 3,831  | 4,419  | 4,094  | 4,331  |
| West    | 762   | 773   | 1,018  | 1,236  | 1,660  | 2,277  | 2,155  | 2,342  | 3,015  | 3,434  |

Source: own presentation using Ukrstat data

**Fruit and vegetable production is already an important element of Ukraine's rural economy given its high labour intensity**. This sector accounts for about 22% of gross agricultural output and 25% of agricultural employment. Fruit and vegetable production and yields significantly increased over the last decade (Figure 2.3.17). The growth is especially pronounced in tomato and cucumber sectors in the south of Ukraine, where farmers achieve high yields by introducing modern irrigation technologies.

**The growth of fruit production is comparable to the growth of vegetable production.** Top individual fruits in terms of production volumes are apples, cherries, pears and plums. The large export potential is almost untapped in Ukraine. But for Ukraine to tap this potential, large investments to increase the productivity and reduce the losses in the value chain are required.

**Figure 2.3.17. Structure of fruit and vegetable production**

**in Ukraine in 2004 and 2018, 1000 tons**

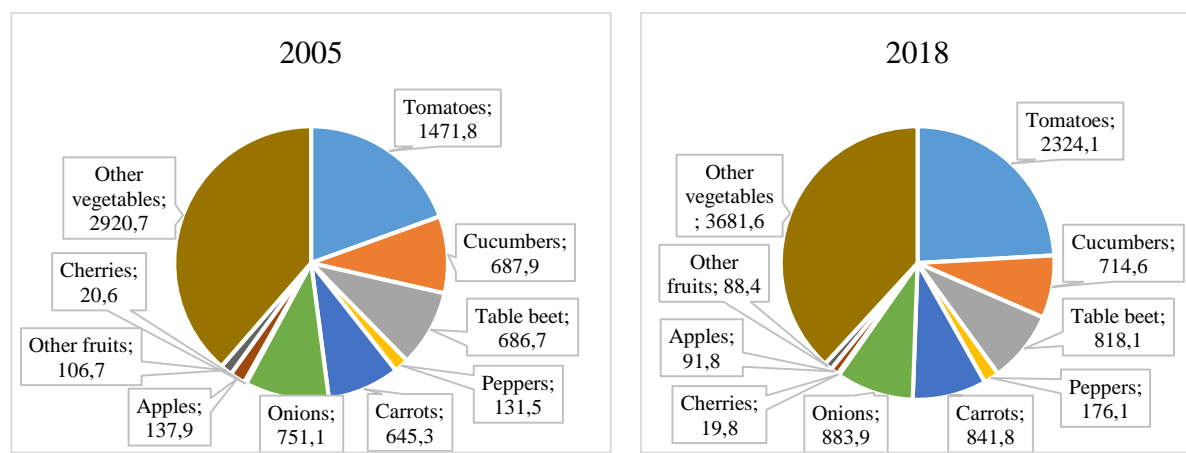

Source: own presentation using Ukrstat data

## 2.3.7 Development and structure of livestock production

**Livestock sector was severely hit after the collapse of the former Soviet Union.** Livestock numbers have dropped drastically by 2000 (Figure 2.3.18). The number of cattle (incl. cows) continues to fall, while pig and poultry sectors were able to recover. Production of meat is dominated by households, this is especially pronounced in beef sector (Figure 2.3.19). Poultry sector is a success story as it recovered quite quickly after the transition collapse and turned into the export-oriented sector. For now, poultry meat takes around 60% of all meat produced in Ukraine (Figure 2.3.20). Production of pork is catching up with the poultry sector. Although Ukraine is a net importer of meat, it has all chances to become a net exporter, mainly due to the domestically abundant feed.

**Figure 2.3.18. Development of livestock numbers**

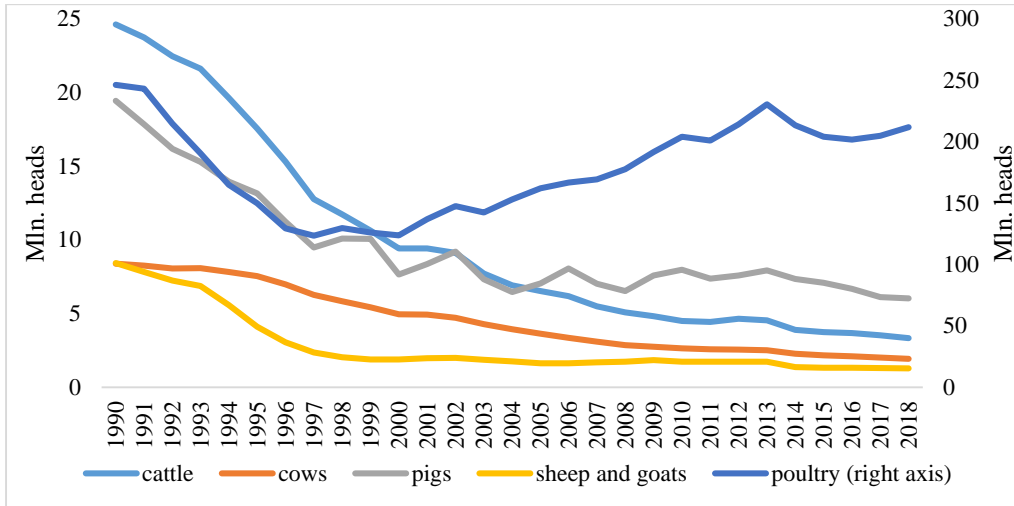

Source: own presentation using Ukrstat data

**Figure 2.3.19. Production of meat by categories of enterprises**

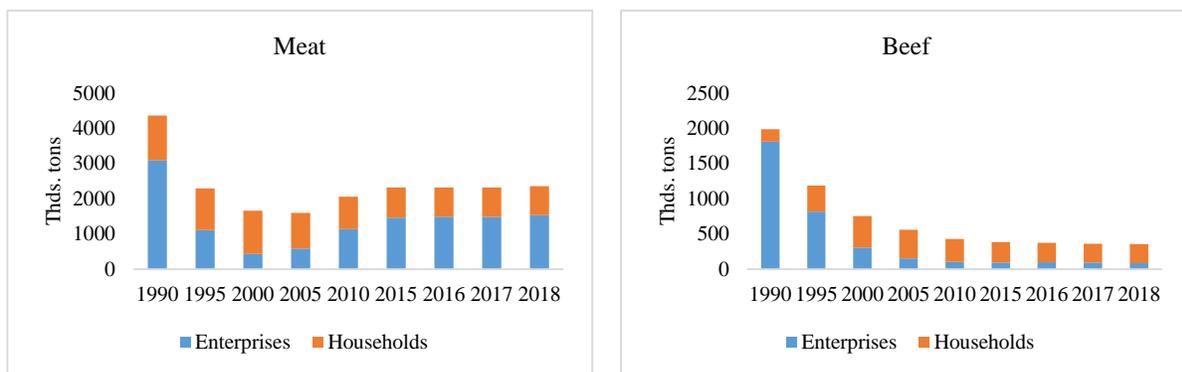

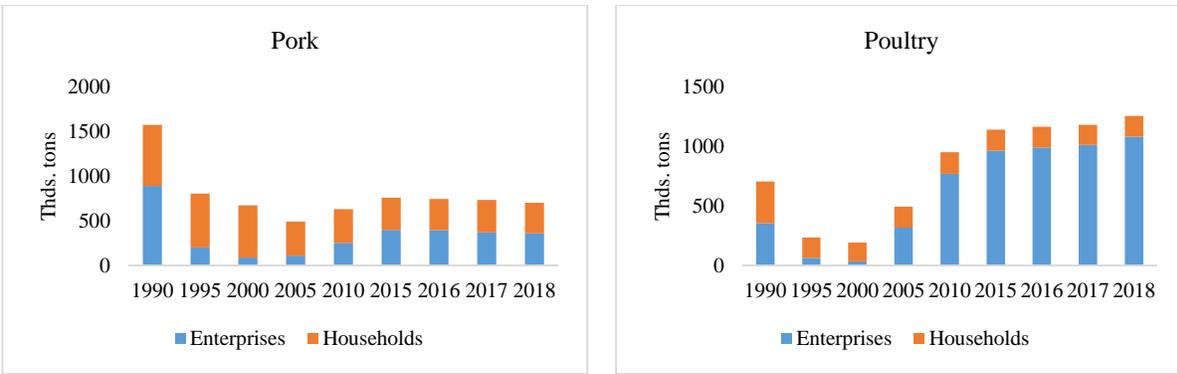

Source: own presentation using Ukrstat data

**Figure 1.20. Structure of meat production in Ukraine in 2005 and 2018, 1000 tons**

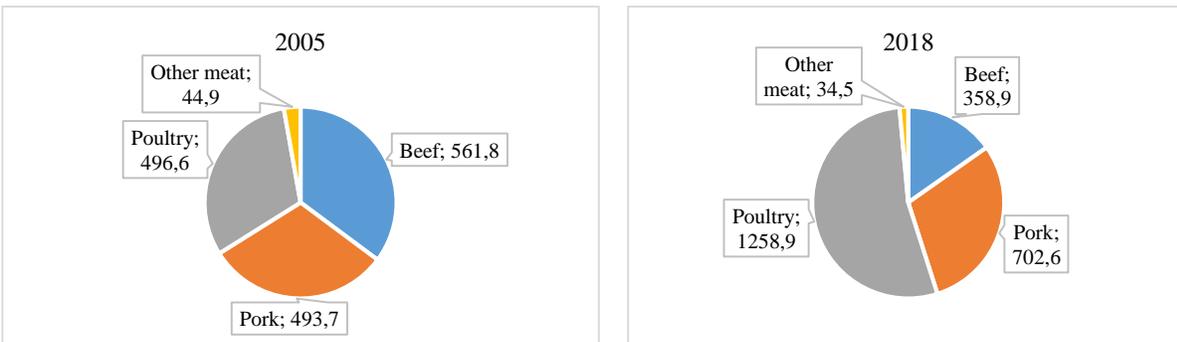

Source: own presentation using Ukrstat data

**Pork production has been on a fragile upward trend with some short-term fluctuations** (Table 2.3.7). As Ukrainian consumers historically favour pork, production volume of the sector has not dropped as drastically as in beef production. At the moment, large corporate pig farms are gaining market share, while pork production of household and peasant farms is stagnating. Regionally, in Poltava (top grain producer) and Ternopil regions pork production increased the most (Table 2.3.7).

**Table 2.3.7. Pork production in Ukraine 2009–2018, 1000 tons**

|  | 2009 | 2010 | 2011 | 2012 | 2013 | 2014 | 2015 | 2016 | 2017 | 2018 |
|---|---|---|---|---|---|---|---|---|---|---|
| Ukraine | 526.5 | 631.2 | 704.4 | 700.8 | 748.3 | 742.6 | 759.7 | 747.6 | 735.9 | 702.6 |
| Center | 119.3 | 147.0 | 164.8 | 159.9 | 180.1 | 181.9 | 186.5 | 182.4 | 180.9 | 160.7 |
| East | 69.2 | 84.1 | 102.3 | 96.3 | 107.0 | 109.8 | 110.3 | 104.1 | 92.3 | 87.7 |
| North | 81.9 | 94.8 | 107.6 | 106.5 | 111.0 | 113.6 | 116.1 | 118.0 | 124.7 | 123.5 |
| South | 106.6 | 120.6 | 140.7 | 136.2 | 132.5 | 99.1 | 97.4 | 89.9 | 82.4 | 76.1 |
| West | 149.5 | 184.7 | 189.0 | 201.9 | 217.7 | 238.2 | 249.4 | 253.2 | 255.6 | 254.6 |

Sources: own presentation using Ukrstat data

**There has been an impressive growth in the large-scale poultry production over the last 15 years.** This was triggered by the short production cycle and the corresponding investment cycle, as well as a heavy state support. Poultry meat can be produced in a short period of time, with high efficiency in transforming feed grain into meat. Large agri-holdings dominate poultry production

(about 77% market share). Two of the largest agri-holdings, MHP and Agromars, dominate the poultry meat market (MHP's share is more than 50%). Whereas, the largest agri-holding, UkrLandFarming, dominates the egg market. Distribution of poultry production across the regions reflects the availability of grains/fodder. The large-scale poultry production is mainly located in the central, northern and western regions of Ukraine (Table 2.3.8).

**Table 2.3.8. Poultry production in Ukraine's regions 2009–2018, 1000 tons**

|  | 2009 | 2010 | 2011 | 2012 | 2013 | 2014 | 2015 | 2016 | 2017 | 2018 |
|---|---|---|---|---|---|---|---|---|---|---|
| Ukraine | 894.2 | 953.5 | 995.2 | 1074.7 | 1168.3 | 1164.7 | 1143.7 | 1166.8 | 1184.7 | 1258.9 |
| Center | 379.8 | 454.1 | 468.1 | 496.0 | 575.6 | 665.9 | 703.8 | 750.6 | 753.5 | 807.3 |
| East | 98.3 | 95.1 | 92.4 | 104.6 | 105.8 | 77.8 | 46.3 | 50.2 | 49.3 | 47.3 |
| North | 166.9 | 150.3 | 160.5 | 174.2 | 186.4 | 183.9 | 163.7 | 148.3 | 172.8 | 170.5 |
| South | 105.9 | 105.8 | 108.6 | 113.0 | 101.4 | 33.9 | 34.2 | 26.3 | 24.1 | 24.5 |
| West | 143.3 | 148.2 | 165.6 | 186.9 | 199.1 | 203.2 | 195.7 | 191.4 | 185 | 209.3 |

Sources: own presentation using Ukrstat data

**Beef in Ukraine is underdeveloped and is mainly produced as a by-product of dairy farming.** Around 75% of beef is supplied by households (Figure 2.3.17). This is very close to the share of households in the total milk supplies (see below). As beef production needs long-term investments, limited access to long-standing financing has been one of the main reasons for the underdevelopment. The largest beef production regions are located in the west and center of Ukraine (Table 2.3.9).

**Table 2.3.9. Beef production in Ukraine 2009–2018, 1000 tons**

|  | 2009 | 2010 | 2011 | 2012 | 2013 | 2014 | 2015 | 2016 | 2017 | 2018 |
|---|---|---|---|---|---|---|---|---|---|---|
| Ukraine | 453.5 | 427.7 | 399.1 | 388.5 | 427.8 | 412.7 | 384 | 375.6 | 363.5 | 358.9 |
| Center | 89.2 | 84.6 | 76.1 | 74.8 | 87.9 | 90.8 | 85.3 | 81.4 | 75.7 | 74.5 |
| East | 49.4 | 47.0 | 43.8 | 43.9 | 48.3 | 45.5 | 45.7 | 44.4 | 38.0 | 34.7 |
| North | 69.2 | 66.4 | 62.2 | 62.2 | 70.0 | 66.9 | 60.4 | 59.3 | 58.6 | 57.6 |
| South | 79.2 | 77.9 | 70.0 | 66.7 | 68.6 | 56.9 | 53.7 | 53.7 | 54.2 | 60.1 |
| West | 166.5 | 151.8 | 147.0 | 140.9 | 153.0 | 144.5 | 138.9 | 136.8 | 137.0 | 132.0 |

Sources: own presentation using Ukrstat data

**Given that meat cattle breads are almost absent in Ukraine, the production of raw milk shows the similar trend as beef production.** Total output decreased from about 24.5 million tons in 1990 to nearly 10 million tons in 2018, i.e. by more than 50%. Milk production is dominated by households. Currently they account for 73% of the total milk production, as compared to 24% in 1990. This adds extra costs to the entire dairy value chain via problems associated with quality, difficulty to capture economies of scale both in dairy farming and in the upstream and downstream industries, problems to guarantee a large and stable supply of high quality milk, seasonality of supplies, high collection costs and other transaction costs. However, the share of raw milk supplies from households is constantly decreasing and commercial dairy farms are putting up their supplies (Figure 2.3.21). This has happened because of the decreasing number and aging of rural population, shrinking areas for grazing (due to the expansion of crop areas), expensive milk collection from households and the need to ensuring high quality of milk. Also, consolidation of dairy farms can be observed, the number of dairy farms has been decreasing against the background of growing herd size.

**Figure 2.3.21. Raw milk production structure**

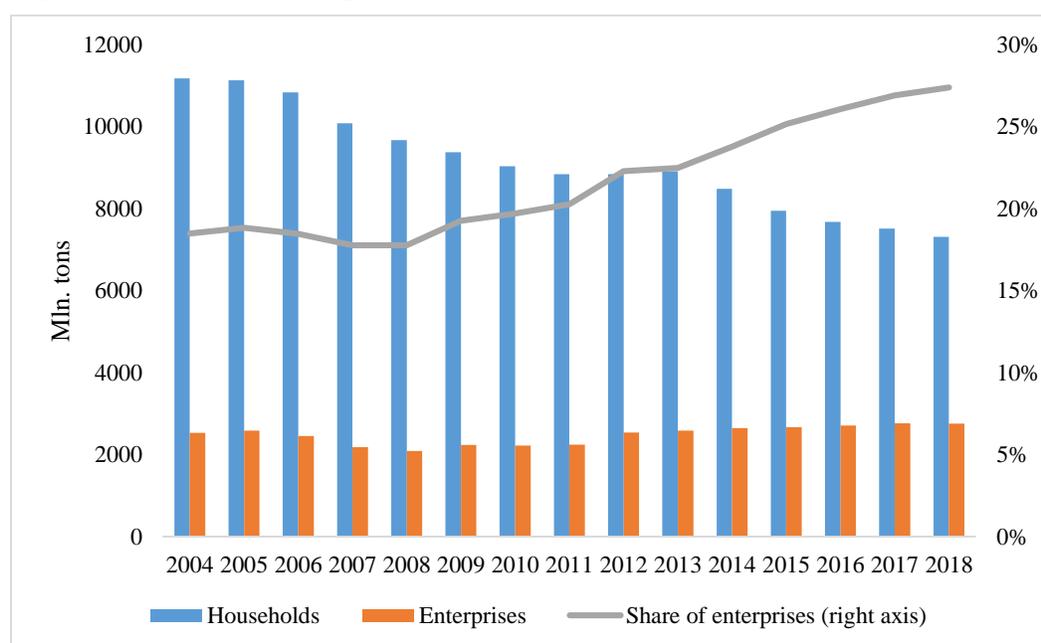

Source: own presentation using Ukrstat data

**Domestic demand for raw milk is very strong due to the high external demand for dairy products.** The falling domestic raw milk supply has been constantly increasing a competition for raw milk among dairy processors. In fact, there is a shortage of raw milk supplies to fully utilize the existing processing capacities. As of 2018, processing capacity utilization was 60-65%. Similarly to beef production, the production of raw milk is higher in western, central and southern regions of Ukraine (Table 2.3.10).

**Table 2.3.10. Raw milk production in Ukraine 2009–2018, 1000 tons**

|  | 2009 | 2010 | 2011 | 2012 | 2013 | 2014 | 2015 | 2016 | 2017 | 2018 |
|---|---|---|---|---|---|---|---|---|---|---|
| Ukraine | 11609 | 11248 | 11086 | 11377 | 11488 | 11132 | 10615 | 10381 | 10280 | 10064 |
| Center | 2756.6 | 2699.3 | 2701 | 2776.3 | 2823.6 | 2877.4 | 2818.2 | 2796.3 | 2757.9 | 2673 |
| East | 1157.7 | 1090.7 | 1076.1 | 1128.2 | 1141 | 1060.1 | 911.1 | 846.1 | 837.5 | 839.9 |
| North | 2126.2 | 2041 | 2004.7 | 2079.8 | 2083.2 | 2056 | 1994.9 | 1968.7 | 1948.4 | 1934.2 |
| South | 1755.2 | 1683.4 | 1643.8 | 1639.2 | 1639.1 | 1345.6 | 1289.8 | 1260.5 | 1244.8 | 1187.7 |
| West | 3813.9 | 3729.2 | 3655.7 | 3749.8 | 3796.1 | 3793.7 | 3601.4 | 3509.9 | 3491.9 | 3429.2 |

Sources: own presentation using Ukrstat data

**So overall, the agri-food sector in Ukraine's economy is growing and its structure is changing.** Agriculture's contribution to economic growth has been resilient and increasing since 2014. Its share in the GDP (including forestry and fishing) has been floating around 10% and the share of agri-food exports in total exports increased to almost 50% in 2020. Ukrainian agriculture has shown significant development, accompanied by market consolidation, vertical integration, and by substantial productivity growth, though productivity gap still remains quite large. This trend is more pronounced in the crop sector, which is based on the cultivation of highly profitable cash crops. In contrast, the livestock sector has stagnated due to low production efficiency and limited export markets. In the next chapter, we will describe the current agricultural policy agenda to show how policy decisions affect the development of the agri-food sector in Ukraine.

## 2.4 Current agricultural policy in Ukraine

### 2.4.1 Contrasting agricultural and rural development policies

**Ignorance of rural development in contrast to agricultural development** has been a common practice by the Government of Ukraine since independence. This resulted in a situation when rural policies where biased towards the benefits of agribusinesses, depriving rural communities from a substantial resources for development (due to, for example, of generous tax breaks to agribusinessnes) and incomes (through, e.g. agricultural land sales moratorium – see Nivievskyi and Deininger 2019), neglecting other economic activities in rural areas, and thereby contributing to the ongoing depopulation and economic degradation of the countryside. The role of rural areas for the well-being of Ukrainians, both villagers and citydwellers is a way much wider than just agriculture. Therefore, rural development policy should be considered separatly from agricultural policy.

**The objective structural changes in agriculture exacerbate the divergence between the policy goals in this sector and rural development agenda.** In market economy, the progress of agricultural policy can be measured in three main dimensions: economic efficiency, food quality and safety, and environmental sustainability. From these perspectives, policymakers have to pursue three main goals: to generate value added through overcoming the productivity gap, to supply high-qulity products and to contribute to preserving nature and landscape (see in details Kuhn and Demyanenko, 2004). In the context of intensive industrialization of agriculture and demographic changes in countryside, these goals are often conflicting with the well-being of villagers. Particularly, the technological progress in both crop and livestock production results in the increased rural unemployment. The tightening of product quality and safety requirments reduces the demand on low-quality foods that are produced primarily in households; this tendency is especially pronounced in the livestock sector. Besides, high safety standards can decrease the number of local inhabitants employed on agricultural enterprises. For example, in order to avoid the propagation of bird flu large poultry producers can forbid the employment of local people that raise poultry on own backyards. The environmental requirments might impose essential restrictions on the production practices in rural households, reducing their profitability (Kuhn and Demyanenko, 2004).

**Rural policy comes down to the achievment of two main goals: providing public services in rural areas and supporting their economic development** (Kuhn and Demyanenko, 2004)**.** The first goal is positively asossiated with the development of agricultutal sector while the second goal often imply the inclusion of the majority of rural population into agricultural production. In this respect, optimal rural policy has to be designed in order to eliminate the conflict of interests of villagers and agricultural producers. Particularly, the inclusiveness of rural development should not be covered fully by agricultural production. There are a number of non-agricultual industries that can be developed successfully in the countryside. The other positive factor for the rural economic development is the growing trend for remote work via internet. Apart from the

stimulation of non-agricultural employment, the rural policy has to be based on providing the physical infrastructure (roads, utility services) as well as social infrastructure (healthcare, education).

**The coordination of rural development policy is more complex comparing to agricultural policy.** The responsibilities of rural policy should be horizontally coordinated between several ministries, namely the Ministry of Infrastructure, the Ministry for Development of Economy and Trade, the Ministry of Social Policy of Ukraine, the Ministry of Communities and Territories Development, the Ministry of Agrarian Policy and Food, the Ministry of Education and Science of Ukraine. Moreover, the international experience indicates the importance of decentralization in the development of rural areas. The transfer of key administrative responsibilities to the local level increases, among other things, the transparency and efficiency of using fiscal resources (see Harus and Nivievskyi 2020). Finishing the decentralization reform that was launched in 2014 would also contribute to the improvement in relations between rural population and agribusiness since the agricultural taxes will be the essential source of local budgets replenishing. As Kuhn and Demyanenko (2004) emphasize, the optimal form of rural policy realization is the project approach implemented on the local level. This approach is more targeted and less bureaucratic comparing to the complex national programmes. The local projects should be applied on the periods of several years in order to ensure higher transparency and avoid the political pressure that can be possible in the case of frequent renegotiations.

### 2.4.2 Overall approach and strategy

**Ukraine's agricultural policy is characterised as unreliable and inconsistent** in addressing key constraints in the sector to reach sector development goals. So far there is no officially adopted and Agricultural Development Strategy that would establish conceptual coherence sector development goals, current development constraints and specific policy instruments that would target the constraints in accordance with economic principles. Agricultural policy making continues to be ad hoc and opportunistic, it lacks transparency in application of policy measures and it creates significantly inequitable distribution of benefits. Governance issues have eroded public trust and confidence in the state initiatives.

The Concept of the State Target Program for the Development of the Agrarian Sector of the Economy for the period of up to 2021 was approved by the Cabinet of Ministers by its Order No1437-p, dated December 30, 2015 (with changes and additions from February 14, 2018, N 102-p). The Concept identifies the Government's priority areas for development of the sector going forward, which include a broad range of instruments for sector support, including in the areas of food safety, taxation, environmental support measures, credits and other financial instruments, and agricultural R&D, among others.

**Agricultural state support policy is characterized by a weakening dominance of the market-distorting measures** (market price support measures), budget payments and generous, although declining over time tax benefits to producers and limited spending on growth-enhancing investments and public goods' provision (General services expenditures), see Figure 2.4.1. Market- and trade-distorting policy interventions are getting less intense and severe. All of these policy interventions alter the domestic market price of a commodity as compared to its reference border price. So the market price support (MPS) indicator allows to assess the combined effect of a variety of government market and trade policy measures. Overall in 2019 budget expenditures and imputed tax benefits made up more than 5% of the value added generated by agricultural sector. The allocation of the state support funds has often been done in a non-transparent way, with large shares of support going to the largest producers. Current budget performance indicators are process-oriented and do not allow for the assessment of how different programs contribute to the strategic goals.

**Overall Ukraine's agricultural policy framework is functioning in a way that exportable agricultural products tend to be taxed, while importable ones tend to receive support**. Figure illustrates that poultry, pork, beef receive the largest state support, although this support is declining. In particular, poultry producers could receive close to 60% of their revenues from various market and trade policies. The most important exportable goods (grains, oilseeds and dairy products) are implicitly taxed. Negative % single commodity transfer (driven primarily by negative MPS) for grains reflects export restrictions in 2010-12 that took place in the form of either quotas or export taxes. In marketing years 2012-16, export restrictions took the form of the export VAT non-refunds, as well as the form of voluntary export quotas, whereby traders bounded themselves to export up to 80% of exportable surpluses. Sunflower seeds exports are taxed at 10%.

**Figure 2.4.1. Agricultural state support in Ukraine**

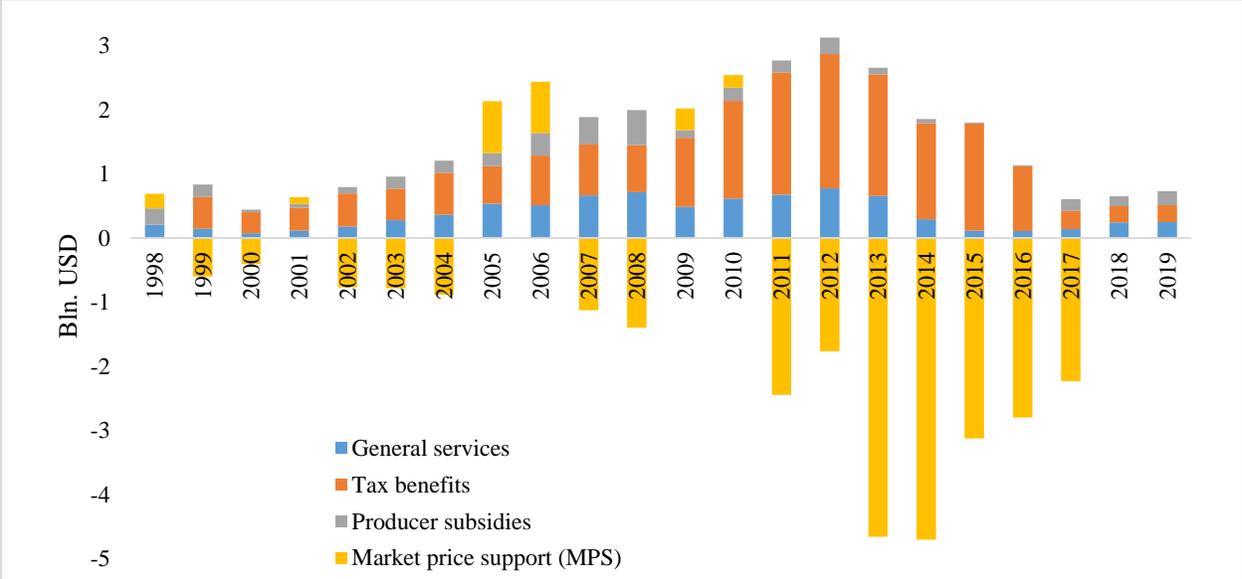

Source: own presentation using OECD PSE data for Ukraine, State Budgets of 2018 and 2019 years. Note: MPS is not calculated for 2018 and 2019, but assumed to be zero

**Figure 2.4.2. Individual products support, % Producer Single Commodity Transfer**

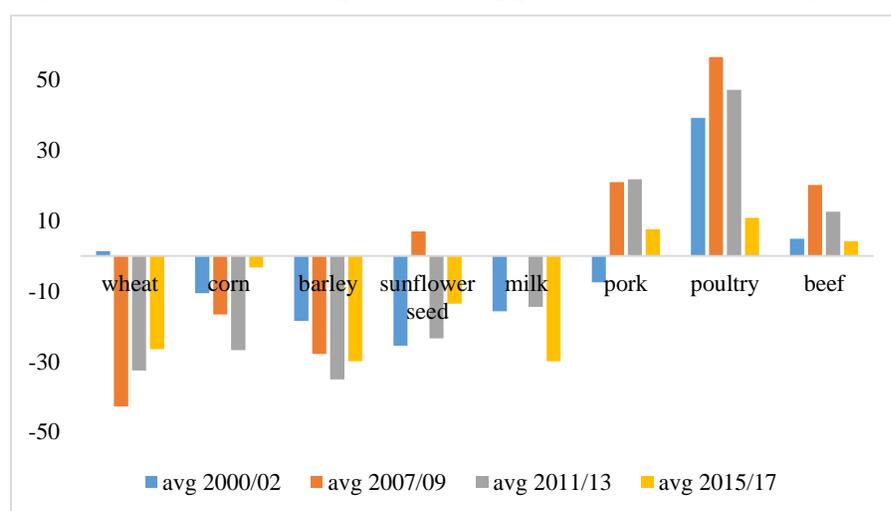

Source: own presentation using OECD PSE tables for Ukraine

**Small farms in the overall agricultural policy development agenda**

Overall small farmers have been side-tracked in Ukraine's agricultural policy agenda over the last 20 years (see a detailed discussion in Nivievskyi et al, 2021). Agricultural support policy in the form of substantial tax benefits and subsidies has been pro-large thus putting small producers at disadvantage in development and growth. This is reinforced by the fact that small farms are disadvantaged in access to financial services due to information asymmetry and transaction costs and by the existing ban on agricultural land sales being in place since 2001 (Nivievskyi and Deininger, 2019).

**Pro-small farms start in the 90s**

In the 90s there was an attempt to turn former and inefficient soviet collective enterprises (ua: "kolhospy" and "radhospy") into a large group of small and medium private agricultural holders by transferring the land of those enterprises to their members and other rural inhabitants. Altogether about 28 mln of agricultural land of collective farms was transferred in shares ("payis"; 3.6 ha on average) into the private ownership of 6.9 mln people or 16.2% of Ukraine's population. This was done in a hope that those people will start cultivating their land plots and develop into a small or medium farming businesses (start cultivating their land plots and develop into a small or medium farming businesses (Demyanenko 2005). This indeed gave birth to more than 40 000 commercial individual farmers – legal entities (ua: "fermerski hospodarstva") and to more than 4 mln family farms – households. Altogether these smallholders produced more than 60% of the gross agricultural output in 2000. Since then, however, due to a following and drastic policy change, precipitated by difficulties of a transition to market economy, they have not developed much and now their contribution to the gross agricultural output has contracted to less than 40%.

**Pro-large farms' policy shift since 2000**

Since 1999, the government of Ukraine introduced a couple of crucial policies at a substantial advantage to large agribusinesses, these are: 1) substantial pro-large tax privileges since 1999, 2) ban on agricultural land sales since 2001; 3) pro-large agricultural subsidies system; 4) glaring underfinancing of public goods, agricultural knowledge and innovation

1. **Pro-large tax privileges since 1999.** In 1999 Government of Ukraine introduced substantial tax benefits for agricultural sector that have been the dominant element of the overall fiscal support to agriculture since then. Tax benefits accrued from a so-called single tax (or Fixed Agricultural Tax before 2015 - FAT) and a special value-added tax regime in agriculture – AgVAT. The FAT is a flat rate tax that now replaces profit and land taxes, but it replaced about 12 other taxes and fees before 2012 (World Bank, 2013). Its rate varies from 0.09% to about 1.00% of the normative value of farmland. In 2010, the FAT resulted in an average tax payment of only roughly 0.75 US$/ha of arable land that left farm profits in Ukraine essentially untaxed. In 2015, due to significant increase of the normative value of land, FAT liabilities increased to roughly $US10/ha, which is also very low compared to what the farmers would have paid on the general tax system.
   According to the AgVAT regime, farmers were entitled to retain the VAT received from their sales to recover VAT on inputs and for other production purposes. In 2016 and 2017 the AgVAT system was gradually eliminated under the IMF and other international donors pressure. In 2015, the benefits from the AgVAT were estimated at UAH 28 bn. In 2017 the AgVAT tax benefit system was terminated and replaced by so-called 'quasi accumulation VAT' regime. This was no longer a tax benefit system, but instead agricultural producers (mainly livestock and horticulture producers) were entitled to receive budget subsidies proportionally to the VAT transferred to the state budget. The total volume of the program was UAH 4 billion. The FAT or profit tax exemption is still in place and is expected to continue.
   Both types of tax benefits are progressive by nature, since they favor or provide disproportionally more support to more productive larger farms thus implicitly favouring large-scale agriculture in Ukraine.
2. **Pro-large ban or moratorium on agricultural land sales since 2001.** Starting 2001, the rights of individual owners to dispose of private land were constrained by the moratorium on sales of agricultural land. The major effect of the moratorium is virtually non-developed rural financing that could use the land as a collateral. And this is despite the fact that smallholders operated mainly on their own land but could not use it to attract financing for their development. And this is on top of the intrinsic disadvantages of small farmers in access to financial services due to information asymmetry and transaction costs. Usually they have no bank-friendly financial reporting (due to the simplified system of taxation and reporting in the agricultural sector – see above) and lack credit history and collateral, making it difficult for banks to assess the risks of extending credit to them. Land sales moratorium is set to expire in July 2021, but it will take quite some time for the rural financing to develop.
3. **Pro-large agricultural subsidies system.** All current producer subsidies generally fall into 5 major programs: 1) concessional credits; 2) individual farmers support; 3) support of horticulture; 4) support of livestock, processing and storage; 5) partial compensation of costs of (domestically produced) agricultural machinery. Despite overall inefficient and ineffective design of the support system (Nivievskyi and Deininger, 2019), the system implicitly focuses on larger agricultural producers and does not focus on family farms at all. First of all, only about 15% of the budget support is channeled specifically to individual small farmers. All other programs are not size specific, but small farmers simply cannot access them due to a specific design. For example, credit concession programs are accessible only to those farms that already have an access to commercial banks' lending. Mostly these are the farms larger than 2,000 ha. Smaller farms (lower than 500 ha size or even below 100 ha) usually do not have access to commercial credits. As a consequence, only a meager share of credit concession subsidies (if at all) ends up with small farmers.

4. **Glaring public goods and agricultural knowledge and innovation under-provision.** Provision of public goods to rural areas including roads, health services, clean water, and schools and investing in agricultural research and extension is considered as a backbone of smallholders supports. This clearly has not been in a priority of agricultural policy since 1998. As a result, advisory and extension services are virtually nonexistent for small farmers, so are the information and knowledge systems (though this is improving to some extent with the development of modern digital technologies), rural infrastructure is in a bad shape inflicting disproportionally larger transaction costs on small farmers, research and development system is virtually nonexistent for small farmers (large producers have and develop their own private systems).

### 2.4.3 Institutional set-up of agricultural policy making

**Institutional setup in agricultural policy making in Ukraine is quite centralized**, with little power delegated to the local governments. On a high level of a decision making process, the following hierarchy exists: 1) the President and Presidential Administration, 2) the Prime Minister and the Cabinet of Ministers (GoU), 3) Ministry of Economic Development, Trade and Agriculture (MEDTA) and Ministry for Agrarian Policy and Food (MoA) [7], 4) other ministries, but primarily the Ministry of Finance (for state support and fiscal issues) and the Ministry for Regional Development, 5) the Parliament of Ukraine (Verkhovna Rada). Draft laws can only be initiated and registered by people's deputies (PD) or a group of PDs, the President, the Cabinet of Ministers, or the National Bank (i.e., the 'subjects of legislative initiative'). The President has a power to veto the laws. Resolutions of the Cabinet of Ministers or/and Decrees of the corresponding responsible ministries are implementing the norms of laws.

On the lower level, the MEDTA via its subordinated agencies implements policies in agriculture. The main MEDTA agencies are:

1) The State Agency for Geodesy, Cartography and Cadastre (SGC in the following) deals with maintaining the State Land Cadastre of Ukraine. The Register of land rights, however, is maintained by the Ministry of Justice.
2) The State Agency for Food Safety and Consumers' Protection is dealing mainly with SPS measures.

Moreover, the GoU and the MEDTA have about 500 their state owned enterprises (SOEs) to affect the agricultural markets. The largest is the State Food and Grain Corporation of Ukraine (DPZKU), Agrarian Fund (market intervention public company) and Ukrspirt (State ethanol producer monopolist). DPZKU is a public joint stock company that was founded in 2010. DPZKU is a relatively large grain trading state enterprise with about 10.5% of the total certified storage facilities, 12% of the total port facilities of the country and significant assets in food processing sector. The DPZKU implements so called 'Chinese credit deal'. In 2012, Ukraine and China agreed on a deal of USD 3 billion. According to this deal, China granted USD 3 billion credit line to Ukraine in the exchange of Ukrainian exports of grain (mainly maize) to China and Chinese agricultural machines, equipment, agrichemicals, and seeds imports to Ukraine in 50-50 proportions, i.e. USD 1.5 billion for Ukraine's exports and USD1.5 billion for the Chinese imports[8].

---

[7] In the fall 2019, the Ministry for Agrarian Policy and Food was merged with the Ministry for Economic Development and Trade. In the fall 2020, the Ministry for Agrarian Policy and Food was singled back from the Ministry for Economic Development and Trade.

MEDTA recently made significant steps towards putting virtually all SOEs (including Agrarian Fund) on privatization and de-monopolization (Ukrspirt). Generally, about 111 agricultural SOEs have to be privatized during 2020-2022.

**Decentralization reform has been unfolding since 2014.** The reform is aimed to delegate the power from the national to the municipal level. The key instrument of the reform is the creation of amalgamated territorial communities (ATCs) that receive rights of self-government, tax collection and public policy. As of 2020, 1470 ATCs have been formed in Ukraine. The territorial consolidation contributed to better performance in terms of local taxes collection and optimization of government expenditures (Harus and Nivievskyi 2020).

### 2.4.4 Agricultural fiscal support policy

**Before 2017, tax benefits were the major source of a fiscal support to agriculture**. Support of general services in agriculture (e.g. sanitary and phytosanitary measures, education and research, food security, extension services etc) or so called «green box» measures (according to the WTO classification) was a second major component of agricultural fiscal support. The weight of direct producers' subsidies (direct budget outlays to producers) has always been the lowest. Substantial tax benefits accrued from the flat rate tax - fourth group of the Single tax (Fixed Agricultural Tax before 2015 - FAT) and a special value-added tax regime in agriculture – AgVAT (described in detailed in the section 2.4.2).

In 2017 the AgVAT tax benefit system was terminated and replaced by so-called «quasi accumulation VAT» regime. This was no longer a tax benefit system, but instead agricultural producers (mainly livestock and horticulture producers) were entitled to receive budget subsidies proportionally to the VAT transferred to the state budget. The total volume of the program was UAH 4 billion. The program was, though, heavily criticized as such that favored primarily large agriholdings and thus was terminated.

**Since 2018 the volume of producer subsidies significantly increased and their structure also changed substantially**. Broadly speaking, now all producer subsidies fall into 5 major subsidy programs: 1) concessional credits; 2) individual farmers support; 3) support of horticulture; 4) support of livestock, processing and storage; 5) partial compensation of costs of (domestically produced) agricultural machinery. Each of these categories contains another 1 to 8 subprograms. So, in one way or another, the above programs could be summarized as so-called input subsidies, i.e. subsidies that decrease the costs of inputs for the farmers (Nivievskyi and Deininger, 2019).

---

[8] Resolution of the CMU #857 as of 13.08.2012.

**Agriculture is a net recipient of a fiscal support**. While agriculture has been one of the most successful and growing sectors in Ukraine, agricultural producers remain net recipients of fiscal support. Figure 2.4.3 below demonstrates that tax benefits (forgone fiscal revenue which is a subsidy) and agricultural subsidies (direct producer subsidies plus expenditures on general services) in total exceed the volume of tax revenues generated/paid by agricultural producers[9].

**Figure 2.4.3. Contribution of agriculture to national economy and to the State Budget of Ukraine, %**

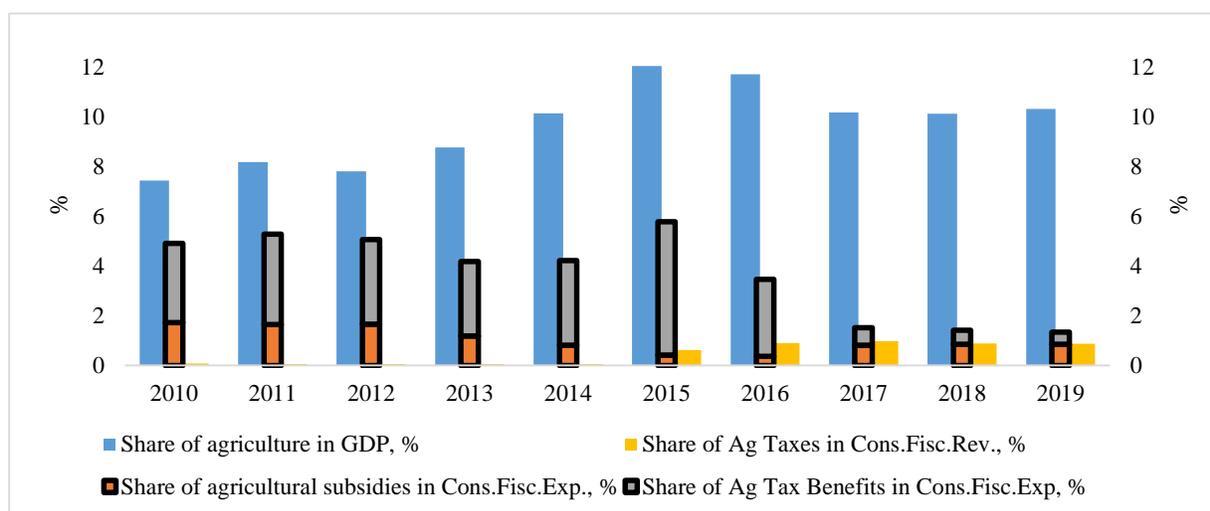

Source: own presentation based on Ukrstat and Treasury data; Notes: Cons.Fisc.Rev. – consolidated fiscal revenues; Cons.Fisc.Exp. – consolidated fiscal expenditures

**Budget support programs were ad-hoc with poor design/implementation without incremental effect on investments and productivity[10].** As a consequence of lacking agricultural development strategy, a design of agricultural support programs has been ad-hoc, without a due public and evidence-based discussion and it was a subject to repeated changes over the last several years. Budgeting and implementation of the programs also has been ad-hoc. Overall level of budget programs execution and its timing remains substantially ad-hoc, 92% of the planned outlays were disbursed in 2017 and only 66% in 2018. This implies the government subsidized investments and purchases that would have made in any case. In order to be effective and to have incremental effect on productivity/output or to change the behavior of producers, budget support programs have to be stable and lasting for a long time so that agricultural producers could develop a trust and include them into their business plans. In other words, one always has to ask a question whether investments, for instance, in a construction of livestock complexes would have been possible without subsidies. If such investments were not possible without subsidies, then these subsidies could be regarded as effective, putting aside a question of their economic effect.

---

[9] We do not account for the revenues from the value added tax (VAT) as well as from the personal income tax (PIT) in agriculture, for these taxes are de-facto paid by either consumers (in case of VAT) or by employees/landowners (in case of PIT) and producers play the role of tax agents only. As it was mentioned above, profits are not taxed in agriculture, but instead a low flat rate tax FAT is levied.

[10] In describing the budget support programs in the following, we heavily rely on Nivievskyi and Deininger (2019).

**Poor design of the programs could be demonstrated with the example of three credit concession programs** tailored for different various farm groups with a total budget of about USD 10 mln in 2019. A fundamental flaw of this program is that a decision about the subsidy (compensation of the interest payments) is made by a special commission (nominated by the MADTA) post factum, i.e. after a decision by the bank on the extension of the credit has already been made. In other words, by applying to the bank for the credit, a farmer does not know whether interest payment will be compensated/or what share of interest payments will be compensated. As a result, the program becomes irrelevant with respect to increasing the volume of agricultural credits. The same design flaw is contained in other major support programs (partial compensation of the cost for construction or reconstruction of livestock farms and partial compensation of agricultural machinery costs), wherein a decision on subsidy is made post-factum, i.e. after a reconstruction and corresponding expenses have been made or after an agricultural machinery has been purchased already. In such circumstances and taking into account a historical distrust among the farmers to budget support programs, one can rather strongly argue that current investments in agriculture (gardens, machinery, seeds, credit etc.) are made irrespective of subsidies and would have made in any case. So current subsidies cannot be considered as effective.

**Fragmented design is reflected in too many budget support programs** (about 24 programs) which dilutes the effectiveness of each of the programs (taking into account limited fiscal resources available for agriculture) and does not address market failures. Apparently this stems from a lack of clearly defined goals of agricultural development and makes an impression that policy makers are trying to tackle a little bit of everything. Fragmented design of the support programs results in

- Overlapping responsibilities: a clear example is the existence of three credit concession programs;
- Conflicting goals: ag machinery costs compensation program is specifically tailored for domestically produced ag machinery which is not necessary in line with increasing agricultural productivity goal. Domestically produced agricultural machinery is often cheaper but of inferior quality/lower productivity compared to the imported one.
- Increased administrative costs: running more than 20 individual support programs certainly requires individual administration and thus increases total costs of administering state fiscal support programs.

**Mounting distortions and poor instruments for achieving higher growth.** As it was mentioned already above, in one way or another, all the producer support measures could be summarized as so-called input subsidies, i.e. subsidies that decrease the costs of inputs for the farmers. According to the WTO classification of agricultural support measures, such subsidies should be classified as distortive (i.e. falling into the amber box measures), for they are to a great extent tailored for specific sectors/products thus distort production structure of agricultural sector by favoring some sectors and ignoring the others.

**The existing documented evidence so far gives a pretty clear guidance on whether coupled support/input subsidies stimulate productivity and efficiency gains in the agricultural sector.** Empirical studies demonstrate that so-called «coupled» subsidies (including input subsidies) have

a major adverse effect on production efficiency and productivity[11]. From a scientific and also practical point of view, we have a competing force of income and substitution effects here. The income effect is that farmers can buy more machinery, for instance, or it is cheaper for them to buy the same number of machinery that can improve their productivity. At the same time, it can lead to so-called soft budget constraints, whereby farmers will invest too much without real economic need. Probably, an extreme example in this case would be Switzerland with the highest tractor intensity in the world. In the United States, for example, tractor intensity is almost ten times fewer tractors per 10 ha. Furthermore, a substitution effect counteracts to the income one because managers may work less hard to achieve a certain level of profitability/efficiency because of the subsidy. Empirical evidence suggest that the substitution effect almost always prevails. An example of the EU, with its heavy support programs, is a clear evidence to that.

**Unintended beneficiaries of inputs subsidies.** Transfer efficiency studies of support programs suggest that the major and ultimate beneficiary of input subsidies are input suppliers – more that 80% of the entire state support ends up in the pockets of inputs suppliers in one way or another. Input subsidies also imply significant dead-weight losses – 5-6% of the state support (Nivievskyi and Neyter, 2018). This is especially evident in case of the program that compensates the costs of domestically produced agricultural machinery. It is widely regarded as a subsidy to domestic producers of agricultural machinery rather than to farmers.

### 2.4.5 Agricultural trade and market policies

**Major external factors that shape agricultural policy making** in Ukraine are its WTO membership and signed Association Agreement with the EU (AA) with its DCFTA. **WTO commitments are substantial.** Ukraine became a WTO member on 16 May 2008 (after almost 15 years of negotiations). This process resulted in a significant liberalization of trade. Ukraine's main commitments under WTO are the following:

- Import tariffs – reduction of binding import duties, including reduction of binding import duties for agricultural goods (see Table 2.4.1) and close to full substitution of specific duties with the ad valorem tariffs. Maximum import duty rates are set for sugar (50%) and sunflower seed oil (30%).

- No quantitative import restrictions of non-tariff measures should be introduced, reintroduced or applied. The only tariff rate quota was set for imports of raw cane sugar (260 thousand tonnes annually and increasing to 267.8 thousand tonnes by 2010).

- Export duty rates – Duty rates should be gradually lowered; moreover, no obligatory minimum export prices should be applied.

---

[11]http://capreform.eu/does-the-basic-payment-make-farmers-lazy/?fbclid=IwAR05NJD6Hgc8We7IK06bwYHI7O1VSWLVKetOmE6ENGSjGOKptxpRWyFlF8g

- No export subsidies in agriculture should be applied.
- Ukrainian national technical regulations, standards and conformity assessment procedures should be based on the international ones. Ukraine committed to reduce number of mandatory certification and to move to the voluntary standards.
- Sanitary and phytosanitary measures should be applied according to the provisions of WTO Agreement, Agreement on Application of Sanitary and Phytosanitary Measures, and Agreement on Import Licensing Procedures.
- Investment regime should be in line with WTO Agreement and Agreement on Trade-Related Measures (TRIMS).
- Ukraine should comply with the Agreement on Trade-Related Aspects of Intellectual Property Rights.
- Ukraine bound itself with commitments regarding market access and national treatment in 11 key sectors and several other sectors
- Government procurement – Ukraine hasn't joined the WTO Agreement on Government Procurement but has observer status in GPA Committee.
- Total Binding AMS is set at UAH 3.4 billion

**Table 2.4.1. MFN structure of import duty rates for agricultural goods in Ukraine after the WTO accession**

|  | At the accession | Final bound rates (2011) |
|---|---|---|
| Zero rate tariff lines (% of all tariff lines) | 9.6 | 10.0 |
| Ad valorem tariff lines (% of all tariff lines) | 62.7 | 95.5 |
| Average arithmetic rate of the applied import tariff* | 13.84 | 11.16 |
| Average weighted rate of the applied import tariff | 18.19 | 10.07 |
| Minimum rate | 0.0 | 0.0 |
| Maximum rate (for ad valorem rates only) | 30.0 | 50.0 |
| International tariff peaks (% of all tariff lines)** | 44.3 | 21.4 |
| Noise' rates (% of all tariff lines)*** | 5.2 | 2.8 |

Source: UNDP (2011); Notes: * calculations by the Ministry of Economy of Ukraine and USAID based on trade data for 2004-2005; ** international tariff peaks are determined as rates higher than 15%; *** 'noise' rates – import rates varying between 0% and 2%

**Table 2.4.2. Changes in export duty rates according to Ukraine's commitments to the WTO (agricultural goods)**

| Commodity | Duty before WTO accession | Duty immediately after accession | Annual rate reduction step | Final rate after WTO accession |
|---|---|---|---|---|
| Sunflower seeds | 17% | 16% | 1% | 10% |
| Livestock | 50%, 55%, 75% depending on livestock type | 50 | 5% | 10% |

Source: UNDP (2011)

**DCFTA with the EU formally came into force in 2017.** It triggered comprehensive institutional and regulatory reforms and created an important opportunity for Ukraine to further boost the value of its agricultural exports. For Ukraine was granted 36 tariff-rate quotas (TRQs) and another set of TRQ expansions, known as Autonomous Trade Measures (ATM).

**According to the DCFTA, the level of tariff protection is higher for the EU countries than for Ukraine.** For protection of domestic market, Ukraine uses mainly ad valorem duties, while EU can impose specific and combined duties, as well as use entry prices. For some categories of agri-food products in Ukraine there is not a complete abolition of the tariff, but its reduction to a certain limit. In the case of EU, import tariffs were usually fully eliminated, but only within the volume of TRQs that were introduced for some goods. Exporters that supply production to EU above the quotas should pay the full amount of import duties. Unlike Ukraine, which has a transition period of 10 years, the EU has pledged to abolish most tariffs from the first year. The establishment of duty-free TRQs is provided for 36 categories (beef, pork, lamb, poultry, milk, cream, yogurt, cereals, bran, honey, sugar, starch, mushrooms, garlic, malt, grape and apple juices, butter, cigarettes, ethanol, eggs and albumin, others). At the same time, tariff quotas for 18 product groups provide an increase in volumes during the first five years from the date of signing the Agreement.

**Table 2.4.3. EU import TRQs for Ukrainian food products**

| Product | UKTZED code | The amount of quota | Over-quota tariff |
|---|---|---|---|
| **Beef** | 0201, 0202 | 12 thds. tons | 12,8% + product-specific duty (from 141,4 to 304,1 €/100 kg) |
| **Pork** | 0203 | 20 thds. tons + additional quota 20 thds. tons for some codes | from 46,7 to 86,9 Euro/100 kg |
| **Lamb** | 0204 | 2250 thds. tons for some sub-codes | 12,8% + product-specific duty |
| **Poultry** | 0207, 0210, 1602 | 20 thds. tons + additional quota 20 thds. tons for codes 0207 12 (10-90) | from 18,7 to 128,3 Euro/100 kg |
| **Milk, cream, condensed milk, yoghurt** | 0401, 0402, 0403 | 10 thds. tons | from 12,9 to 183,7 Euro/100 kg |
| **Milk powder** | 0402, 0403, 0404 | 5 thds. tons | from 12,9 to 183,7 Euro/100 kg |
| **Butter** | 0405 | 3 thds. tons | from 189,6 to 231,3 Euro/100 kg |
| **Eggs** | 0407, 0408, 3502 | 3 thds. tons + additional quota 3 thds. tons for code 0407 00 30 | from 30,4 to 142,3 Euro/1000 units |

Source: Annex I-A to the Association Agreement, Practical guide to EU agricultural exporter

**Despite the liberalization of bilateral trade within the DCFTA, the majority of Ukrainian food products are not allowed in EU countries due to low quality and safety characteristics.**

At the same time, the AA with the European Union (ratified in 2014) stimulates the legislative changes aimed to harmonize Ukraine's SPS regulations with EU requirements. The key document in this area is «Comprehensive Strategy of Implementing Legislation on Sanitary and Phytosanitary Measures» accepted in 2016.

**Current protection of Ukraine's agriculture via import duties is generally modest.** The rest of the economy is, however, more open to international competition. Table 2.4.4 shows that the simple average applied import duty for agricultural commodities is 9.2% — well below the corresponding final bound 11.1%.

**Table 2.4.4. Tariffs and imports – summary, in %**

|  | Year | Total | Agricultural commodities | Non-Agricultural commodities |
|---|---|---|---|---|
| Simple average final bound [committed under WTO] | - | 5.8 | 11.1 | 5.0 |
| Simple average MFN applied | 2018 | 4.5 | 9.2 | 3.7 |
| Trade weighted average | 2018 | 2.8 | 5.5 | 2.6 |

Source: WTO; Note: MFN= most-favored nation

**On a more disaggregated level, sugar is the most protected agricultural product in Ukraine.** Ukraine applies a TRQ on sugar imports at the level of 267.8 million tons (under WTO commitment). Within quota import duty is 2%, while the above quota duty is 50% (Table 2.4.5). Sunflower oil is the second most protected agricultural commodity. The import duty for sunflower oil is at a prohibitive 30% rate. Although the average import duty for cereals is higher than for oilseeds, fats and oils (see Table 2.4.5), it is rather irrelevant as grain exports by far exceed imports. Nominally animal products are moderately protected with an average 10.6% import duty. In reality, however, the gap between domestic and world prices is much higher. To some extent, excessive non-tariff trade barriers in the form of demanding import procedures and regulations are responsible for this state of affairs.

**Table 2.4.5. Tariffs and imports by agricultural product groups in 2018, in %**

| Product groups | Final bound duties | | | MFN-applied duties | | | Imports | |
|---|---|---|---|---|---|---|---|---|
|  | AVG | Duty-free | MAX | AVG | Duty-free | MAX | Share | Duty-free |
| Animal products | 13 | 0 | 20 | 10.6 | 10.1 | 20 | 0.4 | 25.8 |
| Dairy products | 10 | 0 | 10 | 10 | 0 | 10 | 0.1 | 0 |
| Fruit, vegetables, plants | 12.6 | 10.3 | 20 | 9.9 | 19.4 | 20 | 1.5 | 56.7 |
| Coffee, tea | 5.8 | 35.4 | 20 | 5.7 | 35.4 | 20 | 1 | 46.1 |
| Cereals and preparations | 12.5 | 3.4 | 20 | 12.4 | 3.8 | 20 | 1.2 | 39 |
| Oilseeds, fats and oils | 10.5 | 11.5 | 30 | 8.3 | 18.4 | 30 | 1.2 | 87.2 |
| Sugars and confectionery | 19.4 | 0.5 | 50 | 19.4 | 0 | 50 | 0.1 | 0 |
| Beverages and tobacco | 8.4 | 24.6 | 71 | 8.4 | 25.1 | 73 | 1.6 | 24.5 |
| Cotton | 1.4 | 40 | 5 | 1.4 | 40 | 5 | 0 | 34.8 |
| Other agricultural products | 7.9 | 24.9 | 20 | 5.4 | 47.2 | 20 | 0.6 | 23 |

Source: WTO. Note: AVG = average; MAX = maximum; MFN = most-favored nation.

**Export restrictions have been exercised manifold on major export crops allegedly to stimulate domestic processing and value addition.** Export restrictions on grains took place in the form of either quotas or export taxes in 2006/07, 2007/08, 2010/11 and in 2011/12 marketing years. In marketing years 2012/13 and 2013/14, export restrictions took the form of voluntary export quotas. Traders voluntarily agreed to cap their grain exports at 80% of the grain exportable volumes to reduce the uncertainty of restrictions. Also from 2013 till March 2016 export restrictions took the form of the export VAT non-refund (Kirschke et al. 2019). The economic effects from export restrictions were devastating. In the short-term, Ukraine and farmers lost important export revenues due to a dampening effect of export restrictions on domestic prices, resulting in a relatively huge negative MPS and substantial implicit taxation of grains in those periods (Figure 2.4.2). World Bank (2013) assesses (using OECD PSE tables and other studies) that these forgone revenues amounted to: USD 1.3 billion in 2007, USD 3.9 billion in 2008, from USD 1.9 billion to USD 2.6 billion in 2010/11 marketing year. However, in the medium to long-term, export restrictions created major disincentives for domestic and foreign investors to undertake capacity-enhancing investments in production, marketing infrastructure and related services.

**Oilseeds have also been affected by export restrictions.** 23% export duty for sunflower seed was introduced in 1998 to stimulate domestic processing of sunflower seeds (Kuhn and Nivievskyi, 2004). Due to a pressure from international partners during WTO access process, export duty was first decreased to 17% in 2001 and then it was annually decreasing by 1 percentage point from 17% in 2008 to the final agreed 10%. Sunflower seed crushing capacities indeed increased substantially since then, allowing Ukraine to emerge as the largest exporter of sunflower oil in the world. Whether this policy is a success, still remains to be explored in details, for the costs (i.e. forgone revenues of farmers) and benefits (additional value added and processors benefits and tax revenues for the state) of such a policy have never been explored and calculated in details. There is, however, emerging evidence to cast doubts that such a policy experiment is beneficial for Ukraine's welfare. The most recent example of export restrictions is the export VAT non-refund for soy beans that has been introduced from September 2018. This dampened domestic prices relatively to export prices by about 26-29 USD/ton (Nivievskyi et al, 2019).

**In general, agricultural trade policy framework in Ukraine is biased toward livestock producers and processors at the expense of crop producers.** Figure 2.4.2 illustrates that poultry, pork and beef consistently receive the largest state support, although this support is declining. Meanwhile, the most important exportable goods (grains, oilseeds and dairy products) are implicitly taxed. The distortions created in Ukraine's agricultural incentives framework appear to be systemic and consistent.

### 2.4.6 Designing an efficient agricultural support policy

**Guiding principles in designing efficient agricultural support policy[12].** Building upon the stock of knowledge on best practices in agricultural support policies and taking into account Ukraine's political economy/reform content, Ukraine's agricultural support measures should be structured/designed against the following guiding principles:

---

[12] In the following of this section we closely follow Nivievskyi and Deininger (2019b, 2019c).

- **Do not pick up the winners - products/sectors.** If a government explicitly supports a particular sector with subsidies, this implies the government can correctly pick up the right sectors and correctly foresee their future and contribution to the overall economic development. Formally speaking this is so-called infant industry promotion policy. There are, however, two strong arguments against this industrial policy: 1) it is an illusion that a government is in the best position to identify correct industries, products and firms to support, since it requires deep knowledge of the markets and technological processes; this is especially a problem for developing and transition countries where analytical capacity of governments is very limited; 2) in selecting winners, government may be influenced by bribes and lobbying, which generate big distortions and lead to market inefficiencies. Creating top-down list of sectors that require the government support proved to be counterproductive. Instead, it is much better to create an environment where all types of companies in all sorts of industries are able to produce and experiment with different products. The markets will then select good and bad products without any government involvement.
- **Focus on market failures**. There are many cases, however, when governmental intervention is well justified and desirable, i.e. in case of market failures, when market efforts are not enough to provide adequate signal to react. They include all kinds of negative externalities (e.g. environmental problems, climate change). Positive externalities in the form of public goods and services is also a well justified excuse for government intervention, in particular they extend the benefits to all producers. For example, building up the roads, a modern and efficient plant and animal health and food safety systems, efficient land governance infrastructure, information systems, education, research and development, extension services etc. These are all services that benefit all producers but the private sector is not able to supply a sufficient level of public goods output.

  Market imperfections is another case of a market failure. Imperfect financial (credit) markets is perhaps a common one for developing and transition countries. In Ukraine this is magnified by the land sales moratorium, whereby especially small and to some extend medium agricultural producers have no access to credits. This precludes small farmers from making productive investments, increasing their productivity and grabbing higher market shares and incomes. WTO domestic support framework does not restrict its members in expenditures on such activities, which all fall into the green box category of domestic support measures. So implicitly WTO nudges its members to shift support measures from amber to green box measures.
- **Consider fiscal constraints and targeting**. Countries very often face fiscal constraints. This is especially so for countries like Ukraine with its difficult fiscal and macroeconomic situation, significant budget deficit and war in the East. In these circumstances agricultural fiscal support budget is expected to be quite limited and it is important to design farm-income support measures targeting those in a real need.
- **Do not intervene in producer and consumer pricing** to allow farmers fully benefit from international markets and ensure efficient use of domestic resources. Fixing floor prices above the market ones leads to the excessive support of producers at the expense of taxpayers and consumers and to the distortion of market incentives. At the same time, fixing the ceiling prices benefits all consumers (including the rich ones) at the cost of producers. In both cases, targeted support works better than price regulation. This is especially true for food aid programs for poor population which are actively used worldwide.
- **Support to public goods** (e.g. sanitary and phytosanitary measures, food safety, information systems, physical rural infrastructure, education and R&D) is essential to increase return on investments and export potential. Knowledge transfer and financial literacy training should be provided to increase small farmers' awareness and enable them to put together viable investment proposals.
- **Levelling off or Improving access to credit for small farms.** Small farms are disadvantaged in access to financial services due to information asymmetry and transaction costs. They have no bank-friendly financial reporting (due to the simplified system of taxation and reporting in the agricultural sector) and lack credit history and collateral, making it difficult for banks to assess the risks of extending credit to them. This problem is reinforced by the agricultural land sales. While lifting of the moratorium on agricultural land sales provides the preconditions for use of land as collateral for credit markets, credit history by potential borrowers, and familiarity with the sector by banks implies that initially the risk of providing credit to the agricultural sector remains high. A partial credit guarantee (PCG) can reduce such risks without eliminating the responsibility by banks, ideally in combination with other risk management techniques (e.g. crop insurance) to address systemic risk.
- **Investment support to (new and small) agricultural entrepreneurs**. Agricultural support policy in Ukraine in the form of substantial tax benefits and subsidies has always been pro-large thus putting small producers development at disadvantage. To allow small (emerging) farmers compete on an equal footing with larger and established players and correct for a long-lasting policy failure, reshuffling current highly inefficient, distortive and unfair subsidies towards a simple and targeted support to facilitate capital upgrade and diversification may be needed. This could take the form of (i) matching grants; (ii) interest rate subsidies; or (iii) transaction cost subsidies to compensate banks for the cost of serving small clients. Such programs should highly rely on good quality financial intermediaries and could be administered jointly by an entity in charge of providing a partial credit guarantee. Targeting the purpose of financing and clientele is a key element. Targeting capital investments should be a priority, but working capital financing should not be completely excluded either. The target group should be defined carefully. Eligibility criteria should primarily focus on farms turnover and based on the existing evidence. Also, to pursue diversification, oilseed, grains and poultry farms should be excluded from the target farms.

**The outlines of an efficient agricultural support policy based on the above guiding principles are the following.** Rather than trying to 'pick winners', the public support to the agricultural sector should be focusing on achieving high overall sustainable productivity and competitiveness growth and resilience through targeted support to correct for market and policy failures and for more diversified agricultural production structures.

## 2.5 Influence and development of up- and downstream sectors

Although the development of agricultural sector depends on the appropriate agricultural policy, it also depends on the competitiveness of up- and downstream sectors as well as on the policy applied applied to these sectors. The upstream sectors provide resources for crop and livestock production, thereby affecting agricultural productivity and profitability. Downstream sectors use agricultural commodities as the inputs for further processing, thus affecting the demand and output prices for agricultural products. In general, Ukrainian farmers do not compete directly with farmers of other countries. Instead whole national and global food chains are competing against each other on basis of prices, quality, safety and punctual delivery. Hence, the success of Ukrainian agriculture on international markets depends on both own performance and efficiency as well as on the performance and efficiency of up- and downstream sectors, which are addressed in the following subchapters 2.5.1 through 2.5.10 in more detail.

### 2.5.1 Agricultural land market

**The emergence of a full-fledged agricultural land market is yet ongoing in Ukraine**. While 'rental' arm of the land market is functional in Ukraine and constitutes the main channel of farmland transactions for farmers and landowners, its 'sales and purchases' arm is virtually dysfunctional due to the moratorium on farmland sales and purchases that has been in place in Ukraine since 2001. It was introduced as a temporary measure to protect landowners in a situation of underdeveloped land market infrastructure in the process of agricultural land privatization after the breakup of the Soviet Union in 1991 (KSE, 2021).

**Agricultural land privatization and market development run in several stages**. It started in 1991[13] when a state monopoly on lands was abandoned all lands were declared to be subject to land reform. With a new version of the Land Code as of on March 13, 1992, the Parliament of Ukraine introduced collective ownership of land and transferred property rights and management of agricultural land (except some land left in a state land reserve) from the traditional Soviet agricultural enterprises (collective and state enterprises – *kolhospy and radhospy*) into the collective ownership of their transformed peers – collective agricultural enterprises (CAEs). As a result, in January 1993, 99.5 percent of more than 11,000 CAEs received 27.6 million hectares of agricultural land in a collective property/ownership (Demyanenko, 2005).

The transfer of the land into the collective property of CAEs, however, did not make its members real landowners. To strengthen the status of CAE members as co-owners of collective property, a distribution or privatization of the CAEs agricultural land among their members begun in the fall 1994[14]. Each CAE member was given an allotment (share) of land (corresponding to a virtual plot of specified size in an unspecified location), which was identified with a certificate. Land share was 3.6 ha on average, depending on the size of the CAE and the number of workers. Owners of these allotments received the right to manage, physically identify and own them. As a result, 6.92 mn rural residents (about 16% of total population) —members of about 11,000 CAEs—received these certificates for more than 27 mn ha of agricultural land (about 45% of the total territory of Ukraine) in private ownership.

---

[13] I.e. from March 15, 1991, when the Land Code and decree "On Land Reform" came into force and all lands were declared to be subject to land reform
[14] Presidential Decree №666/94 "On Urgent Measures to Speed Up of the Land Reform in Agricultural Production." https://zakon.rada.gov.ua/laws/show/666/94

Distribution of land shares was deterred by a slow process of CAEs restructuring into the new forms of agricultural enterprises, for collective ownership was legally abandoned with a new Constitution of Ukraine that came in force in June 1996. Presidential Decree[15] in December 1999 substantially speeded up the process of CAEs restructuring, so that in March 2000 virtually all CAEs turned into new legal forms of agricultural enterprises (private individual farms, corporate enterprises, limited liability companies, private enterprises etc). By 2013, 96.7 percent of land owners converted these land shares certificates into legally valid land deeds, and so became owners of land plots with specified locations and not just virtual ones. Some 4.5 mn ha of land of other designated use types were privatized by decisions of local governments. The new Land Code was adopted by the Parliament of Ukraine at the end of 2001 and it introduced a moratorium or ban on sales and purchases of virtually all agricultural land in the country (see for more details below).

**So as of today, agricultural land ownership in Ukraine is predominantly private.** Out of 42.7 million hectares of agricultural land (or about 71% of Ukraine's territory) 32million hectares comprise private ownership and 9.2 is in the state and communal ownership. With a launch of the national decentralization reform in 2015, about 1.68 mln ha of agricultural land were transferred from the state into a communal ownership of local communities by the end of 2019. Still there are about 7.1 mln of agricultural land is in the state ownership.

**Leasing of agricultural land is the predominant type of land market transaction in Ukraine.** Agricultural producers operate predominantly on leased agricultural land. Of the 20.5 million hectares of agricultural land used by agricultural producers, 17.4 million hectares (or 84.5 percent) was leased as agricultural land. Land is now typically leased minimum of 7 years, with maximum periods of 49 years. There is another type of a lease contract which is called *emphytheusius* that is not limited in terms of a payment structure and contract duration. So the entire volume of future lease payments could be paid upfront and the contract can last for hundreds or more years, for example. This type of contract is de-facto a quasi-purchase and has been increasingly used since 2016 to go about the farmland moratorium; now more than 1 million ha of agricultural land is operated under this type of contract.

Land lease rates gradually increased to current USD125/ha (Figure 2.5.1) but still are relatively low compared to a return on land (Tavrov and Nivievskyi, 2019) and below the rates that would be expected in the presence of a functioning land market. A high proportion of rental payments is made in kind.

---

[15] Presidential Decree №1529/99 "On Urgent Measures for Speeding Up Reformation of the Agricultural Sector of the Economy» as of 03.12.1999; https://zakon.rada.gov.ua/laws/show/1529/99

**Figure 2.5.1. Value added in agriculture and rental payments**

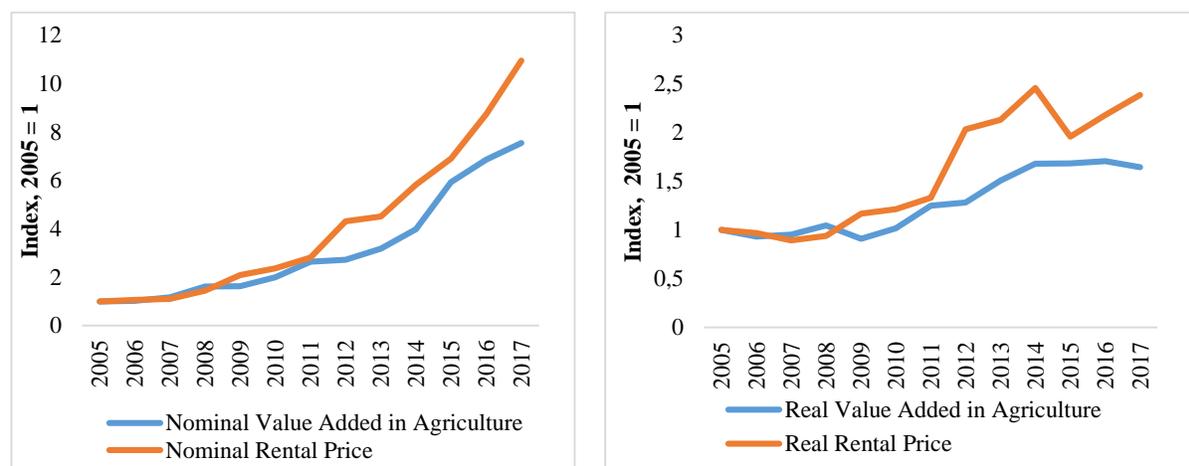

Source: own presentation based on WDI and SGC data

**The scale of the moratorium is impressive**. As it was already mentioned above, buy-and-sell transactions with virtually all agricultural land have not been legally allowed in Ukraine since 2001. Out of more than 42 mln ha of agricultural land, about 38.8 mln ha is under the moratorium. At least 16% of Ukrainian citizens (owners of land plots) are affected by land moratorium as they cannot dispose of their land shares (pai) at their discretion. In 2018 the European Court for Human Rights[16] has recognized the moratorium as a violation of property rights.

Before 2001, land shares certificates and land deeds were traded. Land moratorium was initially adopted for five years, and subsequently was renewed 10 times. Despite the moratorium, however, agricultural land is being 'bought' and 'sold' via a variety of informal arrangements that likely will be formalized whenever the moratorium is finally lifted.

**The moratorium on land sales is one of the extreme examples of tradability restrictions** and unlike other types of restrictions, empirical evidence on the impact of moratorium is very limited in the empirical literature (Nivievskyi and Deininger, 2019a). Among the main negative consequences of the moratorium is less efficient use of land because of barriers for redistributing of land resources to a more efficient owner and producer. This is one of the reasons for a substantial agricultural productivity gap in Ukraine. Despite one of the most fertile soils in the world, Ukraine's agricultural productivity is just a fraction of that in comparable countries. In particular, in 2016 Ukrainian agriculture generated only USD 416 of value added per hectare of arable land, in contrast to USD 2239/ha in France, USD 1414/ha in the US, USD1266/ha in Brazil and USD 773/ha in Argentina.

Also land moratorium prevents from increasing of the investment inflow and expansion of financial capacity especially for small and medium producers, as there is no opportunity for them to use land as collateral (Nivievskyi and Deininger, 2019a). Small and medium producers

---

[16]https://hudoc.echr.coe.int/app/conversion/pdf?library=ECHR&id=003-6089956-7847640&filename=Judgment%20Zelenchuk%20and%20Tsytsyura%20v.%20Ukraine%20-%20ban%20on%20sale%20of%20agricultural%20land.pdf

experience lack of financing as they cannot use one of their most valuable assets as collateral to obtain it because of the ban on land transfers. As a result, financial restrictions undermine opportunities for farmers to grow further and produce more products with higher value added. This is especially relevant for Ukraine, where small farmers work primarily on their own land. This is also empirically supported by the evidence from the CEE transition economies that suggests an increase of the total factor productivity as the farm access to credit increases, partially reflecting that an improvement in access to credit leads to an adjustment of the relative input intensities on farms (Ciaian et al, 2010). Absence of opportunity to purchase land leads to uncertainty regarding the ability of a long-term use of land. This, in its turn, has a noticeable adverse effect on the incentives of sustainable land management to maintain its high quality through crop rotation, investments in irrigation, planting perennials. Also it erodes a stimulus to undertake productivity enhancing investments (Deininger et al, 2017).

**International experience and empirical evidence on the structure of farmland sales markets with land market restrictions[17].** There is a large array of models in the world of how the state interferes with farmland sales markets. There is also a large international economic literature on the consequences of introducing various restrictions on the land market. The literature provides a rather clear suggestion on the most appropriate farmland sales model that is capable of incorporating various peculiarities and circumstances. This suggestion can be summarized in the following way:

- **Farmland sales restrictions, including bans, have rarely achieved their desired results.** There have been many cases where centralized restrictions on land sales seemed justified, but enforcement challenges created distortions that only worsened the situation. Governments' measures to improve farmland sales markets all over the globe have either led to higher transaction costs for participants or have driven land transactions to the informal realm, reducing the welfare of all participants. Universal limitations on farmland sales markets are unlikely to be effective, but may lead to the emergence of large bureaucracies that develop a self-interest in maintaining these restrictions.
- **Fexible forms of economic incentives (e.g. land tax, fees, and tariffs) are preferable to rigid restrictions.** Land tax is one. The most important way in which governments can help to improve the functioning of farmland sales markets is to eliminate distortions; to help reduce transaction costs that would increase the barriers for participation, especially by the poor and smallholders; and to improve the functioning of financial markets. Economic literature implies that the only justifiable interventions are temporary land sales moratoria or limits on accumulating extremely large tracts of land (preventing local monopolies) in situations of rapid transition or emergence of new markets.
- **Decentralized approaches are preferable when defining the model and restrictions on the farmland sales market.** The market for farmland is primarily local and has its own peculiarities. If transparent mechanisms for decision-making are available and local communities bear the costs of their decisions, they may be given the authority to restrict the transferability of land on their local farmland market. This will help strike a balance between the social and economic issues related to introducing and operating the farmland sales market. The expectation is that with changing economic circumstances and territorial competition, restrictions will be relaxed. Where transparent mechanisms are unlikely to prevail, the preferred policy should be to forgo restrictions.

---

[17] See Nivievskyi et al (2016) for a more detailed discussion.

Ex-ante modelling evidence on the outcomes of various designs/options for land market opening is virtually non-existent, except the study of Nivievskyi and Deininger (2019a). In the paper provide the ex-ante estimates of the moratorium impact were simulated using a classical partial equilibrium analysis and existing farm-level performance data. Specifically, various scenarios of the future farmland market design options were simulated subject to how the incomes of various stakeholders and agricultural value added would change. The list of the modelled scenarios includes various combinations of the following options: phasing versus one-go opening of the land market (i.e. allowing transfer of state land first, to be followed by private land later), access of foreigners (yes/no), access of small farms to credits (yes/no), availability of targeted support program for small farmers to allow their productivity improvements (yes/no), multiple land ownership restrictions for physical persons and legal entities, including implicit control against excessive land concentrations. Depending on the scenario setup, the modelled impact ranges from nearly 0% of additional annual GDP growth over the next 5 years time horizon (most restrictive scenario whereby legal entities have no access to the farmland sales market) to 1.86% or USD 10.57 billion of the most liberal market design with access of foreigners, no land ownership ceilings caps and financial support of small farmers to access the capital and increase their productivity.

Based on the results of the modelling scenarios and taking into account global experience, a desired farmland market design for Ukraine could take the following shapes:
- Open markets in one step: Starting to allow transfer of state land first, to be followed by private land later is not recommended for several reasons. Total state land supply at the moment is limited at only about 2 mn ha of the estimated 9 mn ha of registered state land suitable for agriculture, most of which is leased and thus cannot be sold. Markets relying on state land only would thus be very thin, resulting in high prices and outcomes biased towards wealthy individuals/agricultural companies. As state land is expected to be transferred to amalgamated communities (OTGs) who may want to decide for themselves whether to lease or sell, using state land as the motor for land market opening would also conflict with the decentralization agenda. Given state land market low liquidity, banks would not be interested in getting involved so that small farmers already owning land will not be able to access credit for working capital or investment, diversification and job creation. Finally, lifting the moratorium on state land will not allow Ukraine's seven mn landowners to exercise their constitutional rights while tarnishing the country's reputation as it fails to act on the ruling of the European Court of Human Rights that found the moratorium on agricultural land sales indeed violating human rights and required changes to eliminate this. Most importantly, the growth benefits one could expect from lifting the moratorium on state agricultural land are very limited.
- Strictly enforce anti-monopoly legislation: Regulation to ensure competition in the land market is essential to avoid exercise of market power. Strict enforcement of anti-monopoly regulation that limits the share of land owned by one entity to 35% of agricultural area of an amalgamated community (OTG) is necessary to avoid undesirable outcomes*. Lower size limits can be defined in local land use plans and enforced at OTG level. Higher level thresholds currently discussed (i.e. 8% of agricultural area of an oblast, and 0.5% for the nation) are essentially a political decision. Speculative land holding should also be discouraged by increasing land tax rates to realistic levels that can be varied within certain bands at local level and by improving enforcement via electronic link between the Land Cadastre and the State Fiscal Service.
Beyond these, nationally uniform limits on land holding size (e.g. 200 ha for individuals and 1000 ha for legal entities) that neglect the country's regional diversity and the potential of variation and changes in optimum farm size over space and time and that are difficult to enforce and easy to circumvent are not recommended. International experience shows that such restrictions hardly ever worked as intended anywhere but instead created distortions, corruption, and a shadow economy. Moreover, simulations suggest that the costs in terms of foregone growth would be high.
- Allow legal entities to buy land: Restricting land market participants to individuals only will limit demand for land, keep prices low, and limit benefits to landowners as well as economic impact. Commercial banks will not be interested in extending/developing land financing instruments for individuals only, so credit and financial market benefits are unlikely to materialize and the scope for reallocation of land to better producers will be scant. To allow effective implementation of anti-monopoly and anti-money laundering legislation, only Ukrainian legal entities beneficially owned by individuals who are registered in Ukraine should be allowed.
- Provide financial support to small producers: As tradable land is an ideal collateral, functioning land markets can unlock large amounts of mortgage lending. This would benefit small producers who mainly operate own land and who could access credit to invest in intensification and high value-added crops if the moratorium were lifted. Lack of familiarity with and perceived high risk of the SME sector may, however, prevent banks from providing credit in the initial period after market opening, potentially undermining SMEs' competitiveness, market participation, and growth. A partial credit guarantee (PCG) can reduce this risk and allow SME access to finance. The mechanism is establishment and initial capitalization of a commercial agency that, for a fee, assumes part of the risk of default by targeted groups on the credits they get. Preliminary calculations suggest that to capitalize an agency that would cater to initial demand for investment and land purchase about USD 60 would be needed. If a commercially run private entity with majority private participation were set up, donor, IFI, and private sector support could cover all or part of this, potentially supplemented by part of the agricultural subsidy budget.
- Support SME investments through redirecting agricultural subsidies: Most Ukrainian farmers currently produce low margin field crops rather than orchards or horticulture because they lack market links and access to capital for investments, e.g. in irrigation, that could easily double their output per hectare. With developed financial markets and agricultural value chains, credit for such investments would be available. Yet, even after moratorium lifting, these developments will take time. Bridging this gap by providing investment grants to SMEs, possibly administered by banks together with PCG resources, would be a more appropriate use of state subsidies of USD 250 mn per annum which currently mostly go to waste. A more realistic tax regime for the agricultural sector, which will be needed in any case, could then recoup some of this investment in the future. Calculations suggest that such a measure which could be operationalized quickly and could significantly add to GDP growth. A comprehensive farmer registry to verify farmers' land data to establish eligibility and also reduce banks' lending cost would need to be established, and this process has started.
- Do not completely forbid foreign ownership: Land purchases by foreigners imply a host of risks, most importantly money laundering, use of land acquisition for political motives and the fact that foreigners may cause irreversible damage and then just leave the country. Yet few of the key agricultural exporters ban foreign land ownership. Instead, they opt to carefully regulate and scrutinize such investment. The reason is that foreign land ownership provides important benefits that, in the case of Ukraine, would include (i) the ability to tap capital, technological know-how, and access to value chains especially in horticulture and fruits, that are not available locally; (ii) the scope for such investment from the EU to help improve EU market access in return; (iii) the improved transparency associated with FDI from developed countries which are often subject to strict transparency rules; and (iv) higher benefits to landowners in the form of higher land value. The fact that problems caused by foreigners are due to gaps in regulations or enforcement that are also exploited by country nationals further reinforces this.
Instead of banning foreign land ownership altogether, it could thus be prudent to set clear standards in terms of transparency (e.g. no shell companies) or national co-ownership and make foreign land acquisition contingent on criteria, e.g. minimum levels of investment, job creation, or exports of what it produces, to be achieved. As long as these are centrally monitored and enforced, communities could decide if (or under what conditions) foreign land acquisition is allowed in their local development plans.
- Take active measures to minimize risks: Measures to ensure regulations are enforced and the risk of abuse is minimized are greatly facilitated by institutional reforms and technological links. They include:
  - To enforce anti-monopoly laws, registration software should automatically block transactions if restrictions are violated; transaction records and cadastral data should be publicly available; and free messages could be sent to all parties (including local Government) potentially affected by a registered transaction to protect against fraud and allowing to raise objections.
  - Recording and publication, subject to privacy restrictions, of sales price data; using rules to check for transaction prices in force for other real estate and exploring additional seller protections (e.g. a mandatory sign-off on any transactions apparently below-market value) to protect against fraud.
  - Reliance on the free legal aid system already established by the Ministry of Justice with monitoring of outcomes throughout the system to preclude coercion or involuntary dispossession of ill-educated land owners.
  - The scope for negative environmental impacts should be reduced by routinely monitoring land use and compliance with local land use plans using remote sensing data rather than ad hoc mechanisms.
  - To provide funding for SME support, current subsidy programs should be restructured and targeted more effectively and linked to farmers' registry & local land use plans, land tax/fee collection be streamlined and the agricultural tax regime be reviewed to ensure the sector pays its fair share to the state budget. An independently commercially run partial credit guarantee agency involving private & IFI seed capital should be established (Nivievskyi and Deininger, 2019).

**Land market infrastructure has been established and developed to a relatively good level.** Agricultural land registries (registry of land – cadaster and registry of land rights) are separated. The State Land Cadaster is maintained and managed by the State Geocadaster. In 2013 of a unified electronic State Land Cadaster linked to the Registry of Rights was developed and launched (see the Public Cadastral Map – land.gov.ua). Registration of rights to agricultural land is done by the Ministry of Justice of Ukraine, and is part of the Registry for immovable real estate. By 2017, the State Registry of Rights recorded about 2 mn transactions per year with agricultural land, and more than 0.2 mn transactions with non-agricultural land. About 76% of transactions with agricultural land are leases, followed by bequests (18%) and sales (3.1%). For non-agricultural land, 36.8% of transactions are sales, followed by bequests (26.8%) and leases (19.2%). Mortgage transactions are almost non-existent (Nizalov, 2019).

**The legislative framework for lifting the moratorium on agricultural land and establishing a full-fledged and transparent land market is almost completed.** Land market and governance efficiency has been stifled by multiple moratoriums over the last 20 years. The new Land Code that was adopted by the Parliament of Ukraine in 2002, introduced a moratorium or ban on sales and purchases of 38.5 mn ha of agricultural land or 66% of Ukraine's territory and deprived almost seven mln of Ukraine's citizens from their constitutional right to dispose off their private property freely. In 2018 the European Court for Human Rights has recognized the moratorium as a violation of property rights. Moratorium was also introduced on the change of the land use purpose, i.e. it could not be converted from agricultural to industrial use. On top of that, the Land Code also deprived local communities from the right to manage 10.5 mn ha of state land beyond their settlements and transferred it to the oblast level. Ten years later, the management of the state agricultural land beyond the settlements was centralized under the State Land Agency of Ukraine. A big step in the development of the land market was establishment of an electronic State Land Cadaster linked to the Registry of Rights and launch of the open public cadastral map in 2013. The period after the Revolution of Dignity could be marked as very modest in terms of the land reform. With a launch of the national decentralization reform in 2014, about 1.68 mn ha of agricultural land were transferred from the state into a communal ownership of local communities by the end of 2019. In 2019, after the new President and the Government came to power, the land reform got a new momentum and rolled out at an unprecedented scale. A landmark step was the adoption of the land turnover law (No. 552-IX) on March 31, 2020 that established a design for the land sales market to come in on July 1, 2021. A package of complementary laws (including land governance decentralization and deregulation), accompanying secondary legislation would ensure laws' implementation, as well as corresponding institutional arrangements and institutional reform to ensure a transparent, equitable and efficient market for agricultural land (see a detailed description of the package and reform agenda in KSE, 2021).

### 2.5.2 Access to capital/credits

**Ukraine's agriculture is characterized by relatively low capital intensity.** The reasons for that are the poor quality of assets inherited from the Soviet times, the lack of capital investments and relatively cheap labor. The value of capital per ha in Ukraine's agricultural sector lags behind the EU benchmarks (Table 2.5.1), though it is increasing over time (Figure 2.5.2).

**Table 2.5.1. The capital intensity of agriculture in 2017**

| Country | The value of assets per 1 ha, USD | The comparison with Ukraine, times |
|---|---:|---:|
| Germany | 22,198 | 88.4 |
| France | 8,416 | 33.5 |
| Hungary | 7,956 | 31.7 |
| Chech Republic | 6,418 | 25.6 |
| Great Britain | 6,318 | 25.2 |
| Poland | 5,039 | 20.1 |
| Ukraine | 251 | 1 |

Source: Zakharchuk (2019)

**Figure 2.5.2. Fixed assets in agriculture**

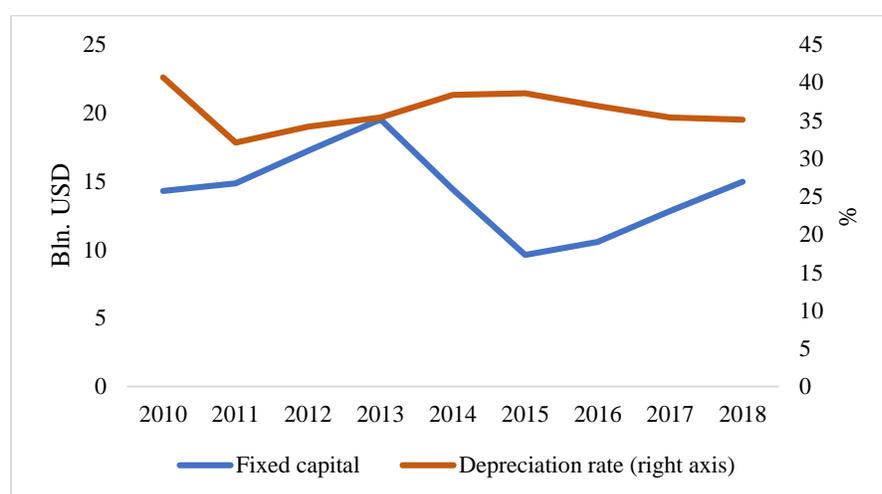

Source: own presentation using Ukrstat data

**The investments in Ukraine's agriculture comes mostly from domestic sources.** While the share of agri-food sector in GDP remains over 10% during the last years, its proportion in direct foreign investments to Ukraine does not reach 2% (Figure 2.5.3). On the other hand, the outflow of foreign investments from agriculture during 2014-2016 crisis was not so intensive as from the rest of economy. As Figure 2.5.4 shows, the amount of capital investments in hryvnias increased after 2014 due to three-times currency devaluation, but in dollars it decreased. The share of agriculture in total capital investments grown up reaching more than 14% in 2017; this is explained by the restricted investments to other sectors.

**Figure 2.5.3. Direct foreign investments in Ukrainian agriculture**

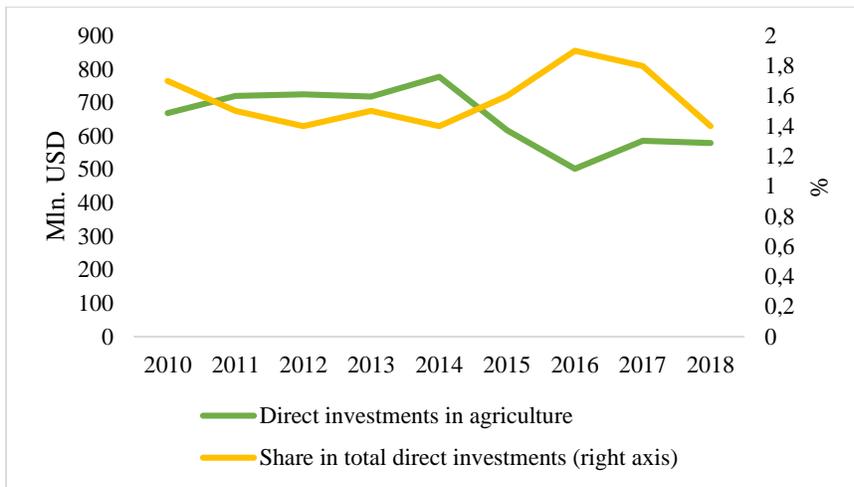

Source: own presentation using Ukrstat data

**Figure 2.5.4. Capital investments in Ukrainian agriculture**

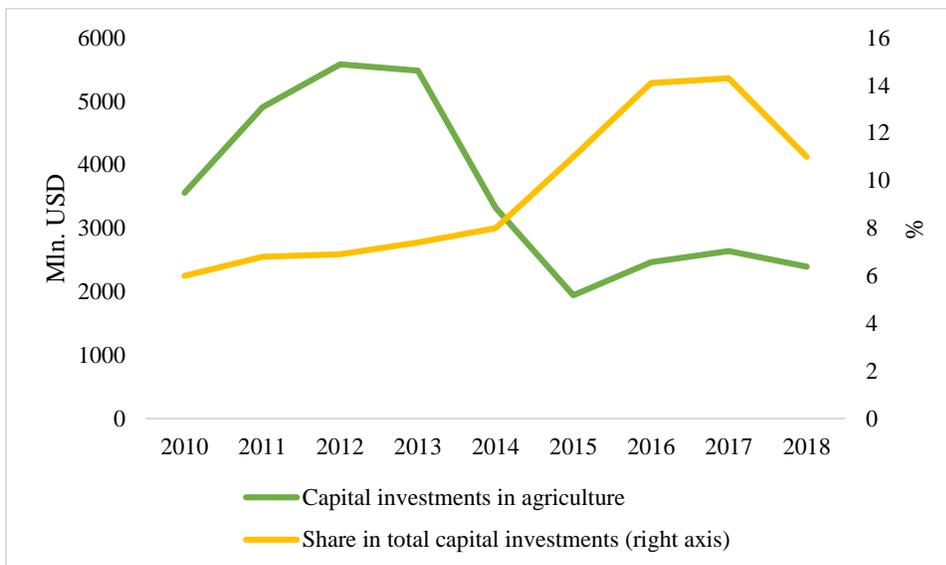

Source: own presentation using Latifundist (2020) data

**The distribution of investments across different agricultural sub-sectors is highly linked to their profitability.** Cash crops production that shows high returns is the most supplied by capital (Table 2.5.2). The livestock sector holds the second place; here, investors prefer to put money into industrial pork and poultry production. The shares of supplementary activities and the production of perennial crops are low, especially in domestic investments. Generally, foreign investments are more diversified.

**Table 2.5.2. The distribution of investments in Ukrainian agriculture in 2018**

| Direction | Direct foreign investments, mln. USD | Capital investments, mln. USD |
|---|---|---|
| Production of annual crops | 297 | 1929.4 |
| Livestock production | 163 | 345.5 |
| Supplementary activity in agriculture | 68 | 44.1 |
| Production of perennial crops | 21 | 51.4 |
| Other | 10 | 18.4 |
| Total | 559 | 2.4 |

Source: own presentation usin Latifundist (2020) data

**The provision of agricultural loans in Ukraine is low by international standards.** The share of agricultural loans in total loans given the contribution of agriculture to GDP is much lower in Ukraine than in the EU. The majority of small and medium firms in the agricultural sector have poor access to finance, which is a major barrier to further expansion and investment. Self-financing in the form of retained earnings and personal savings remains the main source of funding among agricultural enterprises. Particularly, about half of the producers immediately sell 80-100% of their new harvest to finance working capital. The main financial instruments used by Ukrainian farmers are: bank lending, investments, crop receipts, commodity credits, promissory notes financing.

**The sanitation of banking system by NBU in 2014-2016 reduced the number of banks in Ukraine and made the financial sector more resilient.** In 2018, there were 82 licensed banks in the country, 18 of which were with foreign capital, 26 – with Ukrainian capital and 38 with both Ukrainian and foreign capital. The most popular banks that provide agricultural loans are Credit Agricole Ukraine, UkrSibbank, Raiffeisen Bank Aval, Piraeus Bank, Alfa-Bank, Otp Bank, Agroprosperis Bank (UCAB 2018).

**The reasons for limited access to agricultural financing are related to macroeconomic and financial risks, as well as specific risks of the agricultural sector** (see Kirchner and Kravchuk 2012). The major macroeconomic factor is NBU policy of inflation targeting. Launched in 2016, it assumes keeping interest rate at high level in order to anchor the inflation. This makes credits expensive for all sectors of economy including agriculture. With regard to sector-specific risks, the key reasons for limited access are:

1. Significant risk associated with uncertainty and the impact of government interventions with predominantly negative consequences for farmers, leading to the inability to repay the loan.

2. Lack of creditworthy borrowers.

3. The deficit of professional accounting and financial skills among farmers required to draw up business plans acceptable to banks.

4. Shortage of industry-specific knowledge among bank lenders needed to adequately assess the risks associated with agricultural business. With few exceptions, Ukrainian banks today still lack the staff, systems, and experience needed to effectively lend to the agricultural business. At the same time, the Ukrainian banking sector continues to undergo profound changes, including an increase in the presence and market dominance of foreign banks working in Ukraine according to international operating standards and systems (their market share is now about 50%). Most of these banks have more than a century of experience in agricultural lending.

5. Lack of collateral, including the following aspects:

a) Agricultural land cannot be used as collateral. The lifting of the moratorium will not only allow the transfer of resources to the most efficient producers, but will also create a valuable asset that will lead to radical changes in the access of farms to finance.

b) The development of another innovative collateral financing scheme is extremely slow. In Ukraine, financing using warehouse receipts is developing, but due to the underdeveloped legal framework, this tool generally lacks the trust of bankers, and it is used minimally. A key element that is absent in the system of financing using warehouse receipts is the financial protection in the form of Guarantee Fund.

**Crop receipts became a powerful financial tool that helps all groups of agricultural producers to attract resources.** Since 2015, more than 1000 farmers used crop receipts; about 90% of them are small farmers. The total sum attracted via this instrument is about 14.2 billion UAH (7,8 billion UAH of which in 2019). Whereas warehouse receipts use crops stored in a warehouse as a collateral, crop receipts allow farmers to obtain finance secured by agricultural products they will grow in the field. Currently, crop receipts are the most convenient instrument for lending in Ukrainian agriculture.

### 2.5.3 Insurance

**Agricultural insurance is an important tool for managing production risks.** Although Ukraine's agricultural insurance system is gradually improving and the insurance market is growing, the relevant legal framework is still not effective enough. Therefore, government agencies lack the ability to properly administer insurance support programs. The main function of agricultural insurance is the stabilization of farmers' incomes. Ukrainian government previously provided support in the form of partial reimbursement of insurance costs (but withdrew from this practice in 2010). The private insurance sector invested in staff development, a network of agents and assessors in rural areas, as well as in reserves to cover the total sum insured. However, making these investments on an individual basis was problematic for Ukrainian insurance companies.

**Currently, Ukrainian agricultural insurance system is improving, which is reflected in the supply of high-quality products with adequate reinsurance, strict business and procedural standards.** Timely payment of indemnities upon maturity is the main indicator of the reliability of the insurer. Key market players are gradually restoring the confidence of producers and credit institutions, which was largely lost after the drought in 2003, when a large number of insured losses were not compensated. The agricultural insurance market began to grow significantly in 2005 after the adoption of the Law of Ukraine «On State Support of Agriculture of Ukraine» of June 24, 2004 № 1877-IV, which provided insurance subsidies for agricultural producers. In 2009, this growth stopped due to the limited fiscal space. In 2020, however, the government returned to the stimulation of the agri insurance sector. Indeed, the strategy of agricultural support for 2021-2023 includes the coverage of 50% of insurance payments. Although this initiative was positively

perceived by the key stakeholders, the international policy experience as well as empirical literature show the essential drawbacks of such programs. For example, the intensive support of agricultural insurance in the USA leads to the high moral hazard risk, uneven distribution of benefits (Smith et. al 2017), low productivity of production (Roberts et. al 2007).

**The Ukrainian market of agricultural insurance is quite consolidated.** Although about 63 companies are licensed to provide insurance services in agriculture as of 2018, the top-5 companies took about 85% of the total area insured (Figure 2.5.5). Regarding crops insured, about 73.1% of all insured area belonged to winter wheat, winter rapeseed and maize held second and third places with 12.6% and 6.1% respectively. From the perspective of insurance products, the majority of farmers prefer to be protected from the total loss and spring frost (42% of the insured area). Multirisk insurance for future crop is also very popular (21%). Generally, the perspectives for the further increase of the sector are impressive. For now, insurance covered just 3% of risks and 5% of arable land, while in Canada, US and EU these indicators are 90% and 80% respectively. As Table 2.5.3 demonstrates, the sum of insurance measured in hryvnias is gradually growing over the last years; the growth is not impressive in dollar equivalent due to the devaluation of the national currency. The average number of hectares insured by one contract is also growing, indicating higher confidence of clients.

**Figure 2.5.5. Structure of crop insurance market in 2018**

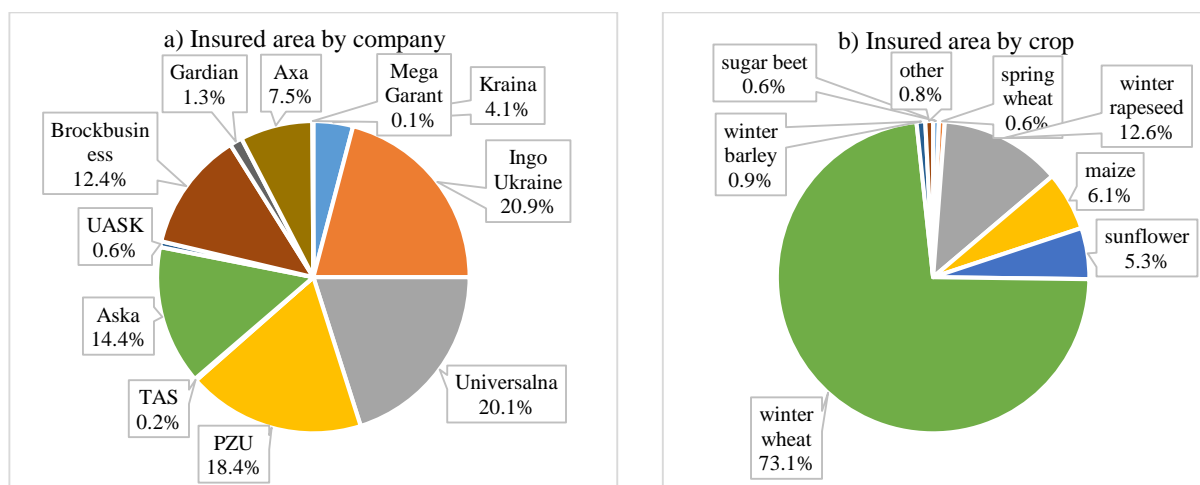

Source: own presentation using MEDTA (2019)

**Table 2.5.3. Crop insurance in 2005-2018**

| Indicator | 2005 | 2010 | 2015 | 2016 | 2017 | 2018 |
|---|---|---|---|---|---|---|
| Number of contracts | 910 | 1217 | 1062 | 793 | 957 | 1207 |
| Area, thds. ha | 390 | 553 | 689 | 700 | 661 | 974 |
| Sum insured, mln. UAH | n/a | n/a | 3969 | 6240 | 5933 | 6675 |
| Insurance premium, mln. UAH | 12.8 | 72.1 | 77.7 | 157 | 204.3 | 208.8 |
| Subsidy, mln. UAH | 5.8 | 0 | 0 | 0 | 0 | 0 |
| Rate of compensation, % | n/a | 50.9 | 12.9 | 44.2 | 4.9 | 4.2 |
| Insurance rate, % | 3.8 | 3.8 | 2 | 2.5 | 3.4 | 3.1 |
| Exchange rate, USD/UAH | 5.05 | 7.91 | 22.91 | 26.02 | 26.54 | 28.27 |

| Sum insured, mln. USD | n/a | n/a | 173.3 | 239.8 | 223.5 | 236.1 |
| Insurance premium, mln. USD | 2.5 | 9.1 | 3.4 | 6 | 7.7 | 7.4 |

Source: own presentation using MEDTA (2019)

**The insurance of livestock is not widespread in Ukraine.** As of 2018, only seven companies deal with these services. The majority of contracts are aimed to insure cattle in households (one or two heads per contract). The pork and poultry sectors are characterized by the small number of contracts signed with large producers, however, the insured sum for them are substantial. Multirisk products are dominant in the portfolios of the livestock producers.

**The major bottleneck of the insurance in pork sector are the risks of the African Swine Fever (ASF).** Private companies avoid to insure the consequences of ASF due to high moral hazard risks associated with the low biosafety level on pork farms. Besides, the losses from the disease are not compensated from the state budget. The main initiator of coping this problem is the Association of Pig Producers of Ukraine (APPU). The organization propose two directions for the insurance of ASF-related risks. The first one is the creation of the system of collective insurance which assumes that many farms are insured by just one contract. This allows to optimize the individual costs for insurance. The second direction is the active involvement of the government into the mitigation of ASF, particularly, the budget compensation of losses from ASF. The proportion of this coverage have to be linked to the biosafety level on the individual enterprise in order to motivate farmers to increase this level.

### 2.5.4 Agricultural inputs

**The distinctive feature of agricultural sector worldwide is the permanent disparity between input and output prices.** Ukrainian agriculture is no exception from this rule. Figure 2.5.6 shows that the growth of agricultural prices lags behind the increase of agricultural inputs prices. The dynamics of output prices differs among industries. Crop prices grow as fast as the total production costs while livestock prices are stickier. This divergence exacerbated after the hryvnia devaluation in 2014-2015 when nominal prices for exportable grains obtained a great incentive to grow. At the same time, livestock products were not so tradable, therefore, their prices became restricted by low domestic demand.

**Figure 2.5.6. Indexes of agricultural prices and total costs for agricultural production (2005=100)**

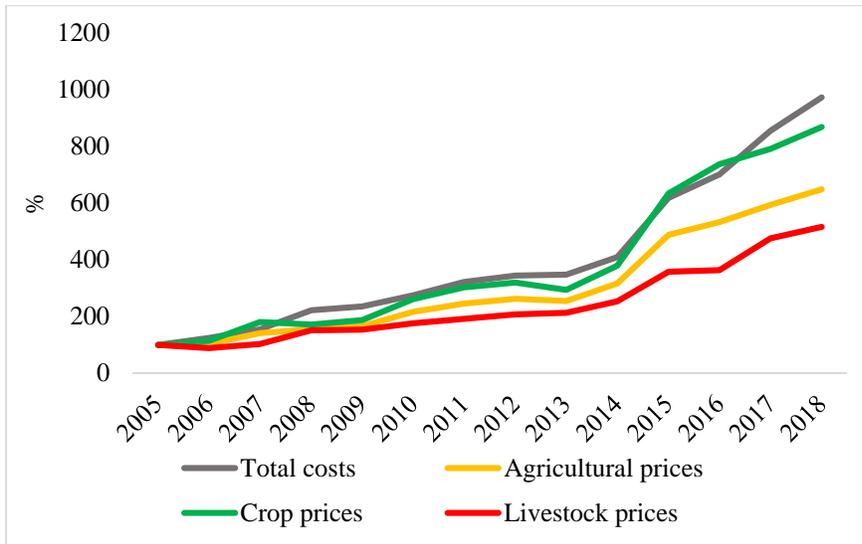

Source: own presentation using Ukrstat data

**Direct material costs take more than a half of total production costs in agriculture.** The largest categories of material costs are fertilizers, feed and fuel (Figure 2.5.7). At the same time, the share of labor costs is small which is explained by the relatively low wages in Ukraine. The proportion of land rental payments is moderate, but has a substantial potential for growth after the launching of land sales market. Since the production costs are highly dependent on inputs markets, these markets deserve to be described more deeply.

**Figure 2.5.7. Costs structure in agricultural enterprises in 2018**

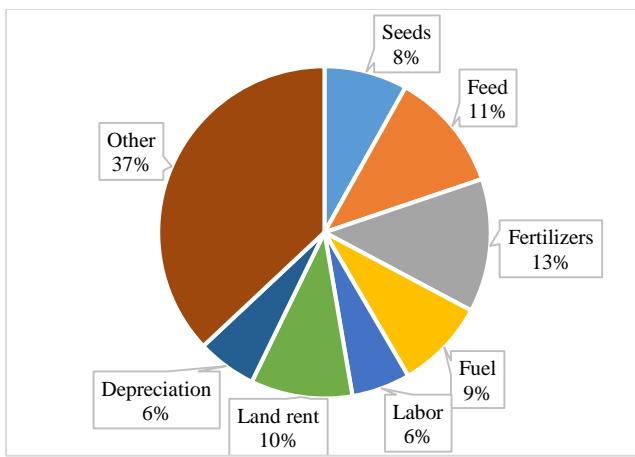

Source: own presentation using Ukrstat data

*Machinery*

**The demand for agricultural machinery in Ukraine is covered primarily by imports.** For some categories of machinery, the share of imported models reaches 100% (Table 2.5.4). This is explained by the low technical characteristics of locally produced machinery. Figure 2.5.8 indicates that tractors, harrows and cultivators are the most demanded by agricultural producers. At the same time, the leasing of agricultural machinery is actively developing. This scheme becomes more popular for the technologically advanced models with clear seasonality of utilization (harvesters, some kinds of tractors).

**Table 2.5.4. The proportion of imported agricultural machinery units in 2017**

| Type of machinery | Share of imported machinery, % |
|---|---|
| Self-propelled sprayers | 100 |
| Grain harvesters | 98 |
| Tractors | 95 |
| Plows | 89 |
| Fertilizer spreader | 55 |
| Seeders | 46 |
| Harrows | 25 |

Source: UCAB 2018

**Figure 2.5.8. The most popular types of agricultural machinery bought by agricultural enterprises in 2018**

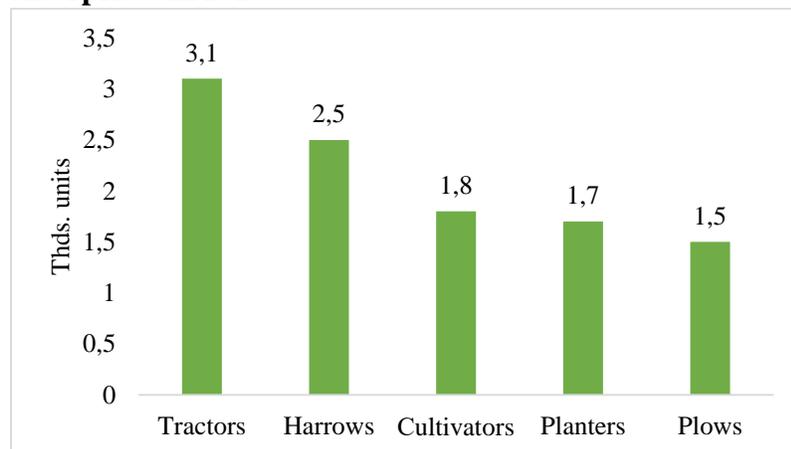

Source: own presentation using Ukrstat data

**Ukrainian agricultural machinery market is characterized by three main trends.** First, the government program of partial compensation of costs of domestically produced agricultural machinery drives the demand for certain models produced in Ukraine. The most required are tillage equipment, trailers, loaders, trailing sprayers and irrigaion systems. However, Ukrainian tractors, harvesters and self-propelled sprayers are not technologically competitive with foreign analogues. Second, agricultural producers tend to increase the demand on tractors with 100-250 horsepower engines while the popularity of 300-400 horsepower tractors declined. Third, the steady growth of production costs fuels the demand on new technologies. The bright example is the technologies of precision agriculture which allow to optimize the usage of main inputs by their differentiated application throughout the fields. The market of agricultural drones is another rapidly developing sector. Besides the agronomical inspection of fields, drones are used for the sprinkling of chemicals, control of fields works and for security measures. Another direction of innovations concerns the minimum and no-tillage systems. The refusal from the traditional tillage practices increases the organic content in the soil, reduces operational costs and allows to reach higher yields.

**The exports of agricultural machinery and equipment is negligibale compared to imports.** Figure 2.5.9 shows that Ukraine actively buys foreign tractors and harvesters. The exports are presented mostly by simpler categories of equipment: planters, harrows and mowers. An essential part of imports comes from Minks Tractor Works (Belarus). The other core suppliers are CNH (Netherlands) with brands New Holland and Case as well as the US companies John Deere and AGCO. The main national producers of agricultural machinery are Kharkiv Tractor Plant and Kherson Machine Building Plant. They export own production to Russia, Moldova, Poland, Romania, Bulgaria, Lithuania, Georgia and other countries.

**Figure 2.5.9. The most popular types of agricultural machinery for export and import in 2018**

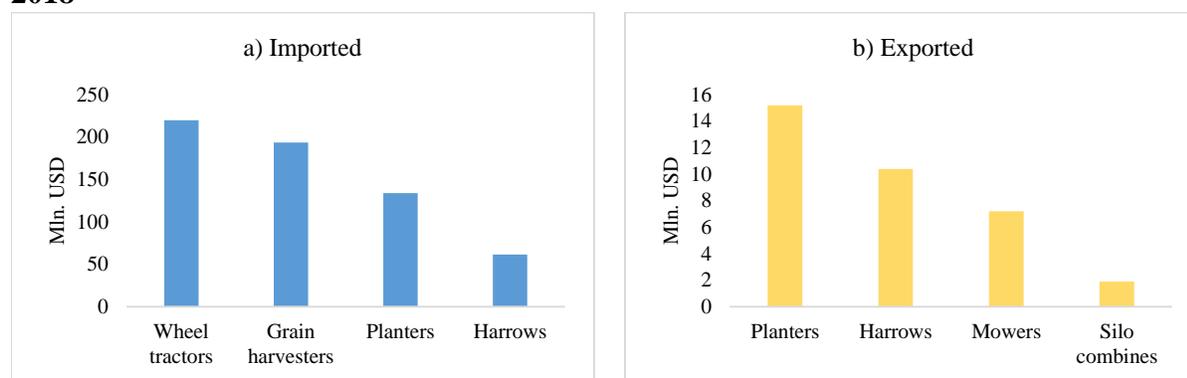

Source: own presentation using Ukrstat data

**The low capital intensity of Ukrainian crop production sector is reflected in concentration of agricultural machinery.** The number of tractors and harvesters per 1000 ha is essentially below than in EU (Figure 2.5.10). However, the EU numbers cannot be considered as benchmarks. European agriculture consists from the large number of small-scale farmers which try to have own equipment even if it is not economically reasonable. The high level of production subsidies implemented by the Common Agricultural Policy allows EU farmers to concentrate excessive capital. Therefore, the optimal concentration of agricultural machinery for Ukraine is much lower.

**Figure 2.5.10. The concentration of agricultural machinery in different countries in 2018**

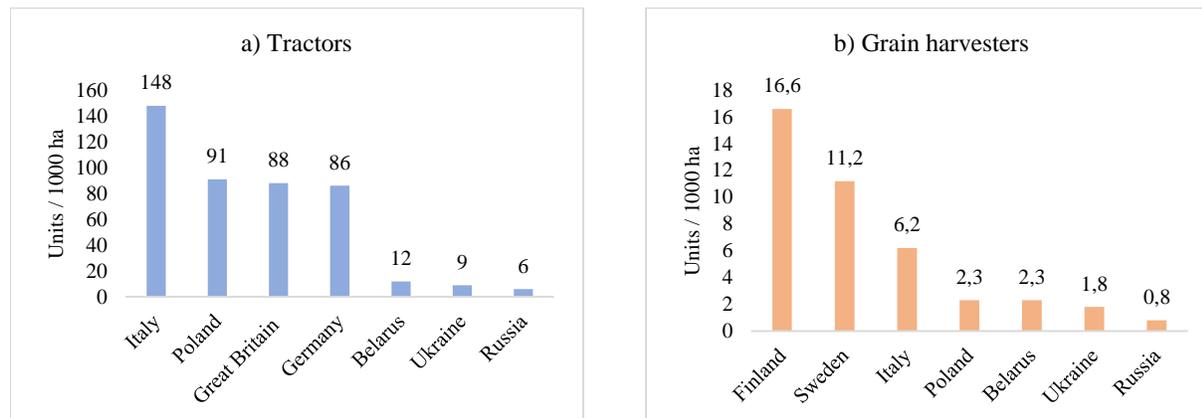

Source: own presentation usin Latifundist (2020) data

*Seeds*

**Ukrainian seed market has been growing.** At the same time, its structure differs for various crops. As Table 2.5.5 shows, the demand for wheat, barley and soybean seeds is almost fully covered by the domestic production. Low import in these segments is explained by high quality of locally produced seeds and weak protection of intellectual rights in seed production (UCAB 2018). Meanwhile, the proportion of imported seeds for maize, sunflower and rapeseed remains high.

**Table 2.5.5. The provision by imported and domestically produced seeds in 2018**

| Crop | Demand | Import | | Domestic production | |
|---|---|---|---|---|---|
| | thds. tons | thds. tons | share | thds. tons | share |
| Wheat | 1586.5 | 2.2 | 0.1% | 1584.3 | 99.9% |
| Barley | 548 | 1 | 0.2% | 546.9 | 99.8% |
| Soybean | 240.3 | 1.6 | 0.7% | 238.7 | 99.3% |
| Maize | 91.6 | 35.8 | 39.1% | 55.7 | 60.9% |
| Sunflower | 30.6 | 24 | 80.8% | 5.7 | 19.2% |
| Rapeseed | 5.2 | 2.7 | 68.4% | 1.2 | 31.6% |

Source: own presentation using UCAB 2018 and Ukrstat data

**Domestic seed demand is met by local production and imports, while exports of Ukrainian seeds are negligible.** Indeed, the export in 2018 was two times lower than in 2014 despite the hryvnia weakening in 2014-2015 (Figure 2.5.11). Ukraine exports primarily maize and wheat seeds to Belarus, Moldova and Israel. The growing imports of maize, sunflower and rapeseed seeds is stimulated by the expanding areas under these crops. The main seed importers to Ukraine are USA, Turkey, France, Canada and Romania.

**Figure 2.5.11. Seed trade in Ukraine**

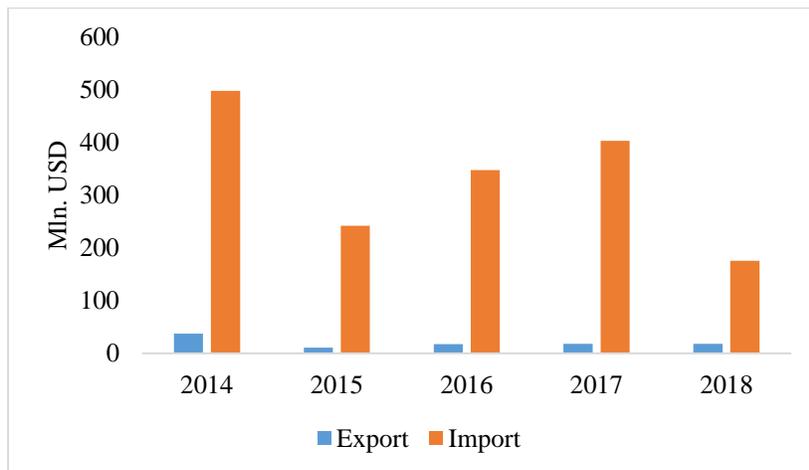

Source: own presentation using the APK-Inform data

*Fertilizers*

**During the last decade, the domestically produced mineral fertilizers are replaced by imported ones.** Despite the workload of local production capacities is decreasing (Figure 2.5.12), import volumes are continuing to grow (Figure 2.5.13). The import expansion is supported by the high price for natural gas in Ukraine which is the main input for the fertilizers production. High production costs restricts the possibilities in price competition for local chemical plants. The main importers of fertilizers are EU, Turkey, Kazakhstan, Belarus. Ukraine buys primarily complex fertilizers with two or three core elements (N, P, K). Meanwhile, the domestic production is focused mostly on the production of nitrogen fertilizers, in particular, ammonium nitrate and anhydrous ammonia. The structure of production is similar to the consumption patterns. In 2017, nitrogen fertilizers took about 68% of total consumption; the shares of phosphorus and potassium fertilizers were 19% and 13% respectively (Figure 2.5.14).

**Figure 2.5.12. Nitrogen fertilizers' consumption and production in Ukraine**

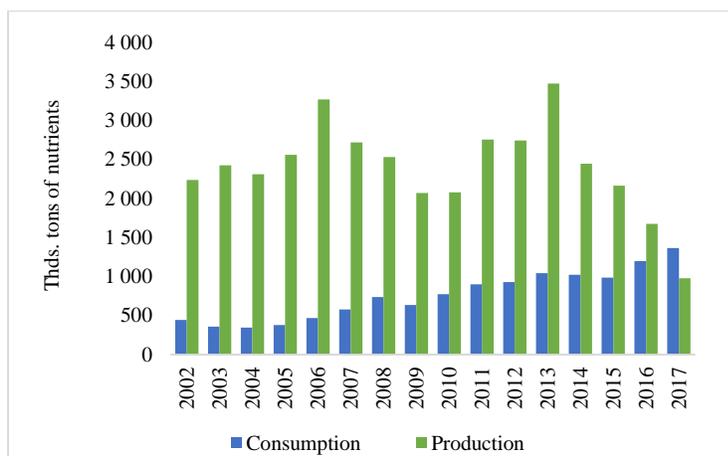

Source: own presentation using Faostat data

**Figure 2.5.13. Export and import of mineral fertilizers in Ukraine**

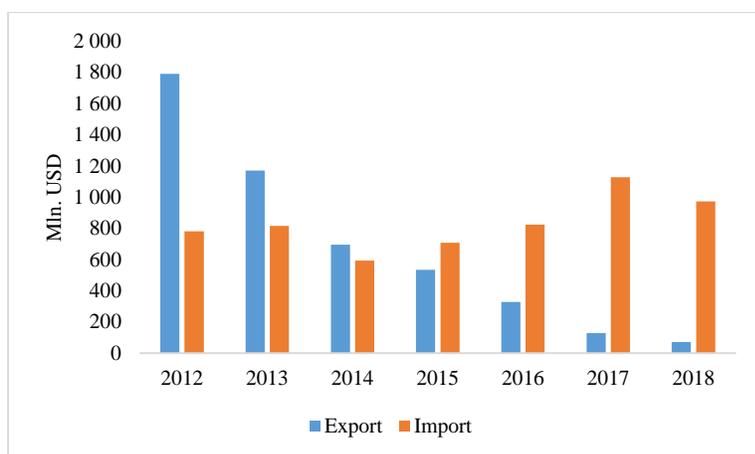

Source: own presentation using Ukrstat data

**Figure 2.5.14. Agricultural use of fertilizers in Ukraine**

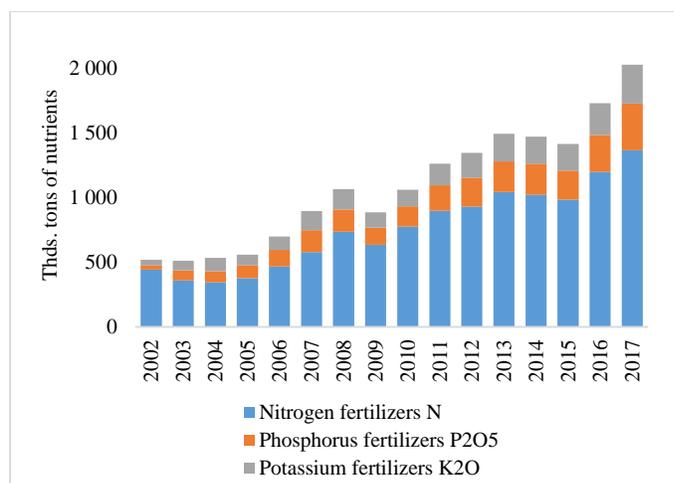

Source: own presentation using Faostat data

**The demand for mineral fertilizers is growing.** In the last two decades, the fertilized area in Ukraine expanded from 4.6 to 16.1 mln. ha (Table 2.5.6), driving the increase of total fertilizers' consumption to more than 2 mln. tons. Currently, around 90% of all arable lands are actively fertilized. The consumption per hectare elevated to 134 kg; this is still about two times lower than in the developed countries (Figure 2.5.15). Therefore, the domestic market has enough space for the further growth. The most fertilized crops are sugar beet, potato, vegetables and maize.

**Table 2.5.6. Fertilizers use in Ukraine**

| Indicator | 2000 | 2005 | 2010 | 2013 | 2014 | 2015 | 2016 | 2017 | 2018 |
|---|---|---|---|---|---|---|---|---|---|
| Total use, thds. tons | 282 | 561 | 1064 | 1494 | 1472 | 1415 | 1729 | 2028 | 2151 |
| Fertilized area, mln. ha | 4.6 | 7.8 | 12.6 | 15.3 | 14.7 | 14.5 | 15.6 | 16.5 | 16.1 |
| Share of fertilized area, % | 22 | 45 | 70 | 81 | 82 | 81 | 87 | 89 | 91 |
| Consumption per 1 ha, kg | 60 | 72 | 84 | 97 | 100 | 98 | 110 | 123 | 134 |

Source: own presentation using Ukrstat data

**Figure 2.5.15. Per hectare consumption of fertilizers in different countries**

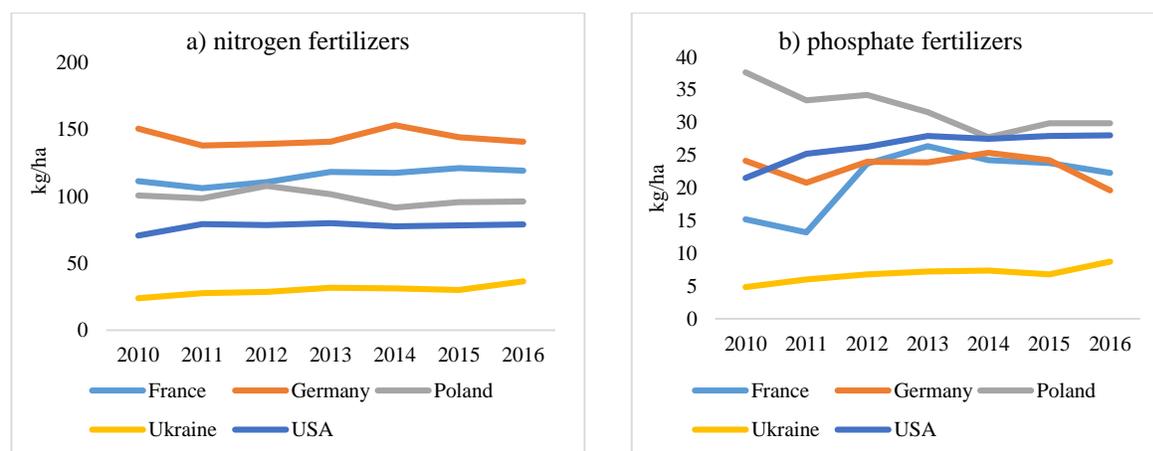

Source: own presentation using Faostat data

**The growth of mineral fertilizers consumption is accompanied by the essential decline of organic fertilizers use.** In the period from 2000 to 2017, the amount of organic fertilizers per ha decreased from 693 to 223 kg. The driver of such falling was the contraction of the cattle numbers and the corresponding drop of manure supply. In contrast to mineral fertilizers which lead to the soil acidification and the high level of nitrate content in plants, organic fertilizers accelerate the recovering of optimal soil structure. Compost is the indispensable component of the sustainable land use practices. In Ukraine, the use of organic fertilizers is very heterogeneous among regions and depend on the concentration of livestock farms. In central regions, the application of organic fertilizers per hectare is 8-10 times higher comparing to south regions. Manure is generally used by farms with diversified production for growing feed crops and maize. At the same time, enterprises specialized on crop production faces the deficit of organic. This situation changes through the increased popularity of no tillage production technologies; the plant residuals received on fields after harvesting partially recover the organic loss in the soil.

*Feeds*

**Despite Ukraine is a large grain producer, local livestock sector often faces problems with feed availability.** This is explained by the fact that feed prices are linked to the world grain prices. The increase of global demand motivates grain growers to sell own harvest to traders for the further export. Thus, livestock producers have to compete with grain traders for buying feed crops on the domestic market. This makes livestock sector vulnerable to the surges on the world grain market.

**Feed structure have changed substantially over the last decades.** The industrialization of livestock sector and limited access to grasslands due to the expansion of areas under the cash crops caused the increased consumption of concentrated feeds. The ration structure in households remained more diversified since rural population try to use more cheap fresh and coarse forages (Figure 2.5.16).

**Figure 2.5.16. Structure of feed consumption in 2018**

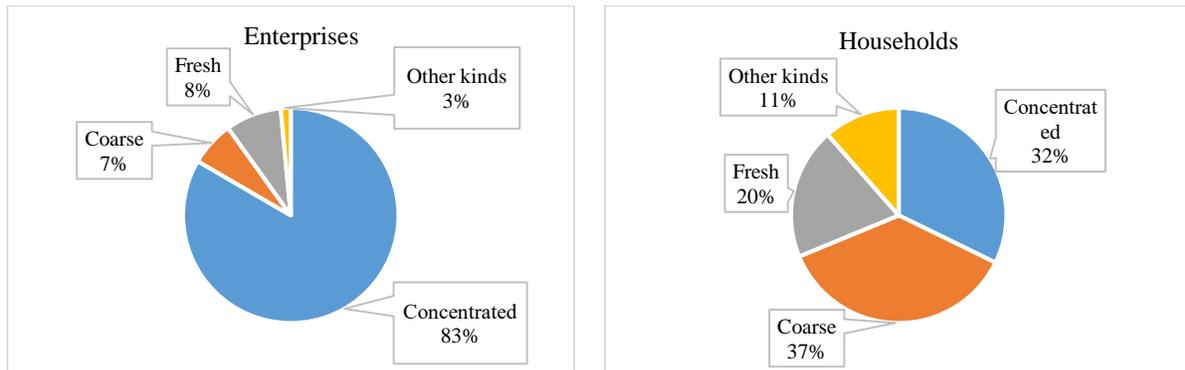

Source: own presentation using Ukrstat data

**Feed demand in Ukraine contracted during the last three decades.** In 2018, total feed demand was around 30 mln. fodder units comparing to 103 mln. fodder units[18] in 1990. The explanation for this is the decline of livestock population over the period of independence. The other reason is the increased productivity of livestock farming reflected in lower conversion rates (Figure 2.5.17).

**Figure 2.5.17. Feed conversion rates in Ukraine's livestock sector**

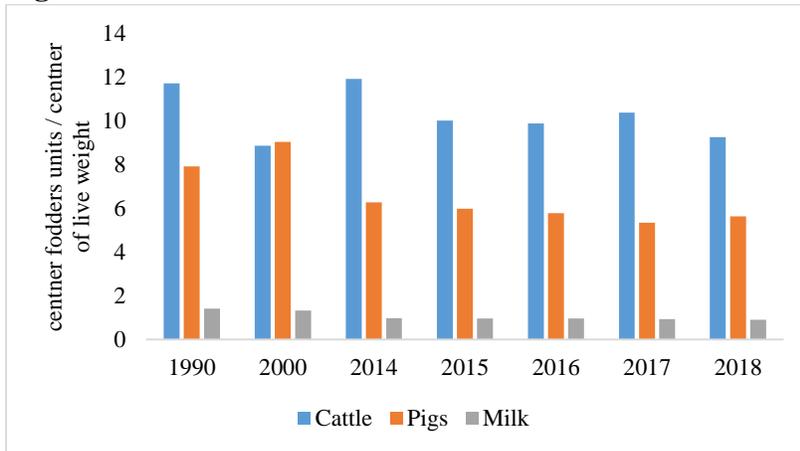

Source: own presentation using Ukrstat data

*Energy and fuel*

**The increase of productivity in agriculture is associated with a gradual decline of energy use.** Particularly, the consumption of electricity has dropped by 7.5 times since the beginning of Ukraine's independence. The use of natural gas remained relatively stable while the volumes of coal used has reduced almost to zero. At the same time, the consumption of non-liquid biofuels has soared. The structure of energy use is presented in Figure 2.5.18[19].

---

[18] Fodder unit is equal in calories to 1 kg of oats. This is a unified measure for the nutritional value of feed.

**Figure 2.5.18. The consumption of energy resources in Ukraine's agriculture**

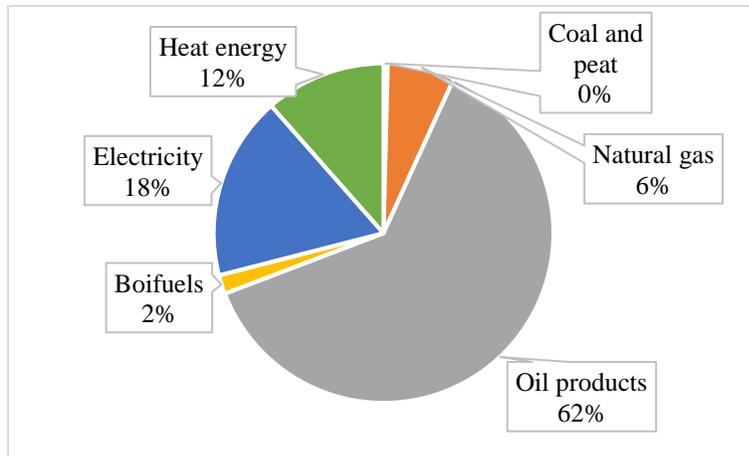

Source: adopted from Vyshnevetska (2020), Ukrstat

**Agricultural sector, especially crop production, is the largest consumer of gasoline and diesel in Ukrainian economy.** The reduction of domestic production of these categories of fuel and increased dependence from import brings serious profitability risks for farmers. At the same time, the consumption of fuel by agricultural enterprises is declining. From 1990 to 2018, the use of gasoline and diesel has dropped in 13.3 and 3.1 times respectively. Currently, the consumption of diesel is 12 times higher than gasoline use. This is because the majority of machinery for crop production use diesel. Besides, farmers tend to outsource logistics or sell own production from the farm-gates, fields or elevators (Vyshnevetska 2020). Given that fuel prices are linked to the crude oil price and exchange rate, farmers have a serious incentive to optimize fuel use in order to avoid costs shocks.

*Pesticides*

**The intensification of Ukraine's agricultural sector permanently increases the demand for pesticides, contributing to the development of this market.** In 2018, the market size was around 1.1 billion USD; this demand was almost fully absorbed by imports. Since 2014, the volumes of imports have been growing (Figure 2.5.19). The main suppliers are France, Germany, China, Spain and Israel.

---

[19] The measurment units of different types of energy are normalized to tons of oil equivalent.

**Figure 2.5.19. Pesticides trade in Ukraine**

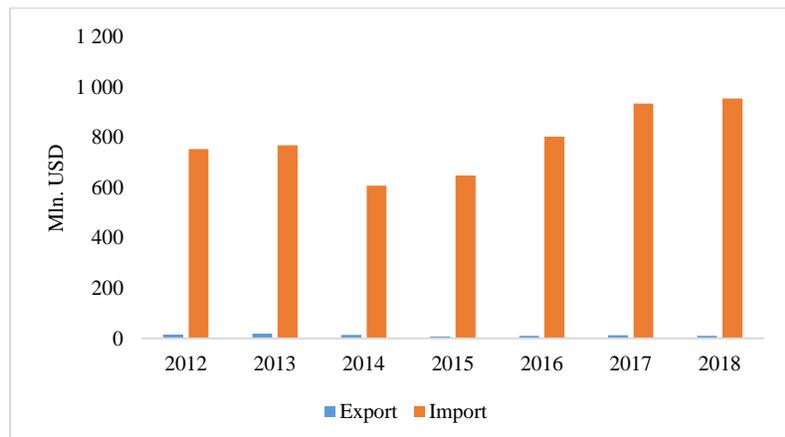

Source: own presentation using Ukrstat data

**The physical volumes of pesticides consumption in Ukraine are declining** (Figure 2.5.20). Therefore, growing import can be explained by the reducing domestic production and increase of world prices. Pesticides are the most actively used for production of the main cash crops – sunflower, maize and wheat. As of 2018, herbicides took 71% from all quantity of pesticides used, fungicides – 19%, insecticides – 7%, other pesticides – 3%. The demand for all categories, especially for fungicides, is projected to increase during the following years. As Figure 2.5.21 displays, the deposition of pesticides per ha in Ukraine has to increase several times to reach the same level as in EU countries. However, the technological innovations such as precision spraying and the introduction of seeds for highly resistant hybrids can reduce the demand for pesticides.

**Figure 2.5.20. Pesticides consumption in Ukraine**

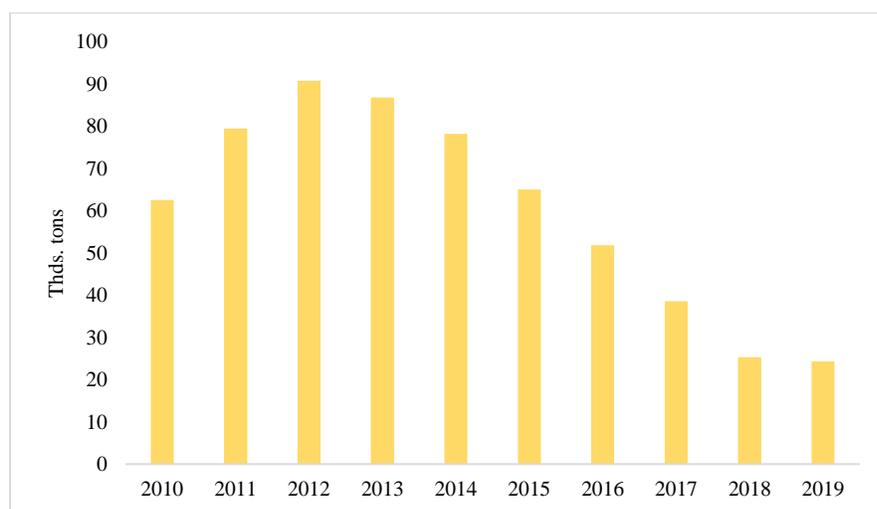

Source: own presentation using Ukrstat data

**Figure 2.5.21. Pesticides use in different countries**

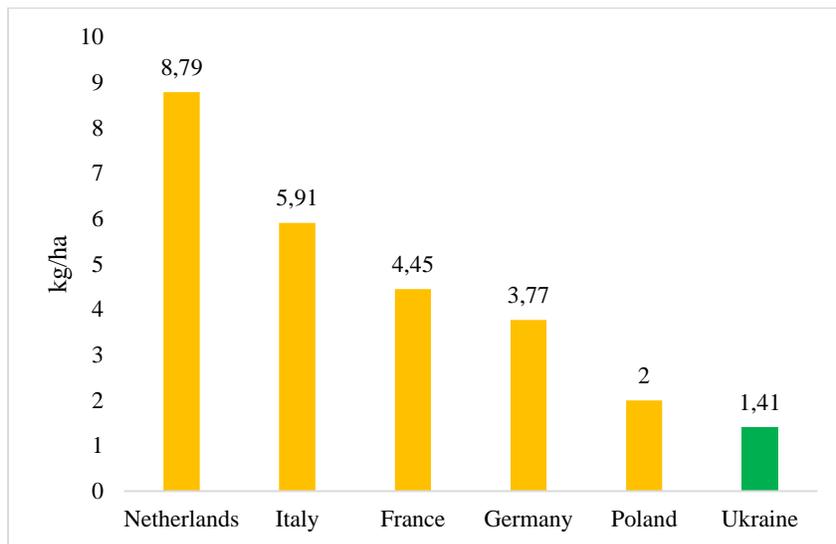

Source: own presentation using Faostat data

## 2.5.5 Human capital

**Current trends in agriculture are associated with the decreased demand in low-skilled labor due to the digitalization of production.** Over the last decades, labor productivity in Ukraine's agriculture constantly increased across all sectors (Figure 2.5.22). This process led to the essential decline of employment in the sector (9% in the period from 2012 to 2019). Note, that the dynamics of labor force in agriculture is pro-cyclical and correlates a lot with the general employment in the national economy (Figure 2.5.23).

**Figure 2.5.22. Labor productivity in Ukraine's agriculture (in 2016 constant prices)**

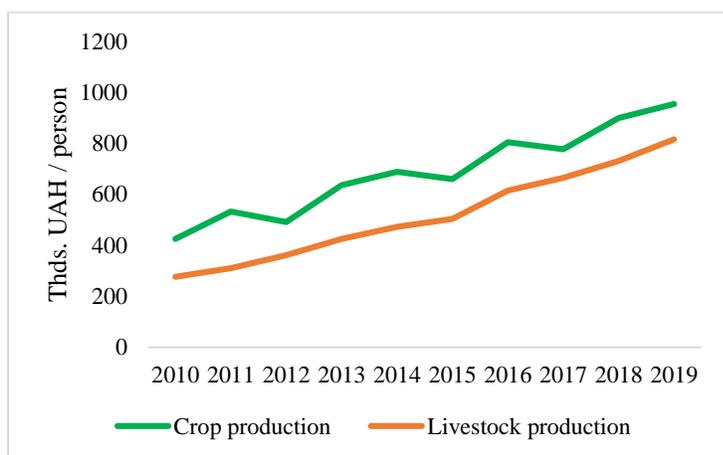

Source: own presentation using Ukrstat data

**Figure 2.5.23. Employment in agriculture and in the whole Ukrainian economy**

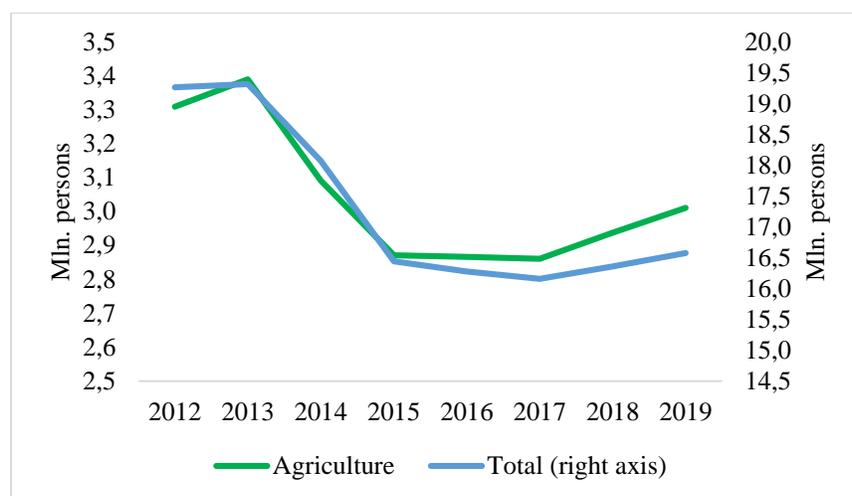

Source: own presentation using Ukrstat data

**At the same time, Ukrainian agriculture faces a shortage of qualified workers which are able to operate modern production technologies.** MEDTA estimates this deficit at the level of 26 thousand persons in 2022. The economic reason for the labor force deficit is low salaries in the agricultural sector comparing to other industries. In 2019, the average salary in agriculture was 7360 UAH versus 9648 UAH in economy as a whole. The wages are also much lower comparing to EU countries which drives labor migration. Besides, the career in agriculture is not in high demand among young Ukrainians. Reasons for that are not only low salary, but also the specificity of technological process and requirements to live in rural areas with poor infrastructure. Social polls suggest that only about 10% of the graduates of agricultural universities in Ukraine want to work in agriculture.

**The persistent deficit of high-quality human capital contrasts sharply with much higher public spending on agricultural education and research over the past decade.** Several factors determine the low quality of education and the lack of incentives to reform educational institutions (Balmann, 2020; Koester et al, 2010; Tillack and Stepaniuk, 2015):

• Universities do not actively hire employees with new knowledge. Experts educated outside of Ukraine must go through a lengthy and costly process of recognizing their degrees in Ukraine.

• The Ministry of Education has a strong influence on the curriculum, and the Ministry of Agrarian Policy and Food dominates the research agenda.

• The process of promotion and hiring does not depend enough on merit.

• The appraisal of working performance is inadequate, and teachers usually continue to work after reaching retirement age.

• Low teachers' salaries force them to work in another job or look for other sources of income.

• Budget allocations to universities are distributed according to the number of students, not performance indicators.

**The Ukrainian higher education system is overloaded with the number of educational institutions and students. At the same time, it produces graduates with insufficient technical and managerial skills.** The system of agricultural education includes about 116 universities of I-IV level of accreditation. As of 2014, the general number of students in these universities amounted to 145.3 thousand persons. Particularly, more than 78.5 thousand persons studied under the state order, about 60% of them are rural youth. The universities produce annually about 9 thousand full-time and 8.8 thousand extramural graduates. Nearly 75% of graduates who studied for state order receive recommendations for work. Educational process is provided by around 12.7 thousand research and teaching staff.

**In addition, there is an extensive network of agricultural postgraduate education institutions**, which includes 34 institutions, where about 25 thousand persons studied annually. 26 vocational schools of the first certification level carry out the training of about 15 thousand workers (Cobets and Puhach 2016). Unlike economists, accounters and managers, the working specializations are demanded in the agricultural sector. According to estimates of Ukrainian Agribusiness Club, the most required positions are agronomists, zoo-technicians, veterinarians and agricultural machinery operators. The other needed category is mid-level managers with a technical background.

**The Ukrainian system of agricultural research is far not only from the practical needs of Ukrainian agribusiness, but also from the international scientific research.** This leads to a very limited exchange of experience with international partners and a lack of locally adapted high-quality knowledge for Ukrainian producers, which puts them at a competitive disadvantage. At the international level, Ukrainian agricultural specialists are not particularly visible – only a few of them publish their work in international journals and make reports at international conferences (see von Cramon-Taubadel and Nivievsky 2011, Koester et al. 2010).

**The Ukrainian agricultural research system suffers from the same institutional and organizational shortcomings as the education system, which undermines creativity, innovation and quality improvement.** The National Academy of Agricultural Sciences (NAAS) of Ukraine is the main institution in the field of agricultural research in Ukraine. Its participation is limited to the training of PhD students. The academy consists of 340 institutions and experimental farms, including 5 national research centers, 52 research institutes and more than 200 experimental farms. One of the institutes, the Institute of Agrarian Economics, conducts research in the field of agricultural economics and has the authority to advise the government on policy-making. NAAS, like other academies, has a special status that gives it full autonomy without independent audit or any control by the state. At the same time, it receives budget allocations to fund its staff and work on research projects. Additional income comes from the management of state property. In order for the Ukrainian agricultural research system to meet the needs of farmers, the National Academy of Agricultural Sciences and its subsidiaries have to implement a comprehensive reform program that includes, inter alia, the following areas: communication with producers and their knowledge needs; connection with the education system; international relations and exchange of experience; general orientation and efficiency of production.

### 2.5.6 Information systems

**Farm management information systems are now being actively integrated in all sectors.** They can be divided into planning and technological systems. The first group encompasses all the software used for the economic planning of farms activity: enterprise resource planning (ERP) systems, accounting software (1C, SAP), CRM systems for sales management and other. The market price information is provided by various open sources. On the local level, crop price data is often supplied by large exporters. For example, agroholding Nibulon posts the daily bid prices for the different regions of Ukraine[20]. On the retail level, the main source of price information are the wholesale markets such as Shuvar market in Lviv[21]. This price data is often listed in the real time in the special software used by agricultural enterprizes. The second group is intended to optimize the technological process along the whole supply chain. The examples of technological systems are automatized milking, precision farming, animal health control etc. The planning and technological systems are often integrated into a common framework, which allows farmers to analyze the relationship between technological and financial indicators.

**Evaluation and data collection system for the purposes of efficient policy monitoring is virtually absent in Ukraine**. This makes current agricultural policy immune to economic rationale and to mistakes committed by other countries (and by the EU CAP in particular) in the past. Such a situation does not hold policy maker accountable for their decision and results eventually in a waste of resources.

Proper and comprehensive data collection system of farms and sector performance would facilitate functioning of the monitoring and evaluation system. In particular, Ukraine does not have a register of farmers and agriltural produsers, thereby a great share of agricultural producers and output that is sold across the country remains poorly accounted (see section 3 for a more detailed discussion) and this includes 4.6 million of small household producers. In that respect, the following measures would facilitate further development of information systems (Nivievskyi et al, 2021):

- The State Agrarian Registry (SAR) that would accumulate information about the universe of agricultural producers in Ukraine and would facilitate their access to the state support, to financing through better exchange of information with the banks and to other important services such as knowledge and information transfers.

- Introduction of a statistical data collection system based on the EU FADN (Farm Accountancy Data Network) model.

### 2.5.7 Irrigation/melioration

**Despite the global warming process and increased need of water for crop production, Ukrainian agriculture faces the decrease of irrigated areas since early 1990-s.** Irrigation infrastructure created during the Soviet era – pumping stations, protective dams, reservoirs,

---

[20] https://nibulon.com/data/zakupivlya-silgospprodukcii/zakupivelni-cini.html
[21] https://shuvar.com/

irrigation systems, trunk and distribution channels – can provide moisture to at least 2 million hectares of land. However, only one quarter of this land is actively irrigated now (Figure 2.5.24).

**Figure 2.5.24. The dynamic of planned and actual irrigated areas in Ukraine**

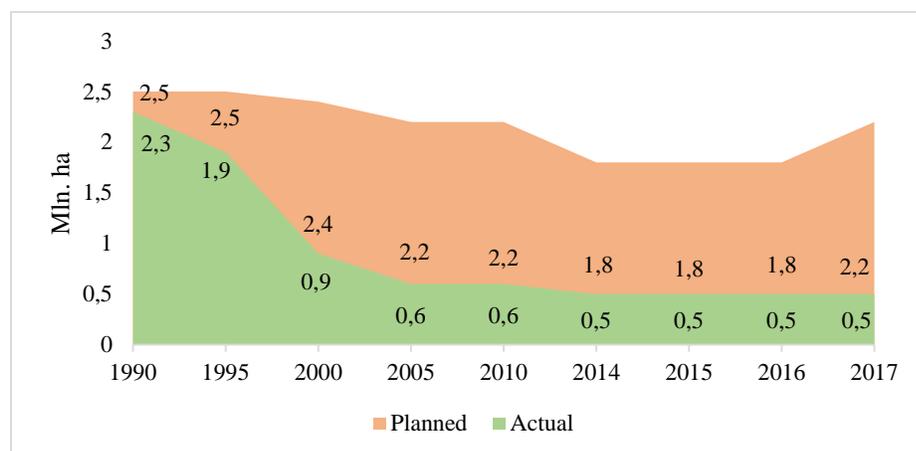

Source: own presentation usin Latifundist (2020) data

**The structure of irrigated areas is different across crops produced.** Currently, large enterprises practice sprinkling mostly for the production of maize, soybeans and sunflower (Table 2.5.7). In contrast, small farms use the drip irrigation systems to grow berries, vegetables and fruits. The irrigation becomes especially popular for seed producers; the constant supply of water allows to enhance the quality of seeds to the international standards.

**Table 2.5.7. The structure of irrigated area by crops planted, thds. ha**

| Crop       | 2010 | 2014 | 2015 | 2016 | 2017 | 2018  | 2018/2010 |
|------------|------|------|------|------|------|-------|-----------|
| Grains     | 125  | 98   | 109  | 102  | 121  | 135.7 | 1.09      |
| Soybean    | 99   | 109  | 108  | 116  | 134  | 126   | 1.27      |
| Sunflower  | 25   | 39   | 48   | 45   | 45   | 61.6  | 2.46      |
| Vegetables | 23   | 21   | 19   | 20   | 19   | 20.5  | 0.89      |
| Potato     | 6.3  | 3.9  | 3.8  | 4.5  | 4.2  | 4.1   | 0.65      |
| Sugar beet | 0.3  | 0.1  | 1.5  | 4.6  | 4.2  | -     | -         |

Source: own presentation using Ukrstat data

**The main reason of the loss of watered areas during the period of independence are disruptions in technological integrity of irrigation systems.** This is exacerbated by the problem of the scattering of irrigated areas, which belong to the large number of farmers. The other problems are the lack of sprinklers and poor state of machinery (Figure 2.5.25). Fundamentally, these issues are caused by the absence of appropriate governmental support. State policy in this field is defined by the laws «On land melioration» from 2000 and «On the state support of agricultural industry of Ukraine» from 2004, the order of the CMU «On approval of the procedure for the use of budget funds for the state support of crop production on irrigated areas» from 2008 and other legislative documents. The practical implementation of the already existing support programs is restricted by the low financial support. The institutional environment of irrigation

sector remains weak. Particularly, the fundamental problem is the farmland moratorium. Given that the majority of producers lease the land, they have no motivation to develop the irrigation infrastructure due to the lack of confidence whether they will operate on this land in future. The launching of sales market for agricultural land and the institutional empowerment of land ownership rights will solve this issue.

**Figure 2.5.25. The reasons of reduction of irrigated areas from the level of 1992, % of total reduction**

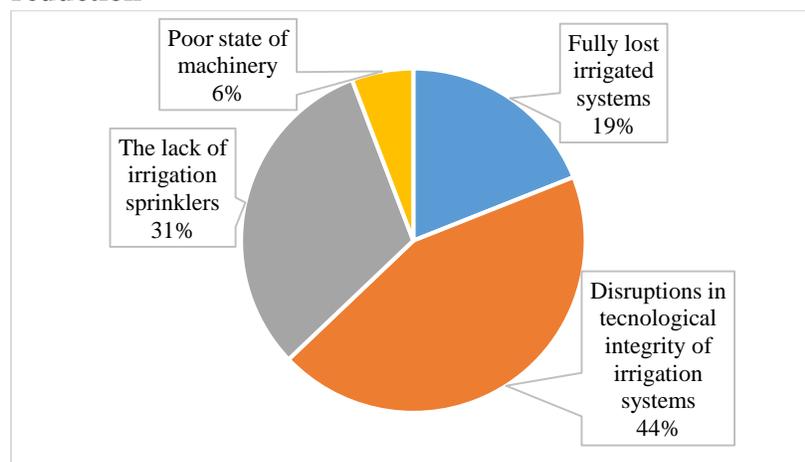

Source: own presentation using Latifundist (2020) data

**The weakness of irrigation and drainage system yields a bunch of economic and environmental issues.** According to EBRD estimates, the annual economic loss from the insufficient level of water supply in Ukrainian agriculture is about 1.5 billion USD. This is reflected in low crop yields, high production costs and accelerated degradation of lands. Indeed, the absence of watering causes soil compaction and salinization, the creation of erosional ravines, soil contamination etc.

**Nevertheless, there are a few strengths of irrigation system in Ukraine.** Currently the system of drip irrigation is being developed. This technology allows to increase crop yields by 20-30% comparing to sprinkling, using much less water. In addition, irrigation water is relatively cheap and can be unlimitedly supplied to the pipelines. Finally, Ukraine has a huge scientific basis in this area. It is presented by Institute of Water Problems and Land Reclamation, Institute of Irrigated Agriculture, Institute of Environmental Economics and Sustainable Development, National University of Water and Environmental Engineering. These institutions are capable to conduct full-fledged research and train qualified specialists in the irrigation and land reclamation sectors (Didkovska 2015).

**The reconstruction of irrigation infrastructure requires a huge amount of capital investments.** The National Academy of Agrarian Science estimates the costs for the modernization of working domestic pipelines at about 1100 USD, restoration of non-working – 2000-2200 USD, construction of new – above 2200 USD per hectare. In 2019, the government approved the Strategy of irrigation and drainage in Ukraine in period to 2030. The strategy assumes the renovation of irrigation infrastructure, the increase of irrigated areas, the development of partnership between state and business in this area. The investments for modernization of irrigation system are 3 billion USD; they will allow to increase irrigated area by 1180 thds. hectares.

**2.5.8 Agricultural logistics and marketing infrastructure**

**Logistic and tranport costs are quite high in Ukraine**. Weaknesses in Ukraine's logistics are reflected in a low ranking in the World Bank Logistics Performance Index (**Figure 2.5.26a, b**). Ukraine is ranked substantially low compared to the LPI champion Germany, but also quite far away from its closest neighbor Poland. Ukraine is especially disadvantaged in the quality of trade and transport infrastructure (e.g., ports, railroads, roads, information technology - 'Infrastructure'), in the efficiency of customs and border management clearance ('Customs') and in the competence and quality of logistics services — trucking, forwarding, and customs brokerage ('Logistics quality and competence'). Moreover, the logistics system does not seem improving over the last 5 years.

In addition, due to low population density, geography and structure of output (heavy reliance on metals, basic industry, and agriculture) Ukraine generates significantly more transport volume per unit of GDP compared to other countries in Europe. This implies that the transport costs make up a proportionately large part of the final price of many goods (Favaro et al, 2019).

**Figure 2.5.26a. Logistics Performance Index and its components in Poland, Germany and Ukraine**

**Figure 2.5.26b. Logistics Performance Index and its components Ukraine in 2014, 2016 and 2018**

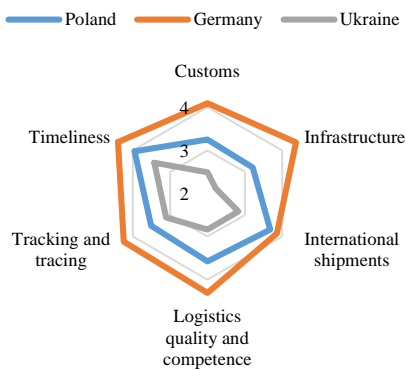
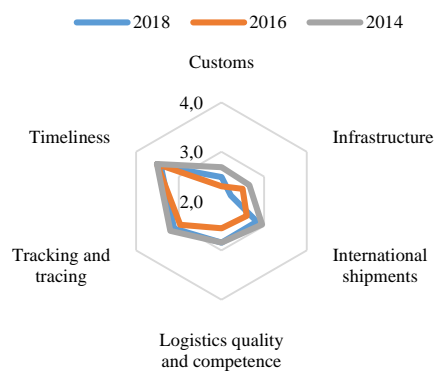

Source: World Bank's Logistics Performance Index.

**Ukraine also scores relatively poor in its relative performance on the Infrastructure and ICT components of the Global Competitive Index 2019 of the World Economic Forum**. Figure 2.5.26c shows the scale of potential improvements in infrastructure with respect to its neighbor Poland and 3d most competitive economy in Europe Germany. For example, airport connectivity in Ukraine is about 60% of that in Poland and only 8% of that in Germany. Quality of road infrastructure in Ukraine is 70% of that in Poland and 57% of that in Germany. On average, Ukraine is only 30% away from Poland and 45% from Germany in terms of its Infrastructure and

ICT performance. Chronic underinvestment in transport infrastructure across Ukraine's regions has been one of the main drivers of such a low infrastructure performance (OECD, 2018).

**Figure 2.5.26c.** Relative performance of Ukraine in infrastructure services and ICT with respect to Poland and Germany

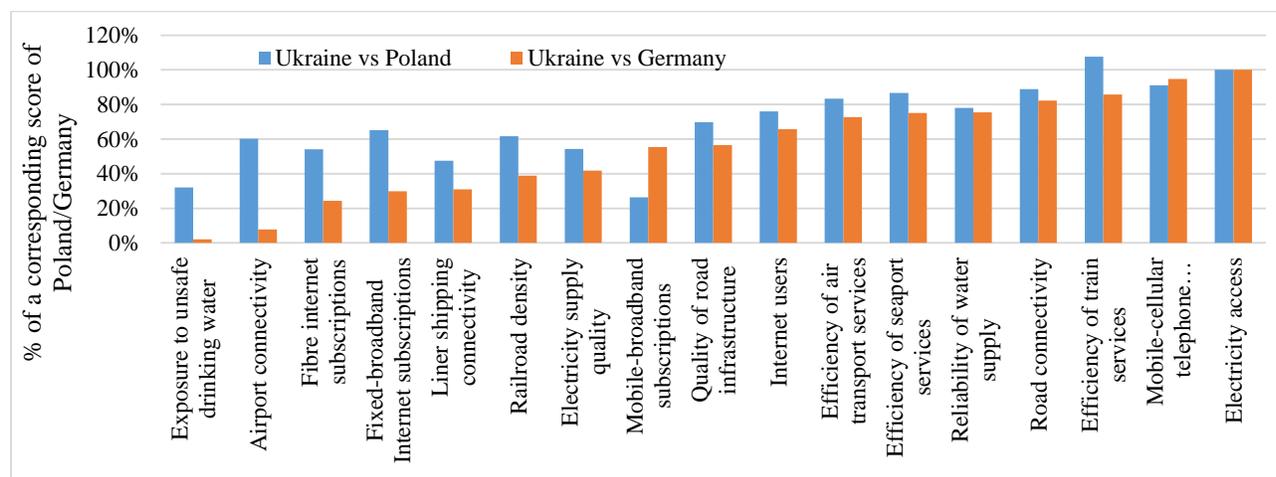

Source: own presentation based on the Global Competitiveness Report 2019 data; Note: Ukraine's scores are shown relative to the corresponding scores of Poland and Germany.

**High logistic and transportation costs due to relatively inefficient logistics and transport infrastructure reduce the farm-gate prices for Ukrainian producers and incomes, as a result.** This could be demonstrated after the first glance on the Figures 2.5.27a, b. The gap between the farm-gate wheat prices and export prices in ports (FOB) is gradually shrinking, but remains essentially high. In USA, transportation costs are much lower even though the distance from elevator to port is longer.

Figure 2.5.27a. Maize export costs in Ukraine and in the US

Figure 2.5.27b. Farm-gate price as % of the export price

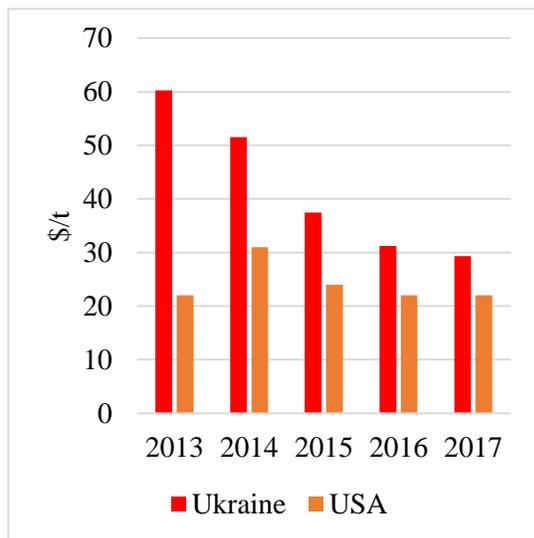
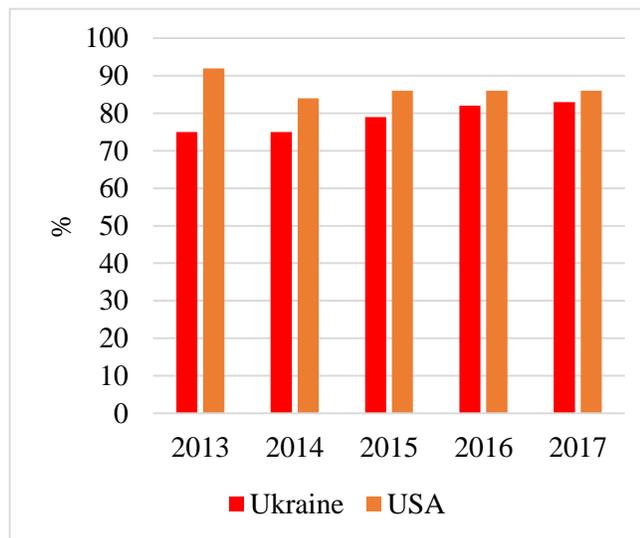

Source: own presentation using Ukrstat, USDA and APK-inform data

**Despite growing storage capacities, they are insufficient for current production.** In 2019/20 MY, the total production of grains and oilseeds was 98 mln. tons comparing to storage capacities of 78 mln. tons. There are 1237 elevators in the country, 800 of them are certified (World Bank, 2015). The certified elevators do not only store grain, but also increase its quality by cleaning and drying. The storage capacity of certified elevators is 42 mln. tons. The main owners of elevators are «Kernel» (60 elevators), the State Food and Grain Corporation of Ukraine (53 elevators), «Nibulon» (24 elevators). Given the increased demand on storage, elevators tend to provide low-quality services for high prices. The storage of grain costs about 1.5 USD/ton per month[22]; this is comparable with EU benchmarks.

**The transportation of agricultural products is carried out by railway, roads and rivers.** Traditionally, railway transport dominates in the grain logistic system, although the share of automobile and river transport increases over the last years (Figure 2.5.27c). The most underestimated branch of grain transportation in Ukraine is river. In the EU and in the US almost half of grain is transported by rivers. The share of water transportation in Ukraine is much smaller and needs further development, including via decreasing its relative cost with respect to other modes of transportation (Figure 2.5.26d). Generally, the transportation by vessels is very cheap comparing to land transport. The water transport system could be improved by dredging, upgrade of the floodgates, stimulation of the investments to private vessels.

---

[22] https://hipzmag.com/tema/tendentsii-v-razvitii-elevatornoj-otrasli-ukrainy/

**Figure 2.5.27c. Intermodal split in grain exporting in Ukraine and in the US**

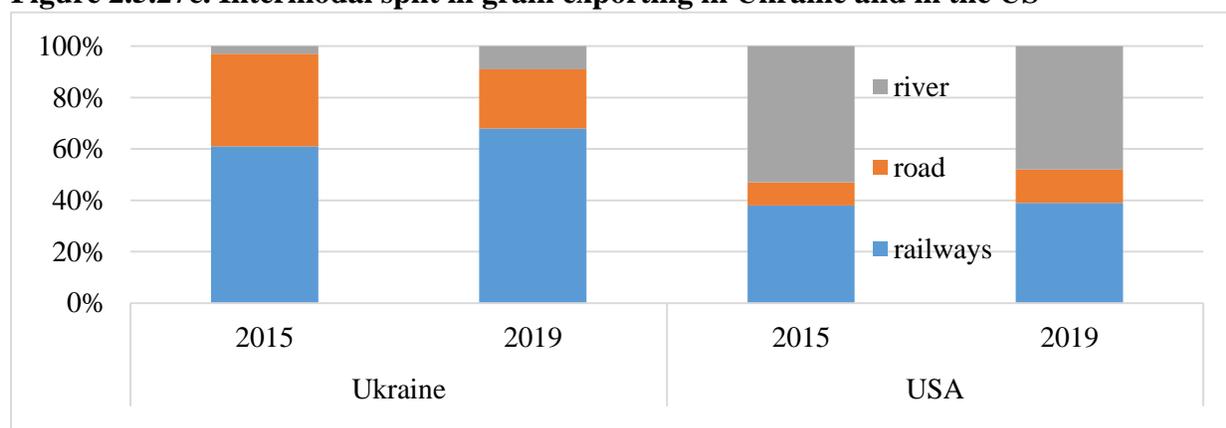

Source: own presentation using Uktstat and AMS USDA data.

**Figure 2.5.27d. Costs of grain exporting in Ukraine, EU, and the US: river vs the other modes of transportation**

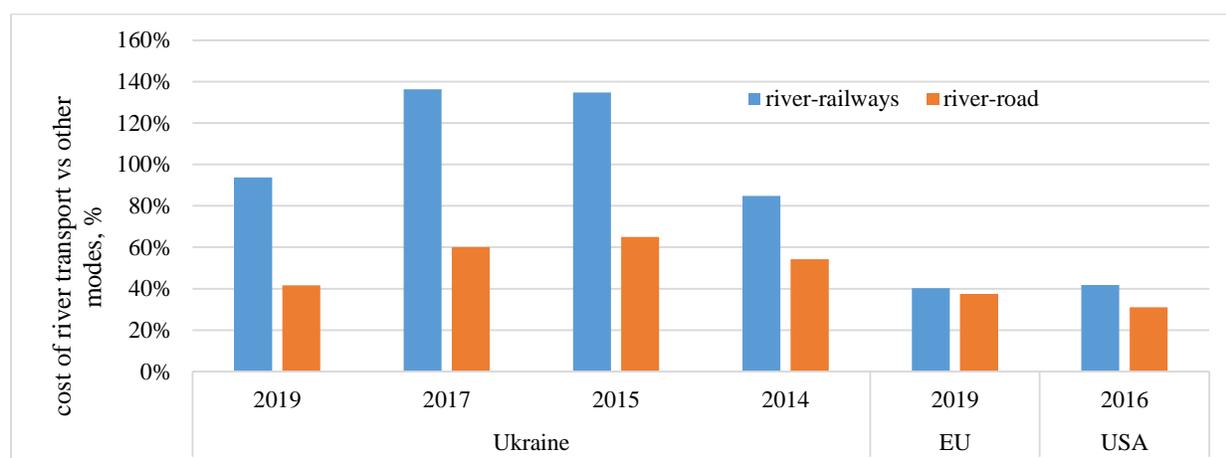

Source: own presentation using Uktstat and AMS USDA data.

**2.5.9 Food processing industry**

**Food industry plays a large role in Ukraine's economy.** As of 2018, the total output of this sector (including agriculture) was 843 bln. UAH or 19% of GDP; export was 22.1 bln. USD (44.2% of total export). Besides, the industry takes around 20% of total employment (3.25 mln. people). Despite the steady growth of domestic agricultural sector, the production of the main finished food products is relatively stable over the last years. As Table 2.5.8 indicates, only sunflower oil sector faces essential increase. At the same time, the production of flour, bakery and milk products is pressured by low domestic demand and restricted export markets. Therefore, food processing industry becomes less diversified and more oriented on basic commodities, namely oilseeds. This tendency is clearly reflected in the food export structure (Figure 2.5.28).

**Table 2.5.8. Production of processed food products by category, thds. tons**

| Category | 2014 | 2015 | 2016 | 2017 | 2018 |
|---|---|---|---|---|---|
| Bread and bakery products | 1357 | 1232 | 1160 | 1073 | 975 |
| Sugar | 2053 | 1459 | 1997 | 2043 | 1754 |
| Sunflower oil | 4401 | 3716 | 4424 | 5355 | 5149 |
| Wheat flour | 2199 | 2056 | 1974 | 1991 | 1746 |
| Pork | 239 | 247 | 245 | 234 | 230 |
| Poultry meat | 738 | 833 | 888 | 852 | 783 |
| Sausages and similar meat products | 260 | 229 | 233 | 247 | 248 |
| Dried fruits | 582 | 474 | 570 | 333 | 486 |
| Milk and cream | 1026 | 933 | 930 | 942 | 940 |
| Butter | 113 | 101 | 102 | 108 | 105 |

Source: own presentation using Ukrstat data

**Figure 2.5.28. Export growth by food and beverages sector**

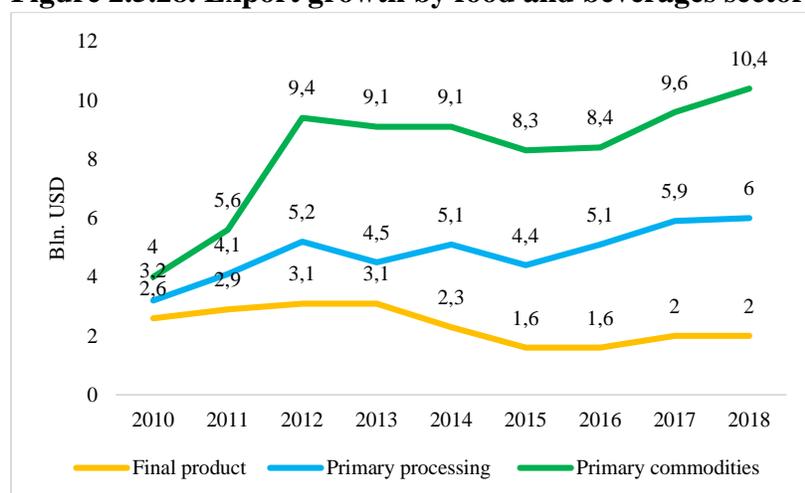

Source: own presentation using Ukrstat data

**Ukraine is a net exporter of the major categories of processed food products.** The exception of this rule are meat and fish products, beverages and some milk products like cheese. Figure 2.5.29 demonstrates that food export is much less diversified than import. The major categories of processed products supplied abroad are sunflower oil, poultry meat, dry milk and butter. At the same time, Ukraine actively imports cheeses, meat products, pork, palm oil and beverages.

**Figure 2.5.29. Trade of processed food products in Ukraine**

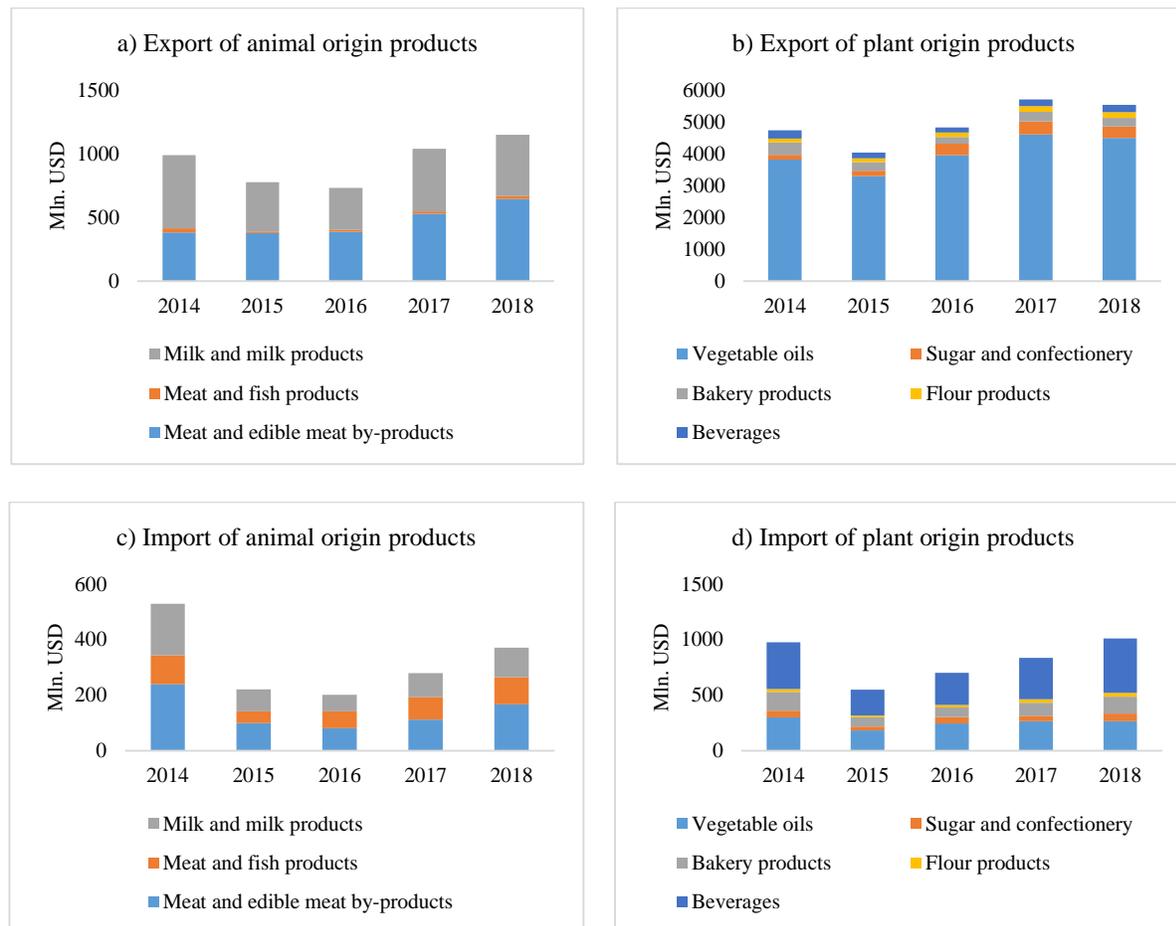

Source: own presentation using Ukrstat data

**Vegetable oils sector is presented mostly by primary sunflower oil processing.** The share of refined oil is increasing, but it still takes just 12-13% of the total production. The development of the sector was fueled by the imposing 23% export tax on sunflower seed in 1999. This restriction helped Ukraine to increase processing capacities and become the largest supplier of sunflower oil. Interestingly, the volume of crushing capacities exceeds the total oilseed harvest. This motivates processors to lobby export restrictions for oilseeds and stimulate their domestic production. Soybean is the second largest crop used for crushing. Processing plants have to compete with exporters to attract the large volumes of soybeans. In contrast, rapeseed is usually not crushed on the domestic market; the whole harvest is exported to EU for further processing. Generally, the oil sector has bright opportunities for the development, in particular: deep refining, branding and labeling, production of high oleic sunflower oil, integration with downstream sectors such as margarine production etc.

**Ukraine is a net exporter of sugar and sugar-containing products.** During the Soviet era, the country was large producer of white sugar. However, since the declaration of independence, the number of sugar plants has decreased from 192 to 42 in 2019. The reasons for this shrinkage were declined export demand and strong consolidation process. Low export is explained by the poor competitiveness of Ukrainian white sugar, especially comparing to the raw cane sugar. The long period of governmental support for domestic sugar price and market protection by tariff-rate quota demotivated producers to increase own efficiency. Nevertheless, since 2015, sugar export became supported by the hryvnia depreciation. As Figure 2.5.30 shows, sugar production is relatively stable over the last two decades and shows volatility due to the high dependence on sugar beet harvest. At the same time, Ukraine is not apart from the global trend of healthy diet and limited sugar intake, therefore, domestic consumption gradually declines. As was mentioned, refinery sector faces strong consolidation. In 2018, top-3 companies («Astarta-Kyiv», «Radekhivskyi Sugar», «Ukrprominvest-Agro») accounted about 53% of total sugar production. The large sugar holdings are highly-effective and vertically integrated with sugar beet production.

**Figure 2.5.30. Production and consumption of sugar in Ukraine**

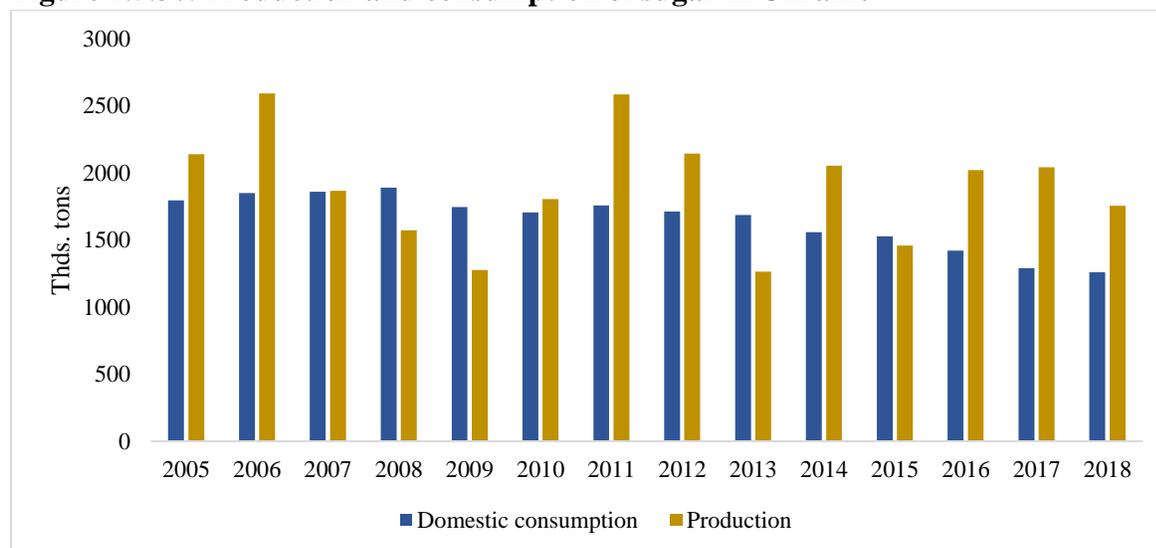

Source: own presentation using Ukrstat data

**Milk processing sector in Ukraine is challenged by the high raw milk prices and strong competition from the imported dairy products.** Low profitability of cattle breading caused to the depopulation of cows and shortage of raw milk. Moreover, milk produced in households does not meet the quality requirements for the further processing. In 2020, the second grade of milk quality was abolished according to DSTU 3662-97 «Whole cow's milk. Purchasing requirements». This regulation exacerbates the deficit of raw milk inside the country. Despite the consolidation of the processing sector, milk plants have not enough market power to decrease milk prices. Low processing margin and weak governmental control foster the existence of shadow market which accounts about 25% of the whole milk processing industry. On the other side, prices for dairy products are pressured by the loss of Russian cheese market and import from EU countries. Considering this tendency, the sector is reorienting from cheese to more profitable dry milk production.

**Meat processing is the largest industry in Ukraine's food sector.** It is specialized on the production of sausages, canned meat and different semi-finished products. The profitability of meat processors is pressured by the highly consolidated retail sector, the major part of this pressure is transferred to the slaughterhouse prices. Indeed, during the last years, the spread between retail and slaughterhouse meat prices for farmers is increasing. The situation with vertical integration is different among the sectors. Beef production is usually not much vertically integrated while large poultry producers encompass all production cycles from feed production to wholesale trade. The situation in pork industry is mixed.

### 2.5.10 Food retail and consumption

**Ukrainian food retail sector experiences fast growth during the last years.** As of 2018, the total income of this segment was around 290 billion UAH or 43.3% from the total retail revenue. As we will see further, the structure of the total FMCG turnover partially reflects the consumption patterns of Ukrainians. About half of revenues consists from the meat and meat products, alcoholic drinks and bakery (Figure 2.5.31). Meanwhile, the proportions of fruits, vegetables and dairy products are moderate.

**Figure 2.5.31. Structure of revenues in food retail**

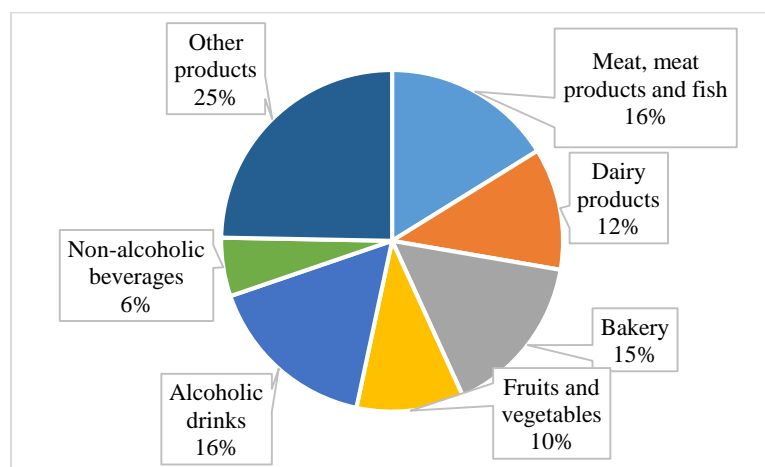

Source: own presentation using Ukrstat data

**Despite the ability for price-making in the upstream sectors, the FMCG retail market is highly competitive and saturated by a large number of companies.** The main market players are Fozzy Group, ATB-market and Auchan Ukraine (Table 2.5.9). Some companies are diversified in terms of consumer type. For example, Fozzy Group includes retailers from both low-income segment (Fora) and premium segment (Le Silpo). At the same time, ATB-market is specialized on the low-income consumers. The attractive prices and small average space of supermarkets made this company the most fast-growing retailer in 2018. Currently, large FMCG retailers follow the common strategy to increase their presence in the periphery regions. Besides, they expand the range of consumer services by upgrading the system of online sales, delivery, different loyalty

programs etc. The other trend is the rapid development of the prepared meals segment. Therefore, supermarkets become serious competitors to restaurants.

**Table 2.5.9. The major FMCG retailers in Ukraine in 2018**

| Company | Retailer | Total commercial space, thds. sq. m |
|---|---|---|
| Fozzy Group | Silpo, Fora, Fozzy Cash&Carry, Le Silpo, Thrash!, Favore | 543 |
| ATB-market | ATB, ATB express | 455 |
| Auchan Ukraine hypermarket | Auchan, My Auchan | 197.2 |
| Metro Cash & Carry Ukraine | METRO, Bery Vezy | 183 |
| Tavria V, Tavria Plus | Tavria V, Cosmos | 142.4 |
| NOVUS Ukraine | NOVUS, NOVUS Express | 113.1 |
| Retail Group | Velyka Kyshenia, Velmart, VK Express, VK Select | 107.9 |
| EKO | EKO-market, Simpatic | 99.2 |
| Furshet | Furshet, Hurman-Furshet | 96.4 |
| Omega | VARUS | 76.6 |

Source: https://rau.ua/ru/analytics/top-10-fmcg-ukrayny-ploshhady/ (accessed in June 2021)

**The economic growth in Ukraine and the development of domestic agri-food market lead to the changes in dietary habits.** Table 2.5.10 indicates that Ukrainians gradually substitute the cheap sources of calories (bakery, sugar) by the products which contain a lot of important nutrients (meat, vegetables and vegetable oils). First of all, the increase of real incomes stimulated the growth of per capita meat consumption through the large demand on relatively cheap poultry meat. The consumption of red meat remains low due to high prices for pork and beef. Besides, the demand on dairy products decreased comparing 2005. The reasons for that are high prices and probability of falsification for the processed dairy products. Generally, Ukrainians prefer to consume raw milk while the intake of cheese and butter remains below the western benchmarks. The demand on eggs is increasing due to high supply of this product proposed by households and industrial poultry companies. The decline of consumption for bread and bakery products is explained by the improved macroeconomic situation and healthier dietary habits. Besides, bread is considered as a social product and is subject to governmental price regulation. This leads to low profitability of bread production and stimulate bakeries to cut costs by lowering the quality of this product. Therefore, bread becomes an inferior good; the increase of incomes encourages consumers to choose other products. The similar picture is for sugar. Like in other countries, the direct intake of this product reduces, but the proportion of sugar consumed through the sugar-contained products grows. Finally, the consumption of vegetables, fruits and oils increases because of the increased payable capacity, more available imported products and changes of preferences.

**Table 2.5.10. Per capita consumption of selected food products in Ukraine, kg/capita**

| Product | 2000 | 2005 | 2010 | 2014 | 2015 | 2016 | 2017 | 2018 |
|---|---|---|---|---|---|---|---|---|
| Meat and meat products | 32.8 | 39.1 | 52 | 54.1 | 50.9 | 51.4 | 51.7 | 52.8 |
| Milk and milk products | 199.1 | 225.6 | 206.4 | 222.8 | 209.9 | 209.5 | 200 | 197.7 |
| Eggs, pcs | 166 | 238 | 290 | 310 | 280 | 267 | 273 | 275 |
| Bakery products | 124.9 | 123.5 | 111.3 | 108.5 | 103.2 | 101 | 100.8 | 99.5 |
| Potatoes | 135.4 | 135.6 | 128.9 | 141 | 137.5 | 139.8 | 143.4 | 139.4 |
| Vegetables | 101.7 | 120.2 | 143.5 | 163.2 | 160.8 | 163.7 | 159.7 | 163.9 |
| Fruits | 29.3 | 37.1 | 48 | 52.3 | 50.9 | 49.7 | 52.8 | 57.8 |
| Fish and fish products | 8.4 | 14.4 | 14.5 | 11.1 | 8.6 | 9.6 | 10.8 | 11.8 |
| Sugar | 36.8 | 38.1 | 37.1 | 36.3 | 35.7 | 33.3 | 30.4 | 29.8 |
| Oil | 9.4 | 13.5 | 14.8 | 13.1 | 12.3 | 11.7 | 11.7 | 11.9 |

Source: own presentation using Ukrstat data

**For the main categories of food, Ukraine has lower per capita consumption comparing to the western countries.** Figure 2.5.32 indicates that the annual consumption of meat and milk products, vegetables and fish products is higher in EU and USA. Interestingly, the per capita demand on vegetables is higher in Ukraine despite much lower real incomes. The consumption of eggs and sugar is almost equal among the reviewed countries. We can conclude that domestic food consumption in Ukraine can be significantly expanded by the red meat, dairy products, fruits and fish.

**Figure 2.5.32. Per capita consumption of selected food products in Ukraine, EU and USA in 2017**

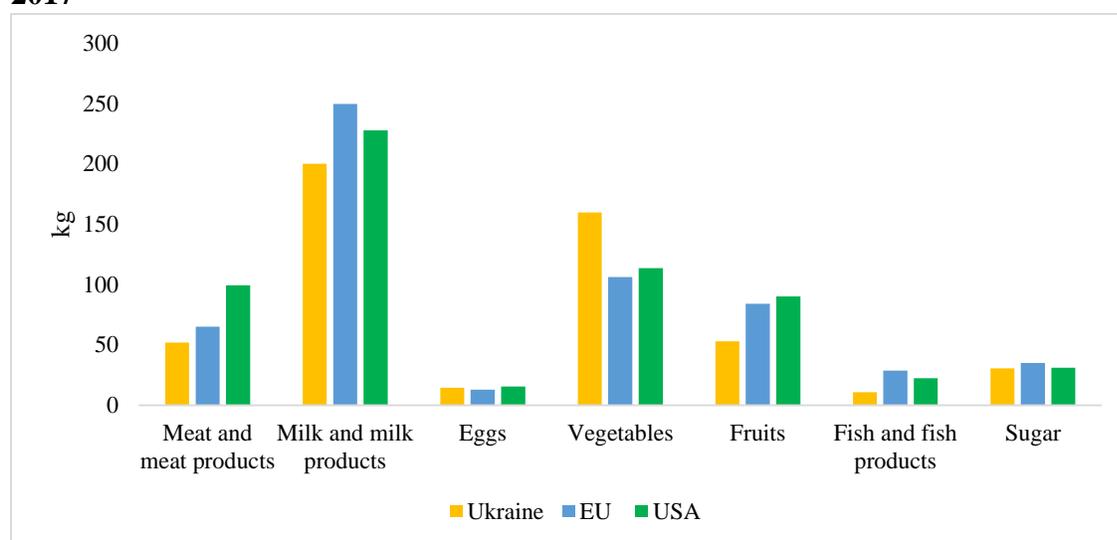

Source: own presentation using Faostat data

## 2.6 Summary and conclusion

In this chapter, we analysed and assessed agricultural policy making in Ukraine since the break-up of Soviet Union till today. We devided the evolution of Ukraine's agricultural policy making into five phases. In the first phase (1991-1994) only a few market reforms were undertaken but most key elements of the Soviet system (state procurement of key agricultural products, state provision of inputs, administrative control of product flows, prices and margins) were maintained. Agricultural production and especially livestock production also declined dramatically in this first phase, albeit at a slower rate than production in the rest of the economy. The second phase (1995-1998) of agricultural policy developments was characterised by an imbalance between macroeconomic and sectoral reforms, in particular macroeconomic reforms were not supported by structural reforms in agriculture and other sectors. These imbalances culminated in a financial crisis in September 1998, triggered by international developments. The third phase (1999-2000) was brief, but crucial for further agricultural sector development. Collective agricultural enterprises were restructured into private one, the state order for grain and other agricultural products was terminated, and substantial tax benefits for agricultural sector were introduced that dominated the overall fiscal support to agriculture since then. The fourth phase (2001-2013) could be characterized as a phase of stop-and-go reforms and dirigistic measures. These measures mainly represented attempts to regulate individual product markets such as those for grains, sugar and oilseeds. Due to the political sensitivity of wheat and bread prices, policy makers reverted to their planning ways and attempted to regulate prices and product flows. Export restrictions in 2006/07, 2007/08, 2010/11 and in 2011/12 marketing years took the form of either quotas, export taxes or export VAT non-refund was cancelled essentially taxing grain exports. Two landmark events happened in that period: multipal bans an agricultural land (sales and purchase, change of land designation and communities governance) that put agriculture and entire economy onto a lower development path, and WTO accession. The fivth or current phase (2014 – today) has been entirely driven by the Assossiation Agreement (AA) agenda with the EU that involved a substantial market and institiution reform agenda.

Then we looked at the resiliently increasing role of agri-food sector in the national economy as well as globally in ensuring food security in the world, although the structure of Ukraine's agri-food exports and sector has been increasingly dominated by agricultural raw products. Agriculture's share in the GDP (including forestry and fishing) has been floating around 10% since 2001: being at 14% in 2001, then dropping to its minimum 6.5% in 2007, bouncing back to 12% in 2015 and stabilizing at 10% since then. The share of agri-food exports in total exports increased from 11% in 2001 to almost 50% in 2020.

Structurally the gross agricultural output in Ukraine is generated by two groups of producers, i.e. legally registered commercial enterprises and not legally registered individual family farms – households. Overall, farms have been in a processes of consolidation. The most shrinking category is the small farmers (both commercial and individual family farms) below 20 ha, which development was essentially stifled by overall pro-large agricultural policy in Ukraine (multiple farmland moratoriums, substantial tax benefits, pro-large direct subsidies and underfinancing of agricultural public goods). The number and total area of large enterprises (agriholdings) with more than 10,000 ha increased twofold over the last decade.

Although Ukrainian agriculture has been experiencing a recovery after almost a decade of the transitional recession, its productivity is still far from its potential. The primary reason for this is that agricultural production in Ukraine increasingly focuses on lower value-added products (such as grains or oilseeds), which in turn is mainly driven by historical and modern agricultiral policy and institutional framework. Generally speaking it is unreliable and inconsistent in addressing key constraints in the sector, to stimulate long-term investment to reach sector development goals. So far there is no officially adopted and Agricultural Development Strategy that would establish conceptual coherence sector development goals, current development constraints and specific policy instruments that would target the constraints in accordance with economic principles. Agricultural policy making continues to be ad hoc and opportunistic, it lacks transparency in application of policy measures and it creates significantly inequitable distribution of benefits. Governance issues have eroded public trust and confidence in the state initiatives. Small farming, that could contribute to diversification of agricultural profile, overall have been substantially side-tracked in Ukraine's agricultural policy agenda over the last 20 years. Agricultural support policy in the form of substantial tax benefits and subsidies has been pro-large thus putting small producers at disadvantage in development and growth. This is reinforced by the fact that small farms are disadvantaged in access to financial services due to information asymmetry and transaction costs and by the existing ban on agricultural land sales being in place since 2001. The opening of agricultural land market in July 2021 is expected to become a structural shock to agricultural development and trigger structural changes in Ukraine's agriculture and rural development. Streamlining of agricultural policy making, making it more long-term, reliable and targeted to make a better value for tax payer money, will only reinforce the expectations from the land market reform.

# Appendix

## Table 1A. Production of grains in Ukraine's regions 2009–2018, 1000 tons

|  | 2009 | 2010 | 2011 | 2012 | 2013 | 2014 | 2015 | 2016 | 2017 | 2018 |
|---|---|---|---|---|---|---|---|---|---|---|
| Ukraine | 46,028 | 39,271 | 56,747 | 46,216 | 62,997 | 63,859 | 60,125 | 66,088 | 61,916 | 70,056 |
| Center | 15,469 | 13,579 | 19,982 | 14,474 | 22,043 | 20,369 | 20,055 | 22,642 | 18,492 | 24,146 |
| Vinnytsia region | 3,092 | 3,111 | 4,244 | 3,625 | 4,844 | 5,063 | 3,768 | 5,563 | 4,889 | 5,911 |
| Dnipropetrovsk region | 2,817 | 2,709 | 3,456 | 1,554 | 3,710 | 3,317 | 3,866 | 3,480 | 3,578 | 3,487 |
| Kirovohrad region | 2,534 | 2,374 | 3,465 | 2,340 | 3,781 | 3,469 | 3,313 | 3,725 | 2,858 | 3,763 |
| Poltava region | 3,830 | 2,854 | 5,055 | 3,645 | 5,637 | 4,821 | 5,363 | 5,783 | 4,241 | 6,341 |
| Cherkasy region | 3,195 | 2,531 | 3,762 | 3,311 | 4,070 | 3,699 | 3,745 | 4,091 | 2,926 | 4,644 |
| East | 5,307 | 3,874 | 7,027 | 5,653 | 7,704 | 8,054 | 6,737 | 7,383 | 7,043 | 6,332 |
| Donetsk region | 1,724 | 1,797 | 2,286 | 1,643 | 2,210 | 2,362 | 1,536 | 1,793 | 1,908 | 1,344 |
| Luhansk region | 1,056 | 811 | 1,269 | 1,294 | 1,292 | 1,226 | 992 | 1,274 | 1,276 | 1,159 |
| Kharkiv region | 2,527 | 1,267 | 3,473 | 2,717 | 4,202 | 4,466 | 4,209 | 4,316 | 3,859 | 3,829 |
| North | 7,774 | 5,932 | 9,296 | 10,492 | 12,140 | 12,900 | 11,527 | 12,977 | 12,404 | 15,884 |
| Zhytomyr region | 1,238 | 1,087 | 1,507 | 1,695 | 2,103 | 1,907 | 1,459 | 2,094 | 1,993 | 2,424 |
| Kyiv region | 2,483 | 2,003 | 2,785 | 3,190 | 3,325 | 3,361 | 2,820 | 3,327 | 2,646 | 4,081 |
| Sumy region | 2,005 | 1,324 | 2,522 | 2,668 | 3,592 | 3,940 | 3,734 | 3,816 | 3,686 | 4,470 |
| Chernihiv region | 2,049 | 1,518 | 2,481 | 2,939 | 3,119 | 3,692 | 3,514 | 3,740 | 4,079 | 4,909 |
| South | 10,866 | 9,953 | 12,428 | 6,319 | 11,034 | 11,114 | 11,735 | 12,014 | 12,390 | 11,492 |
| AR Crimea | 1,663 | 1,404 | 1,931 | 908 | 766 | - | - | - | - | - |
| Zaporizhia region | 2,131 | 1,905 | 2,193 | 1,196 | 2,111 | 2,417 | 2,728 | 2,624 | 2,907 | 2,233 |
| Mykolayiv region | 2,465 | 2,201 | 2,628 | 1,278 | 2,803 | 2,864 | 2,896 | 2,725 | 2,674 | 2,673 |
| Odessa region | 2,839 | 2,929 | 3,194 | 1,880 | 3,670 | 3,677 | 3,489 | 4,403 | 4,264 | 4,319 |
| Kherson region | 1,769 | 1,515 | 2,481 | 1,055 | 1,684 | 2,156 | 2,622 | 2,262 | 2,545 | 2,267 |
| West | 6,613 | 5,933 | 8,014 | 9,279 | 10,076 | 11,411 | 10,064 | 11,061 | 11,580 | 12,194 |
| Volyn region | 640 | 579 | 748 | 869 | 902 | 1,036 | 1,062 | 1,109 | 1,165 | 1,237 |
| Transcarpathian region | 302 | 256 | 322 | 322 | 325 | 343 | 332 | 412 | 390 | 375 |
| Ivano-Frankivsk region | 402 | 347 | 537 | 616 | 678 | 780 | 688 | 772 | 753 | 804 |
| Lviv region | 823 | 623 | 962 | 1,066 | 1,186 | 1,421 | 1,366 | 1,428 | 1,417 | 1,440 |
| Rivne region | 696 | 636 | 791 | 918 | 1,101 | 1,222 | 1,101 | 1,300 | 1,208 | 1,259 |
| Ternopil region | 1,574 | 1,261 | 1,883 | 2,164 | 2,222 | 2,651 | 2,199 | 2,448 | 2,622 | 2,632 |
| Khmelnytsky region | 1,702 | 1,743 | 2,180 | 2,713 | 3,037 | 3,289 | 2,793 | 3,085 | 3,421 | 3,861 |
| Chernivtsi region | 475 | 489 | 593 | 612 | 626 | 669 | 523 | 507 | 604 | 586 |

*Sources: own presentation using Ukrstat data*

## Table 2A. Production of oilseeds in Ukraine's regions 2004–2018, 1000 tons

|  | 2009 | 2010 | 2011 | 2012 | 2013 | 2014 | 2015 | 2016 | 2017 | 2018 |
|---|---|---|---|---|---|---|---|---|---|---|
| Ukraine | 9,281 | 9,921 | 12,454 | 12,074 | 16,232 | 16,214 | 16,849 | 19,058 | 18,330 | 21,377 |
| Center | 3,576 | 3,625 | 4,498 | 4,176 | 5,810 | 5,597 | 6,052 | 6,483 | 5,770 | 7,032 |
| Vinnytsia region | 556 | 512 | 644 | 660 | 980 | 1,137 | 983 | 1,242 | 1,204 | 1,372 |

|  | 2009 | 2010 | 2011 | 2012 | 2013 | 2014 | 2015 | 2016 | 2017 | 2018 |
|---|---|---|---|---|---|---|---|---|---|---|
| Dnipropetrovsk region | 903 | 924 | 1,115 | 841 | 1,342 | 1,052 | 1,297 | 1,352 | 1,338 | 1,489 |
| Kirovohrad region | 905 | 991 | 1,232 | 1,146 | 1,545 | 1,502 | 1,510 | 1,636 | 1,358 | 1,753 |
| Poltava region | 721 | 704 | 906 | 834 | 1,085 | 1,054 | 1,346 | 1,333 | 1,077 | 1,386 |
| Cherkasy region | 492 | 495 | 597 | 692 | 853 | 854 | 916 | 920 | 794 | 1032 |
| East | 1,777 | 1,748 | 2,374 | 2,273 | 2,642 | 2,533 | 2,284 | 2,760 | 2,376 | 2,844 |
| Donetsk region | 683 | 600 | 785 | 746 | 796 | 758 | 536 | 629 | 596 | 568 |
| Luhansk region | 421 | 387 | 582 | 564 | 647 | 530 | 486 | 673 | 566 | 729 |
| Kharkiv region | 673 | 761 | 1,007 | 963 | 1,200 | 1,245 | 1,262 | 1,458 | 1,213 | 1,548 |
| North | 719 | 774 | 1,183 | 1,619 | 2,116 | 2,506 | 2,527 | 3,054 | 3,064 | 3,736 |
| Zhytomyr region | 113 | 113 | 165 | 288 | 330 | 528 | 482 | 550 | 671 | 772 |
| Kyiv region | 303 | 314 | 450 | 563 | 714 | 831 | 743 | 896 | 768 | 1049 |
| Sumy region | 176 | 218 | 346 | 434 | 609 | 651 | 716 | 824 | 835 | 986 |
| Chernihiv region | 127 | 129 | 222 | 334 | 463 | 497 | 587 | 784 | 791 | 930 |
| South | 2,447 | 3,001 | 3,382 | 2,769 | 4,003 | 3,300 | 3,831 | 4,419 | 4,094 | 4,331 |
| AR Crimea | 68 | 92 | 131 | 132 | 166 | - | - | - | - | - |
| Zaporizhia region | 813 | 832 | 1,049 | 785 | 1,021 | 846 | 1,049 | 1,073 | 966 | 828 |
| Mykolayiv region | 630 | 734 | 762 | 738 | 1,062 | 847 | 1,005 | 1,212 | 966 | 1,190 |
| Odessa region | 436 | 605 | 647 | 527 | 989 | 944 | 903 | 1,099 | 1,187 | 1,249 |
| Kherson region | 500 | 737 | 793 | 587 | 766 | 664 | 874 | 1,035 | 976 | 1,064 |
| West | 762 | 773 | 1,018 | 1,236 | 1,660 | 2,277 | 2,155 | 2,342 | 3,015 | 3,434 |
| Volyn region | 38 | 55 | 65 | 90 | 122 | 162 | 161 | 148 | 220 | 306 |
| Transcarpathian region | 4 | 3 | 5 | 9 | 12 | 15 | 20 | 24 | 34 | 25 |
| Ivano-Frankivsk region | 36 | 25 | 55 | 79 | 102 | 155 | 129 | 159 | 241 | 221 |
| Lviv region | 92 | 102 | 101 | 145 | 200 | 235 | 245 | 297 | 383 | 442 |
| Rivne region | 83 | 68 | 87 | 109 | 136 | 206 | 212 | 197 | 289 | 333 |
| Ternopil region | 179 | 150 | 175 | 204 | 288 | 368 | 415 | 428 | 599 | 662 |
| Khmelnytsky region | 254 | 275 | 408 | 485 | 643 | 924 | 827 | 936 | 1072 | 1204 |
| Chernivtsi region | 75 | 97 | 121 | 114 | 158 | 212 | 147 | 154 | 178 | 241 |

Source: own presentation using Ukrstat data

**Table 3A. Pork production in Ukraine's regions 2009–2018, 1000 tons**

|  | 2009 | 2010 | 2011 | 2012 | 2013 | 2014 | 2015 | 2016 | 2017 | 2018 |
|---|---|---|---|---|---|---|---|---|---|---|
| Ukraine | 526.5 | 631.2 | 704.4 | 700.8 | 748.3 | 742.6 | 759.7 | 747.6 | 735.9 | 702.6 |
| Center | 119.3 | 147.0 | 164.8 | 159.9 | 180.1 | 181.9 | 186.5 | 182.4 | 180.9 | 160.7 |
| Vinnytsia region | 14.2 | 21.1 | 25.5 | 20.1 | 20.8 | 20.5 | 22.2 | 23.9 | 26.3 | 23.4 |
| Dnipropetrovsk region | 44.8 | 46.9 | 53.2 | 50.4 | 49.3 | 49.1 | 51.0 | 43.0 | 40.4 | 32.7 |
| Kirovohrad region | 15.8 | 21.3 | 23.1 | 26.9 | 26.9 | 27.0 | 29.2 | 27.5 | 27.3 | 27.4 |
| Poltava region | 17.6 | 25.6 | 28.2 | 28.3 | 46.9 | 48.6 | 48.4 | 53.0 | 48.2 | 43.2 |
| Cherkasy region | 26.9 | 32.1 | 34.8 | 34.2 | 36.2 | 36.7 | 35.7 | 35.0 | 38.7 | 34.0 |
| East | 69.2 | 84.1 | 102.3 | 96.3 | 107.0 | 109.8 | 110.3 | 104.1 | 92.3 | 87.7 |
| Donetsk region | 39.2 | 49.4 | 61.1 | 64.7 | 68.3 | 65.5 | 64.5 | 63.6 | 63.5 | 63.6 |
| Luhansk region | 8.3 | 12.2 | 13.0 | 11.8 | 12.7 | 11.4 | 7.0 | 6.0 | 4.2 | 2.2 |
| Kharkiv region | 21.7 | 22.5 | 28.2 | 19.8 | 26.0 | 32.9 | 38.8 | 34.5 | 24.6 | 21.9 |
| North | 81.9 | 94.8 | 107.6 | 106.5 | 111.0 | 113.6 | 116.1 | 118.0 | 124.7 | 123.5 |
| Zhytomyr region | 19.4 | 21.9 | 24.6 | 24.7 | 26.5 | 25.1 | 26.7 | 26.0 | 24.2 | 25.0 |
| Kyiv region | 39.0 | 44.1 | 53.0 | 58.0 | 58.3 | 56.9 | 58.6 | 57.1 | 66.1 | 65.1 |
| Sumy region | 9.7 | 12.3 | 14.0 | 11.3 | 12.0 | 14.0 | 14.8 | 18.1 | 17.8 | 18.4 |
| Chernihiv region | 13.8 | 16.5 | 16.0 | 12.5 | 14.2 | 17.6 | 16.0 | 16.8 | 16.6 | 15.0 |
| South | 106.6 | 120.6 | 140.7 | 136.2 | 132.5 | 99.1 | 97.4 | 89.9 | 82.4 | 76.1 |
| AR Crimea | 30.1 | 34.3 | 39.7 | 37.3 | 36.6 | - | - | - | - | - |
| Zaporizhia region | 31.3 | 34.4 | 41.0 | 39.9 | 37.6 | 39.0 | 37.4 | 33.9 | 27.9 | 25.5 |
| Mykolayiv region | 5.9 | 7.2 | 10.6 | 11.1 | 9.5 | 11.4 | 11.8 | 10.1 | 9.4 | 9.3 |
| Odessa region | 20.8 | 24.4 | 26.1 | 25.6 | 25.0 | 24.8 | 26.2 | 24.7 | 24.4 | 22.4 |
| Kherson region | 18.5 | 20.3 | 23.3 | 22.3 | 23.8 | 23.9 | 22.0 | 21.2 | 20.7 | 18.9 |
| West | 149.5 | 184.7 | 189.0 | 201.9 | 217.7 | 238.2 | 249.4 | 253.2 | 255.6 | 254.6 |
| Volyn region | 31.1 | 38.2 | 35.3 | 33.6 | 35.2 | 34.2 | 36.3 | 37.5 | 37.7 | 38.4 |
| Transcarpathian region | 19.2 | 21.6 | 23.2 | 28.8 | 29.0 | 31.7 | 29.6 | 29.1 | 30.1 | 29.9 |
| Ivano-Frankivsk region | 14.6 | 20.7 | 25.4 | 28.7 | 30.3 | 33.5 | 36.5 | 36.4 | 37.0 | 35.6 |
| Lviv region | 24.8 | 31.5 | 29.8 | 32.4 | 39.1 | 42.6 | 47.7 | 50.6 | 51.1 | 50.3 |
| Rivne region | 21.1 | 22.0 | 20.6 | 20.7 | 20.9 | 21.8 | 21.0 | 21.3 | 22.0 | 22.5 |
| Ternopil region | 11.8 | 17.7 | 19.0 | 19.9 | 23.6 | 28.4 | 32.3 | 30.6 | 31.5 | 33.4 |
| Khmelnytsky region | 10.7 | 15.6 | 17.8 | 20.4 | 21.7 | 28.8 | 29.5 | 29.5 | 28.8 | 28.3 |
| Chernivtsi region | 16.2 | 17.4 | 17.9 | 17.4 | 17.9 | 17.2 | 16.5 | 18.2 | 17.4 | 16.2 |

Sources: own presentation using Ukrstat data

**Table 4A. Poultry production in Ukraine's regions 2009–2018, 1000 tons**

|  | 2009 | 2010 | 2011 | 2012 | 2013 | 2014 | 2015 | 2016 | 2017 | 2018 |
|---|---|---|---|---|---|---|---|---|---|---|
| Ukraine | 894.2 | 953.5 | 995.2 | 1074.7 | 1168.3 | 1164.7 | 1143.7 | 1166.8 | 1184.7 | 1258.9 |
| Center | 379.8 | 454.1 | 468.1 | 496.0 | 575.6 | 665.9 | 703.8 | 750.6 | 753.5 | 807.3 |
| Vinnytsia region | 18.9 | 16.3 | 18.1 | 40.8 | 132.3 | 219.6 | 263.9 | 280.5 | 276.6 | 309.0 |
| Dnipropetrovsk region | 141.1 | 149.7 | 162.9 | 166.3 | 165.1 | 163 | 167.4 | 186.0 | 206.1 | 203.6 |
| Kirovohrad region | 6.3 | 6.9 | 7.1 | 8.1 | 8.1 | 10.1 | 10.0 | 10.4 | 10.0 | 10.5 |
| Poltava region | 10.0 | 8.2 | 5.6 | 10.1 | 7.6 | 5.7 | 4.6 | 5.8 | 5.9 | 5.8 |
| Cherkasy region | 203.5 | 273.0 | 274.4 | 270.7 | 262.5 | 267.5 | 257.9 | 267.9 | 254.9 | 278.4 |
| East | 98.3 | 95.1 | 92.4 | 104.6 | 105.8 | 77.8 | 46.3 | 50.2 | 49.3 | 47.3 |
| Donetsk region | 40.7 | 37.6 | 41.4 | 47.7 | 44.1 | 28.6 | 13.4 | 11.9 | 12.2 | 12.7 |
| Luhansk region | 22.4 | 20.6 | 23.5 | 24.3 | 25.3 | 14.9 | 2.1 | 3.2 | 1.5 | 1.7 |
| Kharkiv region | 35.2 | 36.9 | 27.5 | 32.6 | 36.4 | 34.3 | 30.8 | 35.1 | 35.6 | 32.9 |
| North | 166.9 | 150.3 | 160.5 | 174.2 | 186.4 | 183.9 | 163.7 | 148.3 | 172.8 | 170.5 |
| Zhytomyr region | 3.5 | 4.5 | 4.2 | 4.5 | 5.3 | 7.2 | 8.1 | 8.7 | 9.7 | 10.2 |
| Kyiv region | 150.6 | 131.5 | 142.6 | 154.4 | 163.7 | 158.4 | 139.4 | 123.6 | 145.9 | 141.9 |
| Sumy region | 6.3 | 9.0 | 8.2 | 9.2 | 9.9 | 12 | 13.1 | 13.2 | 14.7 | 15.4 |
| Chernihiv region | 6.5 | 5.3 | 5.5 | 6.1 | 7.5 | 6.3 | 3.1 | 2.8 | 2.5 | 3.0 |
| South | 105.9 | 105.8 | 108.6 | 113.0 | 101.4 | 33.9 | 34.2 | 26.3 | 24.1 | 24.5 |
| AR Crimea | 81.7 | 80.9 | 84.5 | 88.4 | 72.1 | - | - | - | - | - |
| Zaporizhia region | 8.4 | 10.1 | 9.6 | 10.8 | 13.5 | 13.8 | 13.3 | 12.5 | 12.2 | 11.2 |
| Mykolayiv region | 8.5 | 9.1 | 9.4 | 8.2 | 9.5 | 9.7 | 7.5 | 7.3 | 6.2 | 5.9 |
| Odessa region | 3.5 | 2.4 | 2.4 | 2.2 | 2.1 | 2.7 | 2.6 | 2.4 | 2.4 | 2.3 |
| Kherson region | 3.8 | 3.3 | 2.7 | 3.4 | 4.2 | 7.7 | 10.8 | 4.1 | 3.3 | 5.1 |
| West | 143.3 | 148.2 | 165.6 | 186.9 | 199.1 | 203.2 | 195.7 | 191.4 | 185 | 209.3 |
| Volyn region | 39.7 | 41.4 | 43.9 | 47.1 | 56.0 | 67.9 | 72.9 | 74.2 | 64.8 | 67.2 |
| Transcarpathian region | 8.6 | 7.2 | 7.5 | 8.5 | 8.5 | 8.7 | 6.2 | 5.3 | 5.7 | 5.6 |
| Ivano-Frankivsk region | 16.5 | 16.1 | 19.7 | 20.8 | 21.0 | 20.5 | 20.5 | 15.1 | 14.3 | 21.1 |
| Lviv region | 43.8 | 47.6 | 50.5 | 59.8 | 60.3 | 49.9 | 42.7 | 42.2 | 46.5 | 53.8 |
| Rivne region | 11.9 | 12.2 | 17.6 | 18.3 | 20.3 | 20 | 18.3 | 21.2 | 21.9 | 24.4 |
| Ternopil region | 5.8 | 5.7 | 6.5 | 7.7 | 8.3 | 10.5 | 8.8 | 7.7 | 8.3 | 9.6 |
| Khmelnytsky region | 7.7 | 9.3 | 9.5 | 10.9 | 12.6 | 13.4 | 13.1 | 12.9 | 11.8 | 14.7 |
| Chernivtsi region | 9.3 | 8.7 | 10.4 | 13.8 | 12.1 | 12.3 | 13.2 | 12.8 | 11.7 | 12.9 |

Sources: own presentation using Ukrstat data

**Table 5A. Beef production in Ukraine's regions 2009–2018, 1000 tons**

|  | 2009 | 2010 | 2011 | 2012 | 2013 | 2014 | 2015 | 2016 | 2017 | 2018 |
|---|---|---|---|---|---|---|---|---|---|---|
| Ukraine | 453.5 | 427.7 | 399.1 | 388.5 | 427.8 | 412.7 | 384 | 375.6 | 363.5 | 358.9 |
| Center | 89.2 | 84.6 | 76.1 | 74.8 | 87.9 | 90.8 | 85.3 | 81.4 | 75.7 | 74.5 |
| Vinnytsia region | 25.1 | 23.1 | 19.6 | 19.8 | 22.6 | 22.3 | 21.5 | 18.2 | 15.3 | 16.9 |
| Dnipropetrovsk region | 10.2 | 9.9 | 8.9 | 7.9 | 11.9 | 11.2 | 10.5 | 10.0 | 10.7 | 9.7 |
| Kirovohrad region | 16.8 | 16.6 | 14.1 | 12.6 | 14.0 | 15.0 | 12.3 | 13.0 | 12.9 | 12.1 |
| Poltava region | 15.7 | 15.5 | 15.7 | 19.1 | 21.9 | 23.0 | 22.4 | 20.8 | 19.1 | 19.3 |
| Cherkasy region | 21.4 | 19.5 | 17.8 | 15.4 | 17.5 | 19.3 | 18.6 | 19.4 | 17.7 | 16.5 |
| East | 49.4 | 47.0 | 43.8 | 43.9 | 48.3 | 45.5 | 45.7 | 44.4 | 38.0 | 34.7 |
| Donetsk region | 14.5 | 13.0 | 11.6 | 11.5 | 11.2 | 9.3 | 10.8 | 9.2 | 8.4 | 6.9 |
| Luhansk region | 11.4 | 11.3 | 10.1 | 9.3 | 11.1 | 10.2 | 11.7 | 11.0 | 5.1 | 2.8 |
| Kharkiv region | 23.5 | 22.7 | 22.1 | 23.1 | 26.0 | 26.0 | 23.2 | 24.2 | 24.5 | 25.0 |
| North | 69.2 | 66.4 | 62.2 | 62.2 | 70.0 | 66.9 | 60.4 | 59.3 | 58.6 | 57.6 |
| Zhytomyr region | 18.6 | 18.5 | 17.5 | 18.0 | 19.1 | 15.6 | 15.7 | 15.6 | 15.8 | 15.3 |
| Kyiv region | 14.6 | 15.3 | 13.4 | 13.1 | 16.9 | 18.1 | 16.4 | 16.4 | 19.4 | 19.4 |
| Sumy region | 19.0 | 17.2 | 16.6 | 17.7 | 18.7 | 17.4 | 14.7 | 13.1 | 10.7 | 10.7 |
| Chernihiv region | 17.0 | 15.4 | 14.7 | 13.4 | 15.3 | 15.8 | 13.6 | 14.2 | 12.7 | 12.2 |
| South | 79.2 | 77.9 | 70.0 | 66.7 | 68.6 | 56.9 | 53.7 | 53.7 | 54.2 | 60.1 |
| AR Crimea | 22.2 | 22.2 | 17.6 | 15.0 | 15.6 | - | - | - | - | - |
| Zaporizhia region | 12.7 | 11.5 | 11.2 | 9.9 | 10.1 | 10.5 | 9.5 | 9.6 | 10.5 | 10.4 |
| Mykolayiv region | 9.9 | 11.7 | 10.6 | 11.2 | 11.6 | 14.3 | 12.6 | 12.8 | 13.1 | 18.3 |
| Odessa region | 16.6 | 15.4 | 15.3 | 15.2 | 16.0 | 15.9 | 16.2 | 15.8 | 15.2 | 14.8 |
| Kherson region | 17.8 | 17.1 | 15.3 | 15.4 | 15.3 | 16.2 | 15.4 | 15.5 | 15.4 | 16.6 |

|  | 2009 | 2010 | 2011 | 2012 | 2013 | 2014 | 2015 | 2016 | 2017 | 2018 |
|---|---|---|---|---|---|---|---|---|---|---|
| West | 166.5 | 151.8 | 147.0 | 140.9 | 153.0 | 144.5 | 138.9 | 136.8 | 137.0 | 132.0 |
| Volyn region | 16.2 | 14.0 | 11.6 | 11.2 | 12.8 | 13.7 | 10.2 | 9.6 | 9.3 | 7.0 |
| Transcarpathian region | 18.8 | 17.1 | 16.9 | 16.9 | 17.0 | 16.0 | 14.3 | 13.8 | 15.7 | 17.1 |
| Ivano-Frankivsk region | 28.8 | 26.7 | 25.8 | 25.4 | 26.7 | 27.7 | 27.8 | 27.6 | 28.1 | 27.2 |
| Lviv region | 36.3 | 33.1 | 33.5 | 28.2 | 30.8 | 29.0 | 27.9 | 28.0 | 27.2 | 25.1 |
| Rivne region | 15.9 | 14.8 | 16.0 | 15.9 | 16.6 | 17.3 | 13.7 | 11.9 | 11.2 | 10.4 |
| Ternopil region | 15.9 | 12.4 | 11.6 | 11.8 | 14.4 | 13.6 | 12.9 | 13.3 | 12.6 | 12.4 |
| Khmelnytsky region | 22.5 | 22.0 | 21.2 | 20.9 | 23.7 | 16.2 | 21.8 | 22.6 | 23.1 | 23.0 |
| Chernivtsi region | 12.1 | 11.7 | 10.4 | 10.6 | 11.0 | 11.0 | 10.3 | 10.0 | 9.8 | 9.8 |

Sources: own presentation using Ukrstat data

**Table 6A. Raw milk production in Ukraine's regions 2009–2018, 1000 tons**

|  | 2009 | 2010 | 2011 | 2012 | 2013 | 2014 | 2015 | 2016 | 2017 | 2018 |
|---|---|---|---|---|---|---|---|---|---|---|
| Ukraine | 11609 | 11248 | 11086 | 11377 | 11488 | 11132 | 10615 | 10381 | 10280 | 10064 |
| Center | 2756.6 | 2699.3 | 2701 | 2776.3 | 2823.6 | 2877.4 | 2818.2 | 2796.3 | 2757.9 | 2673 |
| Vinnytsia region | 841.9 | 836.1 | 838.5 | 847.4 | 856.9 | 852.0 | 838.4 | 853.6 | 851.3 | 824.8 |
| Dnipropetrovsk region | 359.2 | 339.8 | 341.7 | 343.4 | 348.0 | 357.2 | 344.6 | 319.5 | 300.7 | 294.3 |
| Kirovohrad region | 346.5 | 343.1 | 331.5 | 309.2 | 322.0 | 324.3 | 310.6 | 307.7 | 305.6 | 307.6 |
| Poltava region | 718 | 701.4 | 725.4 | 777.8 | 785.0 | 814.1 | 794.5 | 796.5 | 792.4 | 762.1 |
| Cherkasy region | 491 | 478.9 | 463.9 | 498.5 | 511.7 | 529.8 | 530.1 | 519.0 | 507.9 | 484.2 |
| East | 1157.7 | 1090.7 | 1076.1 | 1128.2 | 1141 | 1060.1 | 911.1 | 846.1 | 837.5 | 839.9 |
| Donetsk region | 360.3 | 339.1 | 327.4 | 332.9 | 324.8 | 283.0 | 227.9 | 192.8 | 190.2 | 186.4 |
| Luhansk region | 316 | 284.4 | 276.0 | 282.0 | 279.5 | 251.6 | 158.7 | 123.8 | 124.8 | 127.0 |
| Kharkiv region | 481.4 | 467.2 | 472.7 | 513.3 | 536.7 | 525.5 | 524.5 | 529.5 | 522.5 | 526.5 |
| North | 2126.2 | 2041 | 2004.7 | 2079.8 | 2083.2 | 2056 | 1994.9 | 1968.7 | 1948.4 | 1934.2 |
| Zhytomyr region | 602.4 | 578.3 | 569.1 | 594.9 | 597.6 | 589.7 | 578.4 | 566.6 | 547.7 | 553.3 |
| Kyiv region | 475.8 | 451.1 | 438.9 | 476.3 | 475.9 | 467.0 | 446.3 | 438.0 | 435.9 | 433.2 |
| Sumy region | 456.2 | 430.5 | 418.3 | 427.3 | 427.3 | 427.1 | 417.6 | 414.6 | 416.0 | 410.5 |
| Chernihiv region | 591.8 | 581.1 | 578.4 | 581.3 | 582.4 | 572.2 | 552.6 | 549.5 | 548.8 | 537.2 |
| South | 1755.2 | 1683.4 | 1643.8 | 1639.2 | 1639.1 | 1345.6 | 1289.8 | 1260.5 | 1244.8 | 1187.7 |
| AR Crimea | 367.2 | 348 | 330.5 | 305.9 | 292.4 |  |  |  |  |  |
| Zaporizhia region | 287 | 261.7 | 248.1 | 257.5 | 264.9 | 267.5 | 260.7 | 259.5 | 260.7 | 243.9 |
| Mykolayiv region | 367.7 | 364.0 | 365.9 | 367.4 | 370.7 | 369.3 | 343.8 | 341.6 | 342.2 | 324.6 |
| Odessa region | 414.4 | 403.8 | 397.3 | 397.9 | 402.3 | 405.9 | 385.3 | 363.3 | 348.6 | 334.8 |
| Kherson region | 318.9 | 305.9 | 302.0 | 310.5 | 308.8 | 302.9 | 300.0 | 296.1 | 293.3 | 284.4 |
| West | 3813.9 | 3729.2 | 3655.7 | 3749.8 | 3796.1 | 3793.7 | 3601.4 | 3509.9 | 3491.9 | 3429.2 |
| Volyn region | 461.7 | 450.2 | 450.5 | 466.5 | 467.0 | 459.3 | 425.2 | 412.4 | 411.9 | 391.1 |
| Transcarpathian region | 385.2 | 391.8 | 389.3 | 401.1 | 410.3 | 409.6 | 358.1 | 320.4 | 325.2 | 346.0 |
| Ivano-Frankivsk region | 485.5 | 465.4 | 451.8 | 466.3 | 470.5 | 483.3 | 474.0 | 466.8 | 463.5 | 441.5 |
| Lviv region | 682.5 | 656.2 | 629.6 | 620.7 | 619.4 | 601.0 | 571.2 | 543.2 | 528.3 | 506.7 |
| Rivne region | 448 | 432.7 | 420.2 | 442.6 | 453.4 | 458.3 | 436.8 | 437.2 | 433.3 | 396.9 |
| Ternopil region | 418.7 | 416.7 | 418.1 | 459.6 | 485.9 | 480.6 | 460.7 | 453.5 | 451.4 | 449.6 |
| Khmelnytsky region | 620.4 | 608.1 | 598.2 | 594.7 | 591.5 | 602.3 | 581.4 | 589.6 | 596.7 | 624.0 |
| Chernivtsi region | 311.9 | 308.1 | 298.0 | 298.3 | 298.1 | 299.3 | 294.0 | 286.8 | 281.6 | 273.4 |

Sources: own presentation using Ukrstat data